\documentclass[preprint]{aastex}
\usepackage{aas_macros}
\usepackage{amsmath}

\let\captionbox\relax
\usepackage{caption}
\usepackage{subcaption}
\usepackage{wasysym}
\usepackage[english]{babel}
\usepackage[section]{placeins}
\usepackage{natbib}
\usepackage{hhline}
\usepackage{multirow}

\begin{document}
\vspace{5cm}

\title{Searching for nuclear obscuration in the infrared spectra of nearby FR-I radio galaxies}
\author{R.~C. Gleisinger}
\affil{Department of Physics and Astronomy, University of Victoria,\\ Victoria, BC V8P~5C2, Canada}
\author{C.~P. O'Dea}
\affil{Department of Physics and Astronomy, University of Manitoba,\\ Winnipeg, MB R3T~2N2, Canada}
\author{J.~F. Gallimore}
\affil{Department of Physics and Astronomy, Bucknell University,\\ Lewisburg, PA 17837, USA}
\author{S. Wykes}
\affil{Independent Researcher}
\and
\author{S.~A. Baum}
\affil{Faculty of Science, University of Manitoba,\\ Winnipeg, MB R3T~2N2, Canada}

\begin{abstract}
\noindent How do active galactic nuclei with low optical luminosities produce powerful radio emission? Recent studies of active galactic nuclei with moderate radio and low optical luminosities (\citealt{FanaroffRiley74} class I, FR-I) searching for broad nuclear emission lines in polarized light, as predicted by some active galactic nucleus unification models, have found heterogeneous results. These models typically consist of a central engine surrounded by a torus of discrete dusty clouds.  These clouds would absorb and scatter optical emission, blocking broad nuclear emission lines, and re-radiate in mid-infrared. Some scattered broad line emission may be observable, depending on geometry, which would be polarized. We present a wide-band infrared spectroscopic analysis of ten nearby FR-I radio galaxies to determine whether there is significant emission from a dusty obscuring structure. We used Markov-chain Monte Carlo algorithms to decompose \emph{Spitzer}/IRS spectra of our sample. We constrained the wide-band behavior of our models with photometry from 2MASS, \emph{Spitzer}/IRAC, \emph{Spitzer}/MIPS, and \emph{Herschel}/SPIRE. We find one galaxy is best fit by a clumpy torus and three others show some thermal mid-infrared component. This suggests that in those three there is likely some obscuring dust structure that is inconsistent with our torus models and there must be some source of photons heating the dust. We conclude that 40\% of our FR-I radio galaxies show evidence of obscuring dusty material, possibly some other form of hidden broad-line nucleus, but only 10\% favor the clumpy torus model specifically. 
\end{abstract}
\newpage
\clearpage
\pagebreak
\section{Introduction}
\label{sec:intro}

\subsection{AGN unification}
\label{sec:AGNuni}

Active galactic nuclei (AGN) show an array of different spectra which are usually grouped into broad categories based on their radio luminosities, bolometric luminosities, and the presence or absence of broad optical emission lines. AGN unification models attempt to explain many of these different classes with a single type of object which differs in only a few parameters. Its particular focus is to explain the presence or absence of optical broad lines in objects which are otherwise similar (e.g., similar radio and X-ray profiles). These models consist of a central engine and some anisotropic distribution of gas and dust which scatters or absorbs light from the central engine. The central engine is matter accreting onto a supermassive black hole. Our description of this model follows the one by \cite{Antonucci93}. For further reading see \cite{Antonucci12,Tadhunter16,Padovani17}. 

In these models, material accreting onto the central supermassive black hole forms an optically thick but geometrically thin accretion disk which is obscured from some lines of sight by an optically thick, warm ($\sim 200-1000\,\text{K}$) mass of dusty material with a roughly toroidal geometry. This torus would absorb optical emission from the central engine and re-radiate that energy in infrared. \cite{Pier92,Pier93} modelled the torus as a smooth continuous cloud of variable density due to computing constraints but they believed it to be composed of discrete clouds (``clumpy").  The alternative to a torus is that the central engine is intrinsically dim in the optical rather than obscured, resulting in a lower mid-infrared luminosity where warm obscuring dust would radiate. Smooth torus models require fine-tuning in order to match the suppressed low-equivalent-width $10\,\mu\text{m}$ silicate features observed in infrared spectra \citep{Granato94,Granato97}.

\cite{Rowan95} claimed that clumpy torus models can adequately explain silicate feature suppression observed in Seyfert 1 galaxies but, just like \cite{Pier92}, lacked a fully 3D radiative transfer code to fully explore it. \cite{nenkova02,nenkova08a,nenkova08} formalized the clumpy torus model we consider in our decompositions in Section~\ref{sec:clumpy}. This model describes an obscuring torus as being composed of small, discrete, optically thick clouds. The inner radius of the torus is determined by the sublimation radius $R_d$ for the component dust grains i.e., the minimum distance from the central engine at which a grain can survive. This sharp boundary is an approximation since larger dust grains can survive significantly closer to the central engine than the inner edge of the main torus \citep{Pier93,Schartmann05}.
The remaining parameters that \cite{nenkova08a,nenkova08} include in their torus model are the optical depth of a single torus clump at 550~nm, $\tau_V$ (assumed to be the same for all clumps), the mean number of clouds along a radial equatorial line, $N_0$, the radial extent of the torus in units of the dust sublimation, $Y$, the inclination of the torus with respect to the line of sight, $i$ (defined such that $0^\circ$ is pole-on), the angular scale height of the torus (effectively the standard deviation of the normally-distributed clumps around the equator), $\sigma$, and the power law index of the radial distribution of clumps, $q$. From these parameters they show how to calculate the half-opening angle of the torus, the torus height-to-radius ratio $H/R=\tan{(\sigma)}$, and the photon escape probability.

 There are alternatives to the unification model in which the central engines of low-luminosity sources intrinsically differ from those of high-luminosity sources. These are largely beyond the scope of this paper but they often involve an advection dominated accretion flow (ADAF), in which a supermassive black hole is accreting at below 1\% of the Eddington rate  and the infalling material does not radiate much of its energy away. See e.g., \cite{Narayan98,BestHeckman12}.

%

%
%
%
%

\subsection{Existing evidence in FR-I radio galaxies}
\label{sec:evidence}

Fanaroff \& Riley class I (FR-I, \citealt{FanaroffRiley74}) radio galaxies are low-luminosity radio galaxies with sub-relativistic jets that are brightest near their cores. FR-I radio galaxies typically do not show significant broad optical line emission though there are exceptions such as 3C~120 \citep{French80,Leipski09,Torrealba12}. The question of how FR-I radio galaxies fit into AGN models can be reduced to two separate but related parts. These are the nature of the accretion structure that powers the radio jet and the structure of any obscuring material. These are related because if FR-Is are powered by an accretion structure with a high intrinsic optical luminosity such as a Keplerian thin accretion disk then the central engine must be obscured in order to match observations.

\cite{Haas04} analyzed 5 to 200~$\mu$m photometry from the photometer on the \emph{Infrared Space Observatory}, ISOPHOT, of 75 objects in the Revised Third Cambridge Catalog of radio galaxies and quasars \citep{3CRR}, including 5 FR-I sources (of which three overlap with our sample). They clearly detect a thermal component in the FR-Is despite the lower sensitivity compared to \emph{Spitzer} \citep{ISO} but it was too cold to be consistent with the unification models we are considering. \cite{Chiaberge99} detected the central core component in 85\% of \emph{Hubble} Space Telescope observations of FR-I radio galaxies and so put an upper limit on the covering fraction of any obscuring torus with the caveat that the base of an optical jet may appear as an alternate source of optical continuum emission. \cite{Whysong04} found that the MIR core component of M~87 is dominated by synchrotron emission from the base of its relativistic jet and \cite{Perlman07} put further limits on a a warm obscuring torus in M~87. These findings effectively rule out the possibility of a significant amount of obscuring warm dust. \cite{VanderWolk10} found no evidence for warm dust tori in any of the eight FR-I sources in their observations from the Very Large Telescope's (VLT) Imager and Spectrometer for the Mid Infrared. Note that one of the FR-I sources in the \cite{VanderWolk10} sample is 3C~270, which is also in our sample. In \emph{Hubble} Space Telescope (\emph{HST}) images of the optical nuclei of a sample of 54 FR-I and 55 FR-II radio galaxies at $z<0.3$, \cite{Kharb04} found that the optical emission in these sources is primarily non-thermal emission from a relativistic jet.  Their fits also showed no significant improvement from including a thermal accretion disk in the FR-II population as a whole but found some improvement from including a disk component in broad-line emitting FR-II. They claim that this is supported by the presence of a ``big blue bump" which they attribute to an accretion disk. \cite{Kharb04} additionally found that the luminosity of the optical nuclei in FR-I radio galaxies show strong dependence on the radio core prominence $R_c$, which is a common statistical indicator of orientation, and from this they conclude that their results are consistent with the lack of a torus in FR-I radio galaxies. In a study conceptually similar to ours, \cite{Gonzalez15} searched for AGN obscuration in \emph{Spitzer}/IRS (see Section \ref{sec:inst}) spectra and \emph{Chandra} photometry of a sample of low-ionization nuclear emission region (LINER) AGN. The dusty torus present in brighter sources disappeared below bolometric luminosities $L_\mathrm{bol} \simeq 10^{42}\,\mathrm{erg}\,\mathrm{s}^{-1}$, consistent with \cite{Elitzur06}, and that the MIR spectra of such sources is dominated by host, jet, and/or ADAF emission.

\cite{Leipski09} found a \emph{Spitzer}/IRS band thermal excess attributed to warm nuclear dust emission in four of the fifteen galaxies with detected optical compact cores in their sample of 25 FR-I radio galaxies. Six of our ten sources --- 3C~31, 3C~66B, NGC~3862, 3C~270, M~84, and M~87 --- overlap with their sample. Of those, they only found thermal MIR excess in the nuclei of 3C~270 and M~84, which they tentatively attribute to an obscuring torus and starburst activity respectively. However, \cite{Wu06} give a star formation rate estimate of $0.20\,\mathrm{M}_{\odot}\,\mathrm{yr}^{-1}$ for M~84 which is inconsistent with a starburst. They also found a MIR excess in 3C~66B, which they attribute to non-thermal emission.  \cite{Leipski09} also found a very low ratio between the $7.7\,\mu\mathrm{m}$ and $11.3\,\mu\mathrm{m}$ PAH lines, which is not typical for starburst galaxies.

In contrast, \cite{Bianchi19} found red and blue wings on line emission in NGC~3147 consistent with the profile of a mildly relativistic thin Keplerian accretion disk. NGC~3147 is an unobscured AGN with an accretion rate far below the Eddington rate ($\sim 10^{-4} L/L_\mathrm{Edd}$) which lacks a broad line region. The detection of an accretion disk in such an object suggests the possibility that extremely low accretion rates can still support thin accretion disks.

Overall, recent studies favor the two-mode models in which some, but not all, FR-I radio galaxies have some sort of obscuring material around their central engines. The remainder are believed to have central engines that are too dim in infrared (IR) and optical to distinguish from their host galaxies. These dim central engines would still produce jets which may appear in the IR as a synchrotron spectrum. This is likely due to environmental factors such as the source of the accreting material. E.g., cold source material may be more likely to form a disk while hot material may be more likely to form an ADAF. The structure of this obscuring material is, however, not necessarily a torus like that described by \cite{nenkova02,nenkova08a, nenkova08}.

\subsection{Overview}
\label{sec:over}

To determine what powers the central engine in FR-I radio galaxies and how they fit into the broader picture of AGN we search for the signature of warm obscuring dust in the infrared spectra of ten nearby FR-I radio galaxies from a well-studied sample (see Section \ref{sec:samp}). Our primary focus is on continuum emission in the band covered by the \emph{Spitzer} Space Telescope's Infrared Spectrograph. We analyse archival spectra using a pair of specialized fitting codes based on Markov chain Monte Carlo techniques. We seek to determine whether there is evidence for obscuration consistent with the clumpy torus model by \cite{nenkova02} in a sample of ten nearby low luminosity radio galaxies. We used Markov-chain Monte Carlo fitting algorithms to fit spectral lines and and decompose the underlying spectra to IR observations. 

We provide an overview of the concepts involved and previous evidence surrounding the possibility of nucleus obscuration in this section. We then describe our sample of FR-I radio galaxies in Section \ref{sec:samp} followed by an overview of the archival data used in this research in Section \ref{sec:inst} along with the processing of our broadband ancillary data. In Section \ref{sec:spectFit} we provide our analysis of the narrow spectral features in the \textit{Spitzer}/IRS spectra for each source, with particular focus on AGN and star formation tracers and the 10 and 18 $\mu$m silicate features. In Section \ref{sec:clumpy} we show the results of fitting various spectral energy distribution models to our entire broadband data set for each source and provide upper limits on any possible torus component (or equivalent mid-infrared thermal component) in those sources which do not require one. In Section \ref{sec:discuss} we discuss the implications of the models which best agree with the observations and compare our results to those of previous authors.

\section{Sample}
\label{sec:samp}
Our sample is the subset from a list compiled by \cite{VerdoesKleijn99} for which there exists low resolution CASSIS spectra (\citealt{cassis}, see Section~\ref{sec:inst}). We list basic information about the sample galaxies in Table~\ref{tab:sample}. The \cite{VerdoesKleijn99} sample consists of all nearby ($v_r<7000\,\mathrm{km}\,\mathrm{s}^{-1}$) radio galaxies in \cite{Nilson73Book}'s Uppsala General Catalogue (UGC) with declination $-5^\circ < \delta < 82^\circ$ that are $\geq 10"$ at $3\sigma$ in VLA A Array maps (to exclude compact sources) and brighter than 150 mJy at 1400 MHz. This sample was extracted from a catalogue of 176 radio galaxies \cite{Condon88} constructed through position coincidence between their Green Bank 1400 MHz sky maps \citep{Condon85,Condon86,Condon88} and the UGC. \cite{Condon88} distinguished between starburst sources and AGN-powered ``monsters" based on radio morphology, infrared-radio parameter $\log (S_{60 \, \mu\mathrm{m}}/S_{1400 \, \mathrm{MHz}})$, and infrared spectral index. \cite{Xu99} provides a detailed description of this sample's radio properties.

\begin{table}[htp] 
\centering
\begin{tabular}{cccccc}
UGC & NGC & Alternate & RA &  Dec & Redshift \\ 
Number & Number & Name & (J2000.0) & (J2000.0) &  \\
\hhline{======}
\vspace{-0.4cm}\\
597 & 315 & & 00h 57m 48.9s & +30d 21m 09s & 0.016485 \\

689 & 383 & 3C 31 & 01h 07m 24.9s & +32d 24m 45s & 0.017005 \\

1004 & 541 &  & 01h 25m 44.3s & -01d 22m 46s & 0.018086 \\

1841 & & 3C 66B & 02h 23m 11.4s & +42d 59m 31s & 0.021258 \\

6635 & 3801 & & 11h 40m 16.9s & +17d 43m 41s & 0.011064 \\

6723 & 3862 & 3C 264  & 11h 45m 05.0s & +19d 36m 23s & 0.021718 \\

7360 & 4261 & 3C 270 & 12h 19m 23.2s & +05d 49m 31s & 0.007378 \\

7494 & 4374 & M 84, 3C 272.1 & 12h 25m 03.7s & +12d 53m 13s & 0.003392 \\

7654 & 4486 & M 87, 3C 274 & 12h 30m 49.4s & +12d 23m 28s & 0.004283 \\

11718 & 7052 & & 21h 18m 33.0s & +26d 26m 49s & 0.015584 \\
\hline 

\vspace{-0.7cm}
\end{tabular} 
\caption{Basic sample information. Redshifts, Right Ascension (RA) and Declination (Dec) by \cite{Anderson05} and accessed from the NASA/IPAC Extragalactic Database. \label{tab:sample}}
\end{table}

\begin{table}[htp]
\centering
\begin{tabular}{ccccc}
Source & $L_{\nu,\,1.4\,\mathrm{GHz}}$ & $M_\mathrm{BH}$ & $L_\mathrm{Edd}$ & $\log_{10} R_c$ \\
 & $(10^{31}\,\mathrm{erg}\,\mathrm{s}^{-1}\,\mathrm{Hz}^{-1})$ & $(10^8\,\mathrm{M}_\odot)$ & $(10^{47}\,\mathrm{erg}\,\mathrm{s}^{-1})$ & \\
 \hhline{=====}
 NGC 315 & 1.26 & 14.6 & 1.84 & -0.39 \\
  3C 31 & 3.24 & 9.24 & 1.16 & -1.34 \\
 NGC 541 & 0.871 & 2.03 & 0.226 & - \\
 3C 66B & 8.71 & 18.6 & 2.34 & -1.29 \\
 NGC 3801 & 0.309 & 1.53 & 0.193 & - \\
 NGC 3862 & 5.62 & 4.66 & 0.587 & -1.00 \\
 3C 270 & 2.51 & 7.75 & 0.976 & -1.44 \\
 M 84 & 0.224 & 7.30 & 0.920 & -1.18 \\
 M 87 & 7.94 & 22.5 & 2.84 & -1.24 \\
 NGC 7052 & 0.110 & 3.62 & 0.456 & -0.22 \\
 \hline
\end{tabular}
\caption{$1.4\,\mathrm{GHz}$ radio luminosity density, black hole mass estimate, Eddington luminosity, and logarithm of core-extended brightness ratio (an indicator for AGN orientation in unified models) for FR-I AGN in our sample. $L_{\nu,\,1.4\,\mathrm{GHz}}$ by \cite{Condon88},  $M_\mathrm{BH}$ by \cite{Noel07} based on a relation by \cite{Merritt01}, and $\log_{10} R_c$ by \cite{Kharb04}. \label{tab:radioSamp}}
\end{table}

\begin{table}
\centering
\begin{tabular}{cccc}
Source & Redshift & Luminosity & Scale\\ 
  &  & Distance (Mpc) & (kpc/arcsec)\\
\hhline{====}
\vspace{-0.4cm}\\
NGC 315 & 0.016485 & 71.50 & 0.3466\\

3C 31 & 0.017005 & 73.78 & 0.3577\\

NGC 541 & 0.018086 & 78.54 & 0.3808\\

3C 66B & 0.021258 & 92.53 & 0.4486\\

NGC 3801 & 0.011064 & 47.79 & 0.2317\\

NGC 3862 & 0.021718 & 94.57 & 0.4585\\

3C 270 & 0.007378 & 31.78 & 0.1541\\

M 84 & 0.003392 & 14.56 & 0.07059\\

M 87 & 0.004283 & 18.40 & 0.08921\\

NGC 7052 & 0.015584 & 76.54 & 0.3711\\
\hline 
\end{tabular}
\caption{Distance and scale information for sample. We calculated luminosity distances from redshifts using conversion software by the \cite{astropy} for a flat-$\Lambda$ cold dark matter cosmology with Hubble constant $H_0 = 70\,\text{km}\,\text{s}^{-1}$ and an energy density composition of 30\%  ordinary baryonic matter and dark matter ($\Omega_\mathrm{M}=0.3$) and 70\% dark energy ($\Omega_\Lambda = 0.7$). \label{tab:sampDist}}
\end{table}

\begin{table}
\centering
\begin{tabular}{cccc}
Source & Angular Extent & Physical Extent & Reference\\ 
  &  & (Mpc) &  \\
\hhline{====}
\vspace{-0.4cm}\\
NGC 315 & 52' & 1.1 & \cite{Mack97}\\

3C 31 & 45' & 0.96 & \cite{Laing08}\\

NGC 541 & 3' & 0.068 & \cite{Bogdan11}\\

3C 66B & 11.5' & 0.310 & \cite{Hardcastle96}\\

NGC 3801 & 40" & 0.0093 & \cite{Heesen14}\\

NGC 3862 & 8.7'& 0.239 & \cite{Bridle81}\\

3C 270 & 7.8' & 0.072 & \cite{Kolokythas15}\\

M 84 & 3.142' & 0.01331 & \cite{Laing87} \\

M 87 & 13.9' & 0.0744 & \cite{Owen00}\\

NGC 7052 & 2' & 0.044 & \cite{Morganti87}\\
\hline 
\end{tabular}
\caption{Largest extent of radio source. \label{tab:sampRadExt}}
\end{table}

The original paper on this sample by \cite{VerdoesKleijn99} investigated broad- and narrow-band images of 19 of the 21 radio galaxies in its sample from the Wide-Field Planetary Camera 2 aboard \emph{HST}. \cite{VerdoesKleijn02} searched the same sample for relations between radio and optical continua and optical H$\alpha+[\mathrm{N II}]$ emission. \cite{Noel03} analyzed mid-resolution spectra of emission-line-emitting gas in FR-I nuclei from the Space Telescope Imaging Spectrograph aboard \emph{HST}. \cite{Noel07} followed up on this with further analysis of the gas kinematics in the same spectra in order to assess the viability of using gas disk kinematics to estimate central black hole masses. \cite{VerdoesKleijn05} compared dust distributions and their relation to radio jet orientation in the cores of the UGC FR-I sample to two comparison samples. \cite{Capetti05} looked at \emph{HST} optical narrow band images of emission line regions in a sample of 47 galaxies which includes the FR-I UGC sample. The most recent study on the UGC FR-I sample was by \cite{Kharb12}. They examined 1.6 and 5 GHz images of 19 out of the 21 radio galaxies from the Very Large Baseline Array as well as archival \emph{Chandra} X-ray images of 14 sources and new \emph{Chandra} X-ray images of UGC~408. 

\section{Archival data}
\label{sec:inst}

All data for this analysis are archival observations. The central dataset is a set of ten low-resolution \emph{Spitzer}/IRS spectra which cover the waveband where a clumpy torus is expected to be the brightest. Since the nucleus is not resolved by \emph{Spitzer}/IRS, we also need to constrain the contribution from the host galaxy. The stellar population models are constrained by photometry from the Two Micron All Sky Survey and \emph{Spitzer}/IRAC, which extend the band from the IRS short-wave limit of $\sim 5\,\mu\mathrm{m}$ out to $\sim 1\,\mu\mathrm{m}$ where stellar emission is more prominent. To constrain emission from the interstellar medium we use photometry from \emph{Spitzer}/MIPS and \emph{Herschel}/SPIRE (where available) to extend our coverage from the IRS long-wave limit of $\sim 35\,\mu\mathrm{m}$ out to $\sim 500\,\mu\mathrm{m}$. In this section we describe the processing of each of these datasets in order of increasing wavelength.

The Two Micron All Sky Survey (2MASS), as described by \cite{skrutskie06}, was a J, H, and K-band survey covering 99.998\% of the sky performed between June 1997 and February 2001. Observations were performed by two $1.3\,\text{m}$ telescopes in Mount Hopkins, Arizona, USA and in Cerro Tololo, Chile. The point source extractions used an instantaneous point-spread function width of 4" with sky background subtraction in an annulus with inner radius of 14" and an outer radius of 20". Extended source extractions used a fiducial ellipse determined by the 20 mag/$\text{arcsec}^2$ isophote for each source. We list the semi-major axes for these ellipses in Table \ref{tab:2MASSext}. Both sets of extractions are described in detail by \cite{2MASSexplSup}.

For 2MASS data we use point source extractions of our target AGN as a lower bound to the estimated flux which would pass through an aperture the size of the \emph{Spitzer}/IRS SL module's aperture and extended source extractions as an upper bound. This is because the nucleus itself should be an unresolved point source whereas emission from the host galaxy will dominate on larger scales.  3C 66B is not in the 2MASS Extended Source Catalog so we replaced it with extended source photometry by \cite{3C66Bext}.

\begin{table}[htp]
\centering
\begin{tabular}{ccccc}
Source & Semi-major axis & $J, \,1.235\,\mu\mathrm{m}$ & $H, \,1.662\,\mu\mathrm{m}$ & $K_s,\,2.159\,\mu\mathrm{m}$ \\
 & (arcsec) & (ap. mag) & (ap. mag) & (ap. mag)\\
\hhline{=====}
NGC 315 & 65.6 & $9.044\pm 0.012$ & $8.301\pm 0.016$ & $8.050\pm 0.017$\\
3C 31 & 49.9 & $9.619\pm0.014$ & $8.894\pm 0.018$ & $8.599\pm 0.023$ \\
NGC 541 & 37.8 & $10.462\pm 0.017$ & $9.772 \pm 0.026$ & $9.484\pm 0.034$ \\
3C 66B* & - & - & - & - \\
NGC 3801 & 49.9 & $9.999\pm 0.013$ & $9.264\pm 0.020$ & $9.004\pm 0.020$ \\
NGC 3862 & 30.7 & $10.707\pm 0.021$ & $9.976\pm 0.025$ & $9.720\pm 0.034$ \\
3C 270 & 79.9 & $8.355\pm 0.016$ & $7.652\pm 0.016$ & $7.407\pm 0.017$ \\
M 84 & 115.0 & $7.260 \pm 0.015$ & $6.587\pm 0.016$ & $6.222\pm 0.023$ \\
NGC 7052 & 61.9 & $9.628\pm 0.012$ & $8.934\pm 0.015$ & $8.653\pm 0.020$ \\
\hline
\end{tabular}
\caption{2MASS extended source extraction information. Semi-major axes refer to the $K_s$-band 20 mag/$\text{arcsec}^2$ isophotal fiducial ellipse used in the extraction. *not in extended source catalog \label{tab:2MASSext}}
\end{table}

Our \emph{Spitzer}/IRAC data are archival images from the \emph{Spitzer} Heritage Archive. We list the astronomical observation request key (AORKEY) for each source in Table \ref{tab:AOR}. For each source we performed a point-spread function fit to provide a lower bound to the flux and an aperture extraction with the same 4" aperture as 2MASS as an upper bound. Both extractions were centered on the CASSIS source centroid. For this purpose we used the Overlap, Mosaicking, and Multiframe aperture extraction of the Mosaicking and Point Source Extraction (MOPEX) software by \cite{mopex} to combine the IRAC frames (and later MIPS frames) into one mosaic and perform aperture photometry on the result.  All ten sources had usable images for channel 2 (effective wavelength $\lambda_{eff}=4.493\,\mu$m) and all but M 84 had usable images for channel 1 ($\lambda_{eff}=3.550\,\mu$m). We list the IRAC photometry in Table 
\ref{tab:phot2}.

\begin{table}[htp]
\centering
\begin{tabular}{cccc}
Channel & Isophotal wavelength* & Bandpass  & Mean FWHM\\
 & ($\mu$m) & ($\mu$m) & (arcsec)\\
\hhline{====}
1 & 3.550 & 3.19-3.94 & 1.66\\
2 & 4.493 & 4.00-5.02 & 1.72\\
3 & 5.731 & 4.98-6.41 & 1.88\\
4 & 7.872 & 6.45-9.34 & 1.98\\
\hline 
\end{tabular}
\caption{IRAC instrument specifications. Based on the work of \cite{Spitzer} and \cite{IRAC}. *see definition by \cite{IRAC} --- in short, the isophotal wavelength is the wavelength assigned to the flux density of the broadband measurement.\label{tab:IRACspecs}}
\end{table}

\begin{table}
\centering
\begin{tabular}{ccc}
Source  & IRAC AOR & MIPS AOR \\
\hhline{===}
\vspace{-0.4cm}\\
NGC 315  & 19169024 & 4427776 \\

3C 31  & 14805760 & 21681920/4691968* \\

NGC 541  & 19170048 & 4345088 \\

3C 66B  & 14806016 & 10927360 \\

NGC 3801  & 19170304 & 19168512 \\

NGC 3862  & 14807552 & 4692480 \\

3C 270  & 14807808 & 4692736 \\

M 84 &  4671744 & 4692992 \\


NGC 7052  &18257664 & 19168768 \\
\hline
\end{tabular}
\caption{\emph{Spitzer} AORKEYs 
 for ancillary data used in our analysis. M 87 photometry came from a different data set (see Table \ref{tab:M87phot}) and so that object is excluded from this table. *Due to an issue with channel 1 in 4691968, we use the channel 1 data from 21681920 for our analysis of 3C 31. } 
\label{tab:AOR}
\end{table}

\begin{table}
\centering
\begin{tabular}{ccccc} 
Wavelength  & Flux & Aperture & Limit & Reference \\
($\mu$m) & (mJy) & (arcsec) & & \\
\hhline{=====}
\vspace{-0.4cm}\\
 $1.25$ & $227. \pm 3.$ & 14 & Upper & \cite{skrutskie06} \\
 $1.25$ & $78. \pm 2.$ & 7.5 & Lower & \cite{M87_Low125} \\
 $1.64$ & $284. \pm 4.$ & 14 & Upper & \cite{skrutskie06} \\
 $1.65$ & $98. \pm 3.$ & 7.5 & Lower & \cite{M87_Low125} \\
 $2.17$ & $237.\pm 3.$ & 14 & Upper & \cite{skrutskie06} \\
 $2.20$ & $77.\pm 2.$ & 7.5 & Lower & \cite{M87_Low125}\\
 $4.49$ & $24.2$ & 5.8 & Lower & \cite{SpitzerList} \\
 $71.3$ & $480 \pm 40$ & 18 & Upper & \cite{M87MIPS71} \\
 $156$ & $580 \pm 10$ & 40 & Upper & \cite{M87MIPS156} \\
 $250$ & $900\pm 200$ & 18 & Upper & \cite{M87Spire} \\
 $350$ & $1000 \pm 300$ & 25 & Upper & \cite{M87Spire} \\
 $500$ & $1400\pm 400$ & 36 & Upper & \cite{M87Spire} \\ 
\hline
\end{tabular}
\caption{Ancillary photometry data for M 87.} 
\label{tab:M87phot}
\end{table}

\begin{table}
\centering
\begin{tabular}{llccccc} 
Source  & Channel & Limit & IRAC flux & MIPS flux & SPIRE flux\\
  &  &  & ($\text{mJy}$) & ($\text{mJy}$) & ($\text{mJy}$)\\
\hhline{=======}
\vspace{-0.4cm}\\
\multirow{5}{2cm}{NGC 315} & \multirow{2}{1cm}{1} & Upper & $23.20\pm 0.01$ & - & - \\
 & & Lower & $10.02\pm 0.02$& - & - \\
& \multirow{2}{1cm}{2} & Upper & $14.93\pm 0.01$ & $342.\pm 6.$ & - \\
 & & Lower & $7.65\pm 0.02$ & - & - \\
 & \multirow{1}{1cm}{3} & Upper & - & $471.\pm 3.$ & - \\
 \hline
\multirow{5}{2cm}{3C 31} & \multirow{2}{1cm}{1} & Upper & $21.99$  & - & $954.\pm 5.$ \\
 & & Lower & $8.69 \pm 0.01$& - & - \\
& \multirow{2}{1cm}{2} & Upper & $13.45$ & $439.\pm 2.$ & $423.\pm 5.$ \\
 & & Lower & 5.046 & - & - \\
 & \multirow{1}{1cm}{3} & Upper & - & $931.\pm 5.$ & $170.\pm 6.$ \\
 \hline
 \multirow{5}{2cm}{NGC 541} & \multirow{2}{1cm}{1} & Upper & $12.26\pm 0.01$ & - & $80\pm 10$ \\
 & & Lower & $5.87\pm 0.01$ & - & - \\
& \multirow{2}{1cm}{2} & Upper & $7.437\pm 0.006$ & $125.\pm 3.$ & $40\pm 10$ \\
 & & Lower & $3.56\pm 0.01$ & - & - \\
 & \multirow{1}{1cm}{3} & Upper & - & $82.\pm 3.$ & $60\pm 60$ \\
 \hline
\multirow{4}{2cm}{3C 66B} & \multirow{2}{1cm}{1} & Upper & $10.72$ & - & - \\
 & & Lower & $4.70\pm 0.05$ & - & - \\
& \multirow{2}{1cm}{2} & Upper & $6.545\pm 0.003$ & - & - \\
 & & Lower & $2.21 \pm 0.004$ & - & - \\
 \hline
\multirow{5}{2cm}{NGC 3801} & \multirow{2}{1cm}{1} & Upper & $15.79$ & - & - \\
 & & Lower & $8.40\pm 0.01$ & - & - \\
& \multirow{2}{1cm}{2} & Upper & $9.417 \pm 0.005$ & - & - \\
 & & Lower & $4.787 \pm 0.008$ & - & - \\
 & \multirow{1}{1cm}{3} & Upper & - & $1154.\pm 3.$ & - \\
\hline
\end{tabular}
\caption{Ancillary photometry for first five sources sorted by right ascension. Continued for four out of five remaining sources in Table \ref{tab:phot1}.} 
\label{tab:phot1}
\end{table}

\begin{table}
\centering
\begin{tabular}{llccccc} 
Source  & Channel & Limit & IRAC & MIPS & SPIRE\\
  &  &  & ($\text{mJy}$) & ($\text{mJy}$) & ($\text{mJy}$)\\
\hhline{=======}
\vspace{-0.4cm}\\
\multirow{5}{2cm}{NGC 3862} & \multirow{2}{1cm}{1} & Upper & $14.09$ & - & - \\
 & & Lower & $7.838\pm 0.006$ & - & - \\
& \multirow{2}{1cm}{2} & Upper & $9.265 \pm 0.003$ & $192. \pm 3.$ & - \\
 & & Lower & $5.878\pm 0.006$  & - & - \\
 & \multirow{1}{1cm}{3} & Upper & - & $233. \pm 2.$ & - \\
 \hline
\multirow{5}{2cm}{3C 270} & \multirow{2}{1cm}{1} & Upper & $43.32 \pm 0.02$ & - & $200 \pm 10$ \\
 & & Lower & $17.57 \pm 0.02$  & - & - \\
& \multirow{2}{1cm}{2} & Upper & $25.45 \pm 0.01$ & $144.4\pm 0.7$ & $200 \pm 20.$ \\
 & & Lower & $10.50 \pm 0.01$ & - & - \\
 & \multirow{1}{1cm}{3} & Upper & - & $155. \pm 2.$& $200\pm 20$ \\
 \hline
\multirow{4}{2cm}{M 84} &
\multirow{1}{2cm}{1} & Upper & - & - & $250 \pm 20$ \\
 & \multirow{2}{2cm}{2} & Upper & $41.53 \pm 0.02$ & - & $150 \pm 20$ \\
 & & Lower & $21.49 \pm 0.02$ & - & - \\
 & \multirow{1}{2cm}{3} & Upper & - & - & $140\pm 20$ \\
 \hline
\multirow{5}{2cm}{NGC 7052} & \multirow{2}{1cm}{1} & Upper & $20.12 \pm 0.01$ & - & - \\
 & & Lower & $8.38 \pm 0.01$ & - & - \\
& \multirow{2}{1cm}{2} & Upper & $12.07$ & $400.\pm 7.$ & - \\
 & & Lower & $5.23 \pm 0.01$ & - & - \\
 & \multirow{1}{1cm}{3} & Upper & - & $812. \pm 5.$ & - \\
\hline
\end{tabular}
\caption{Ancillary photometry for all remaining sources not included in Table \ref{tab:phot1} except for M 87. We show the ancillary photometry for M 87 in Table \ref{tab:M87phot}.} 
\label{tab:phot2}
\end{table}

All of our IRS spectra, which cover a total wavelength coverage from 5.3-38 $\mu$m, are tapered column extractions from the low-resolution catalog of the Combined Atlas of Sources with \emph{Spitzer}/IRS Spectra (CASSIS, formerly the Cornell Atlas of
Spitzer/IRS Sources) by \cite{cassis}.  

Similar to our \emph{Spitzer}/IRAC data, our \emph{Spitzer}/MIPS data are archival images from the \emph{Spitzer} Heritage Archive for which we list the AORKEY in Table \ref{tab:AOR}. The photometry extraction pipeline for MIPS in MOPEX is also similar to that for IRAC. However, the aperture extraction module in MOPEX failed so we used the point-source flux only. The point-spread function for the 70 and 160 $\mu$m bands is wider than for IRS so the point source flux provided an upper bound on the far-infrared flux of our fits.  Only M 84 had no usable MIPS photometry and NGC 3801 had no usable 70 $\mu$m data. Unfortunately, although MIPS had spectroscopic capabilities, there were no MIPS spectra for the galaxies in our sample. 

The Spectral and Photometric Imaging Receiver (SPIRE) aboard \emph{Herschel} has imaging bands at wavelengths of 250, 350, and 500 $\mu$m with full width at half maximum (FWHM) beam widths of 18", 25", and 36" respectively.  SPIRE also has an imaging Fourier transform spectrometer  but, as with \emph{Spitzer}/MIPS, only photometry observations are available of any of our sample galaxies.

We used archival \emph{Herschel}/SPIRE point source extractions from the SPIRE Point Source Catalog by \cite{SPIREpoint}. We list the observation identification number (obsid) for the two sources in Table \ref{tab:AOR}. The SPIRE Point Source Catalog reports fluxes and not magnitudes so no further processing is necessary.

\FloatBarrier
\section{Spectral feature fitting}
\label{sec:spectFit}

We have developed a modified version of PAHFIT \citep{pahfit}, pahfitMCMC, to measure narrow spectral features and subtract them from {\em Spitzer}/IRS spectra. These narrow line features include mainly fine structure lines and ro-vibrational transitions of H$_2$. The challenge is that narrow lines are undersampled in low-resolution IRS spectra and may sit atop broad spectral features such as PAH emission or silicate (10 and 18 $\mu$m) absorption or emission, which makes continuum subtraction difficult. PAHFIT uses well-established models for PAH and silicate features, in addition to models for starlight and thermal dust continuum, to provide an estimate of the spectral baseline for narrow line features.

In pahfitMCMC, there are two primary modifications over the original PAHFIT. Firstly, we include additional spectral features that may contribute to the infrared spectra of AGN, namely, high-ionization fine structure lines and a simple model for optically thin, warm silicate emission \citep{Gallimore10}. Secondly, we adopted a Markov Chain Monte Carlo (MCMC) approach to parameter estimation. Fitting and {\em Spitzer}/IRS spectrum involves allowing the code to take random steps through parameter space; for this purpose, we used the DREAM(Z) algorithm of \cite{dreamzs}, which efficiently samples high-dimensional parameter spaces. Solutions were accepted or rejected using the Metropolis algorithm \citep{Metropolis}. Initial, out-of-equilibrium solutions were identified by inspection of posterior log-likelihood chains and discarded. The best fit solutions for parameters such as line strength and equivalent width are determined by averages and standard deviations of the posterior chains. 

In summary, pahfitMCMC reports fitted line strengths and equivalent widths for narrow emission lines,  fitted line strengths and equivalent widths for PAH and silicate features, a line-subtracted spectrum for each source (to which we fit continuum models in Section \ref{sec:clumpy}, the simple continuum models at this step are not diagnostic), and Bayesian marginalized error estimates for all fits. We show pahfitMCMC fits for our sample in Figure \ref{fig:pahFits} and then discuss important diagnostics.

\begin{figure}
\centering
\begin{subfigure}{.5\textwidth}
  \centering
  \includegraphics[width=.95\linewidth]{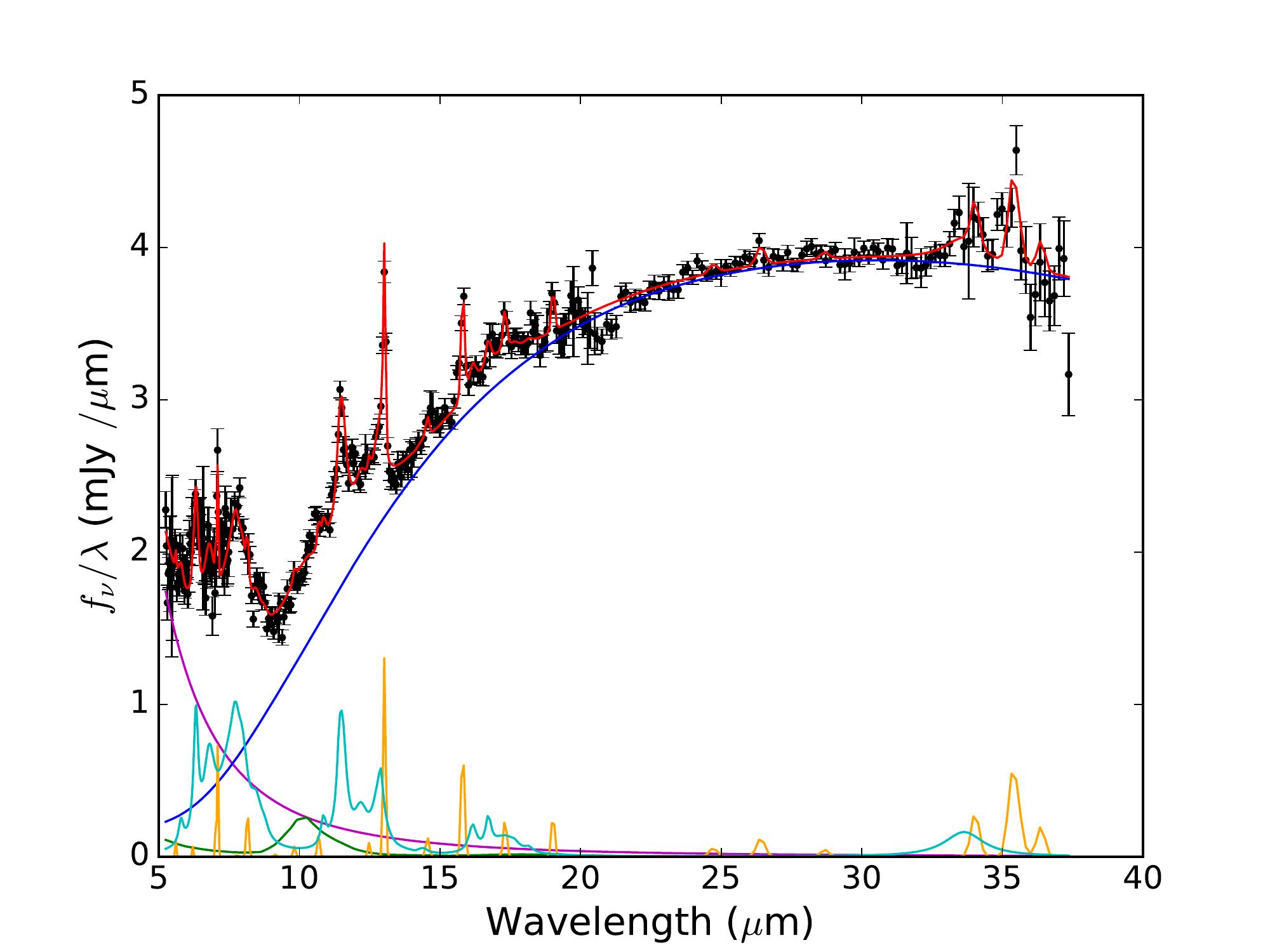}
  \captionof{figure}{NGC 315}
  \label{fig:PAH-NGC0315}
\end{subfigure}%
\begin{subfigure}{.5\textwidth}
  \centering
  \includegraphics[width=.95\linewidth]{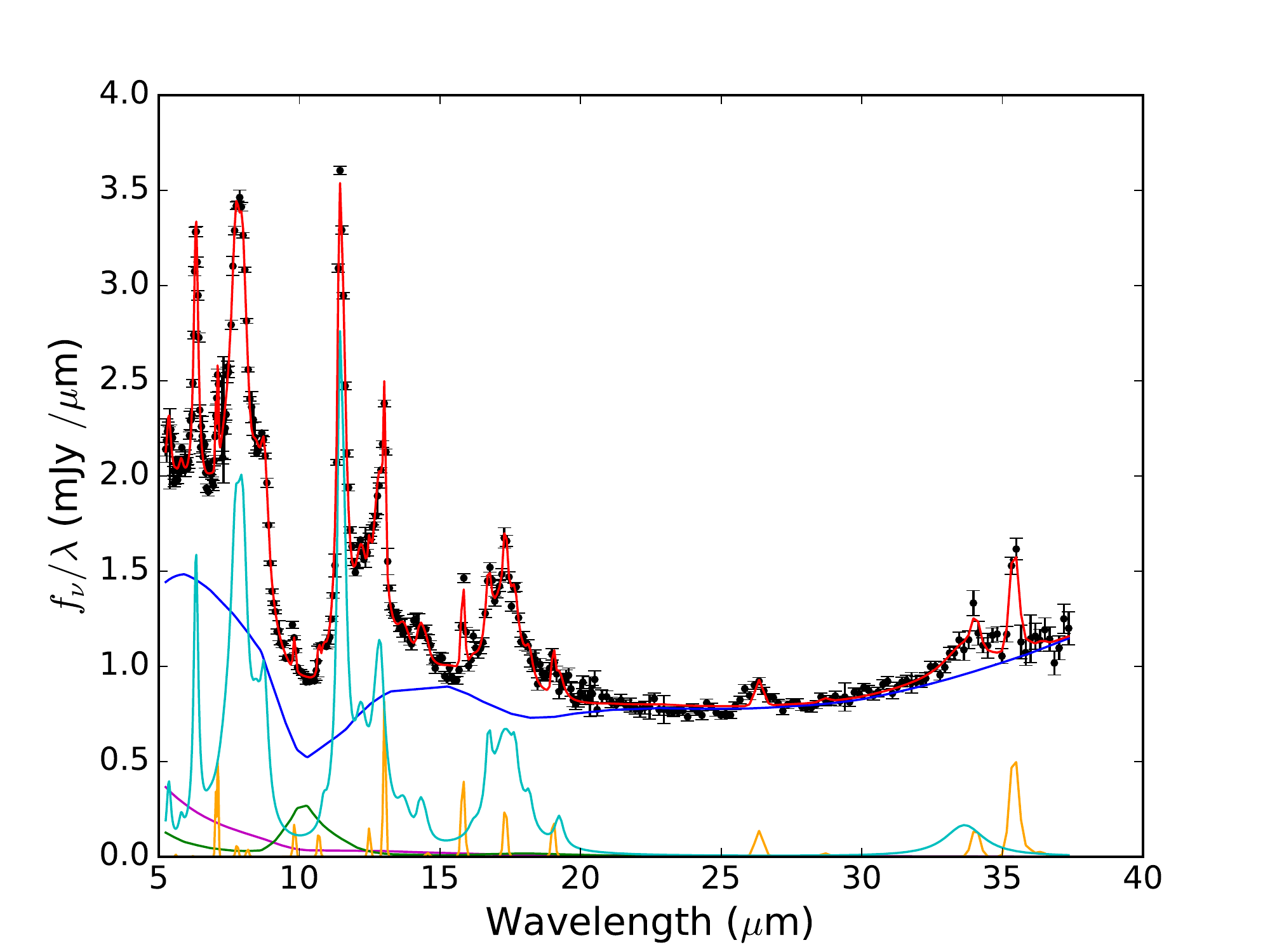}
  \captionof{figure}{3C 31}
  \label{fig:PAH-3C31}
\end{subfigure}%

\centering
\begin{subfigure}{.5\textwidth}
  \centering
  \includegraphics[width=.95\linewidth]{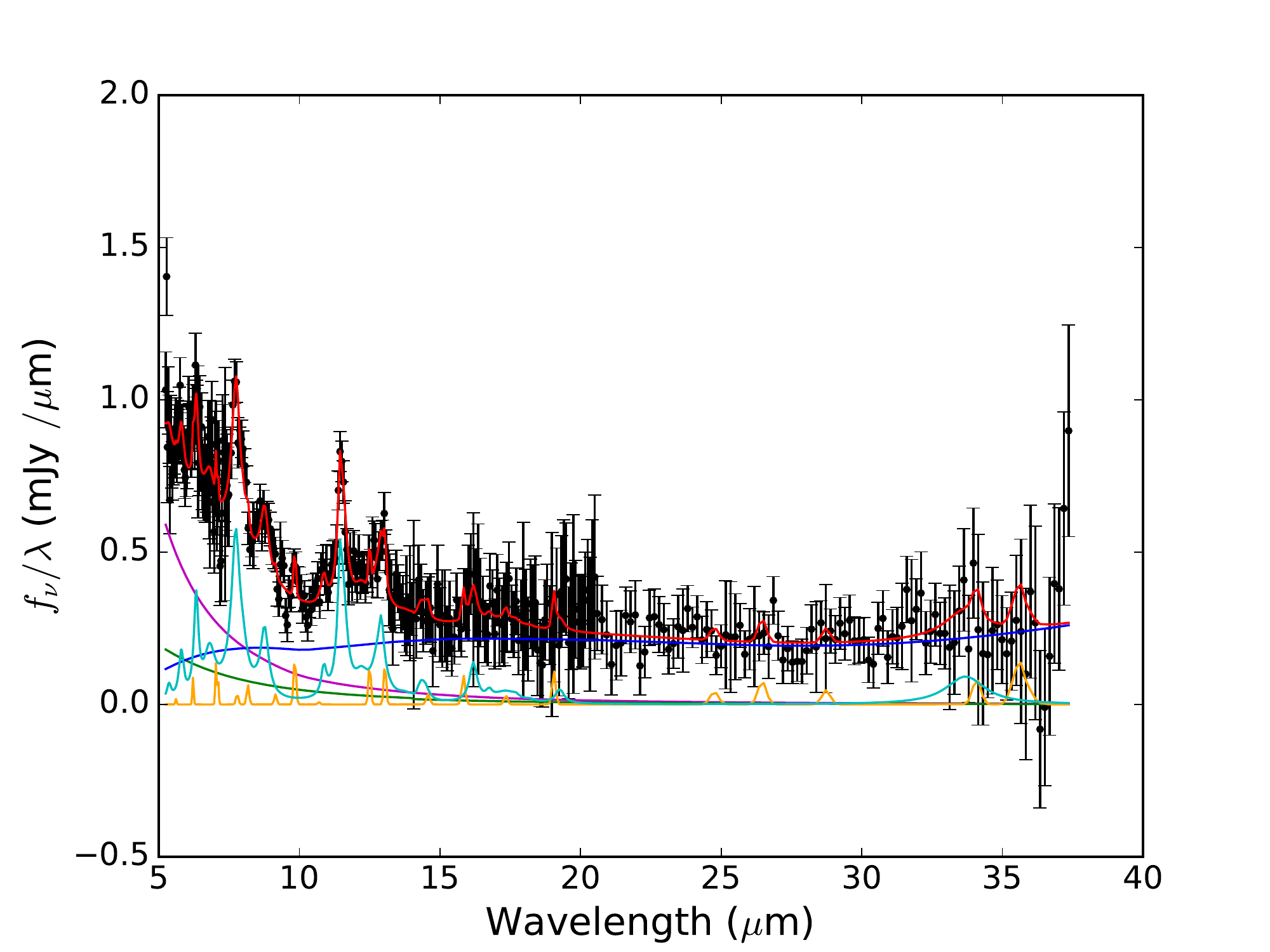}
  \captionof{figure}{NGC 541}
  \label{fig:PAH-NGC0541}
\end{subfigure}%
\begin{subfigure}{.5\textwidth}
  \centering
  \includegraphics[width=.95\linewidth]{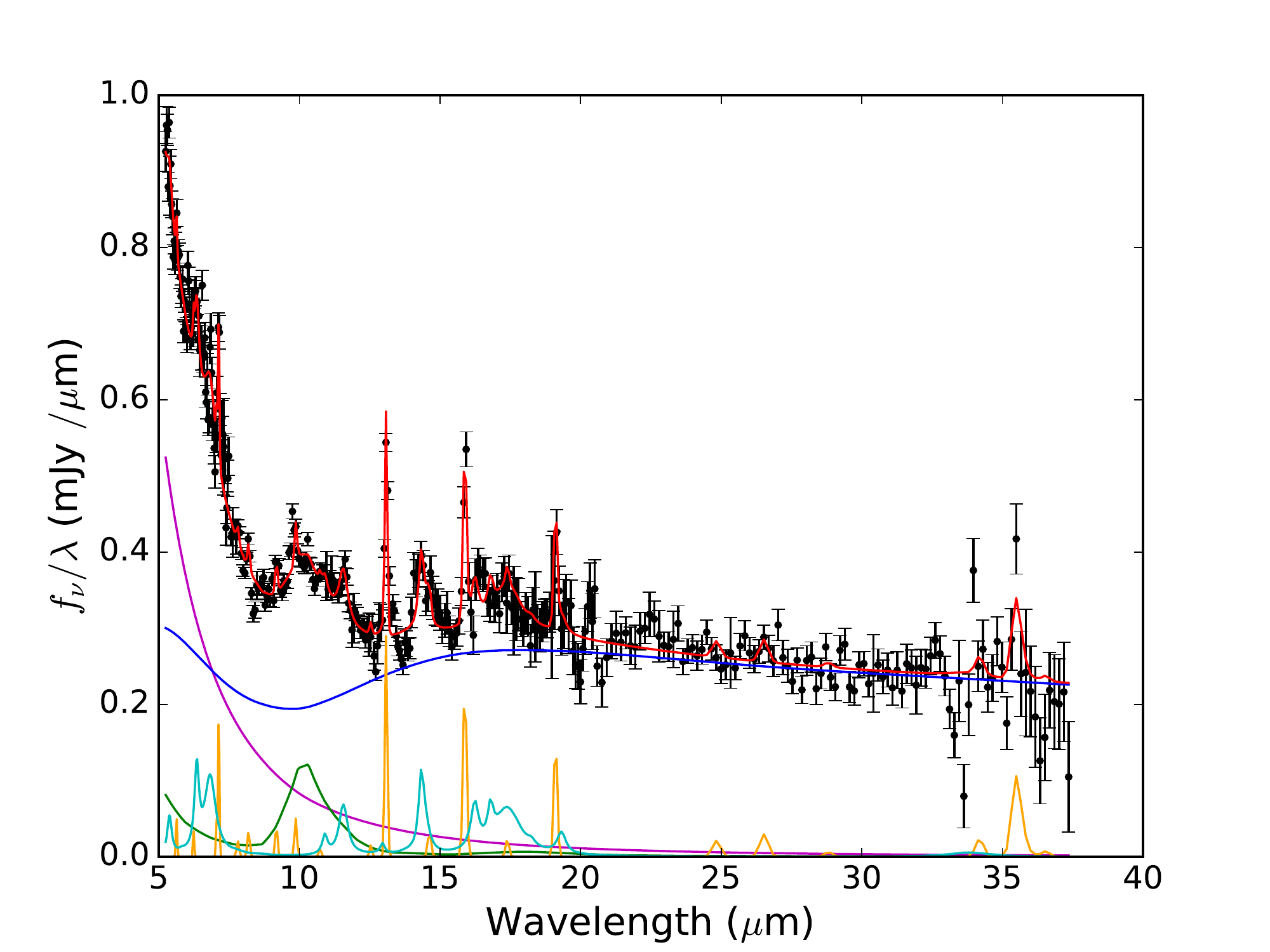}
  \captionof{figure}{3C 66B}
  \label{fig:PAH-UGC01841}
\end{subfigure}%

\centering
\begin{subfigure}{.5\textwidth}
  \centering
  \includegraphics[width=.95\linewidth]{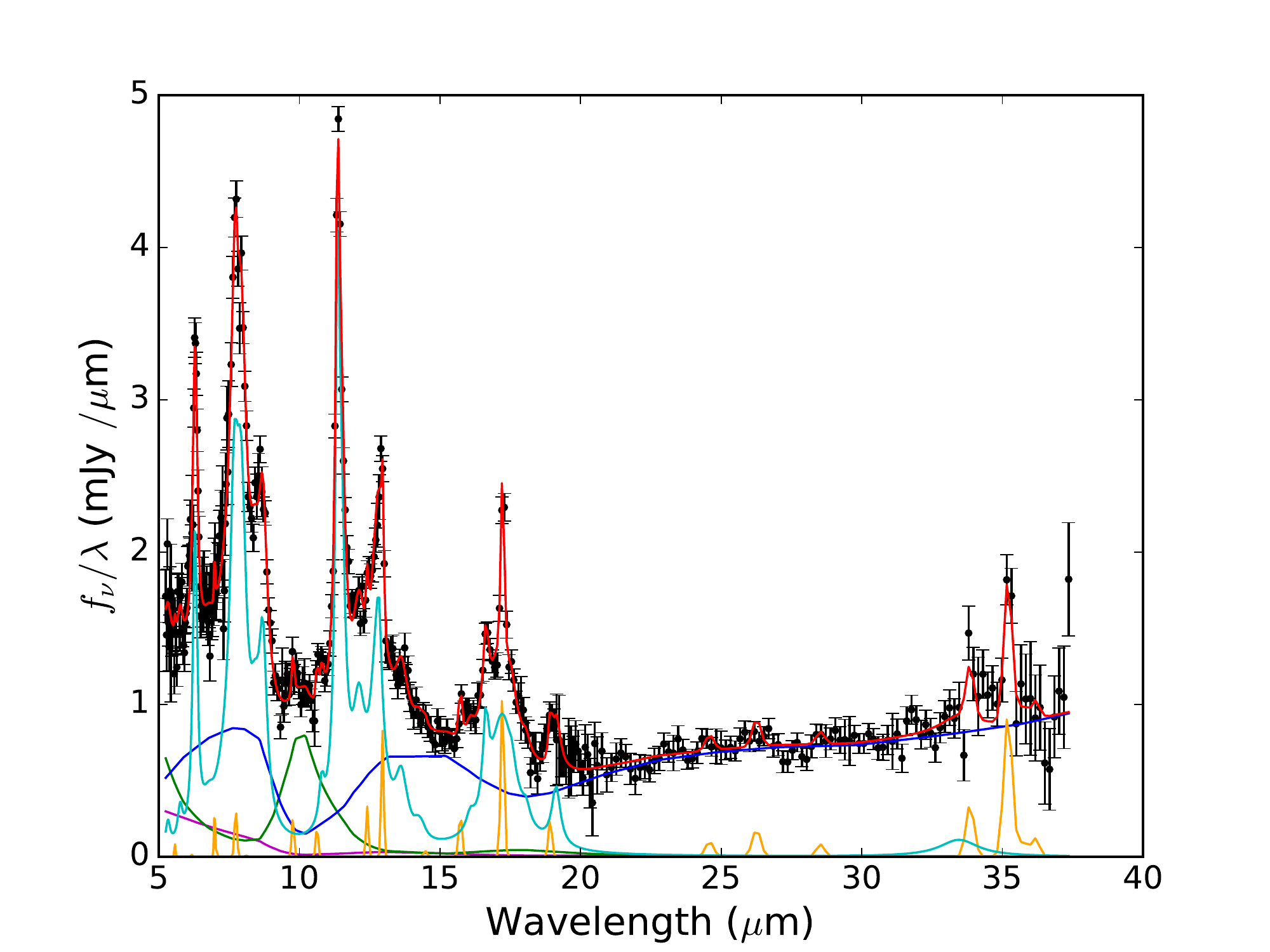}
  \captionof{figure}{NGC 3801}
  \label{fig:PAH-NGC3801}
\end{subfigure}%
\begin{subfigure}{.5\textwidth}
  \centering
  \includegraphics[width=.95\linewidth]{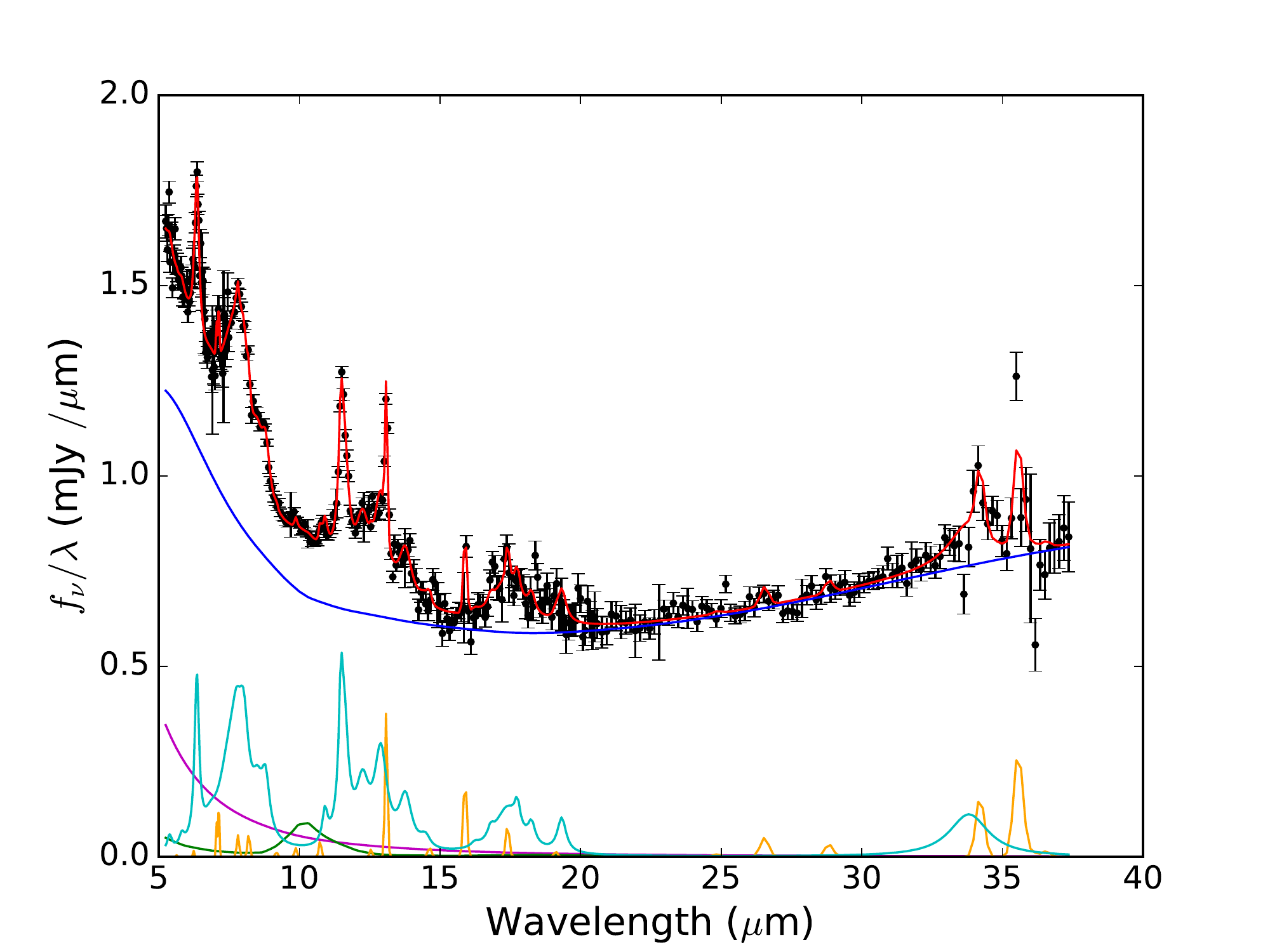}
  \captionof{figure}{NGC 3862}
  \label{fig:PAH-NGC3862}
\end{subfigure}%
\addtocounter{figure}{-1}
\end{figure}

\begin{figure}
\centering
\begin{subfigure}{.5\textwidth}
  \centering
  \includegraphics[width=.97\linewidth]{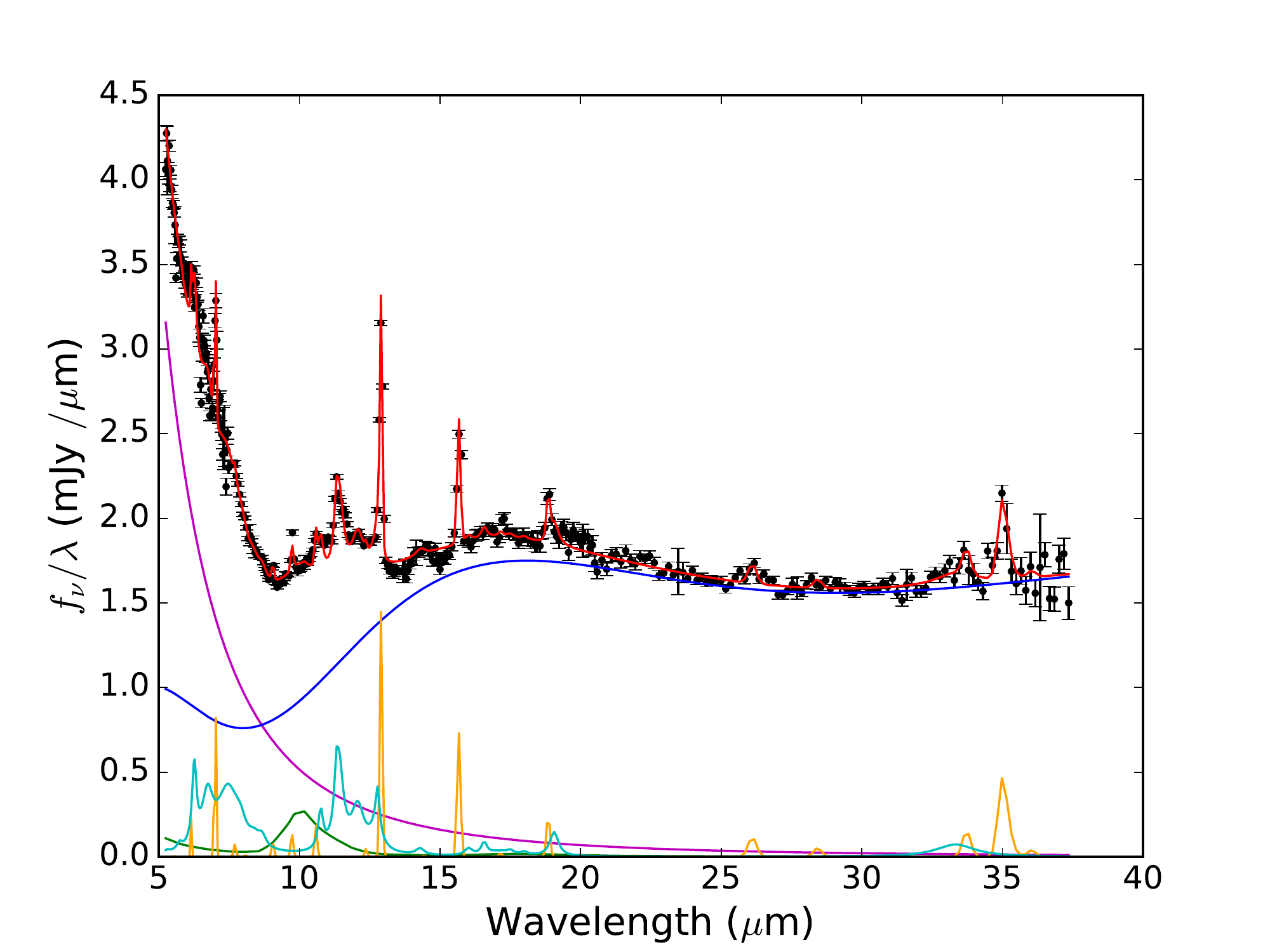}
  \addtocounter{subfigure}{6}
  \captionof{figure}{3C 270}
  \label{fig:PAH-3C270}
\end{subfigure}%
\begin{subfigure}{.5\textwidth}
  \centering
  \includegraphics[width=.97\linewidth]{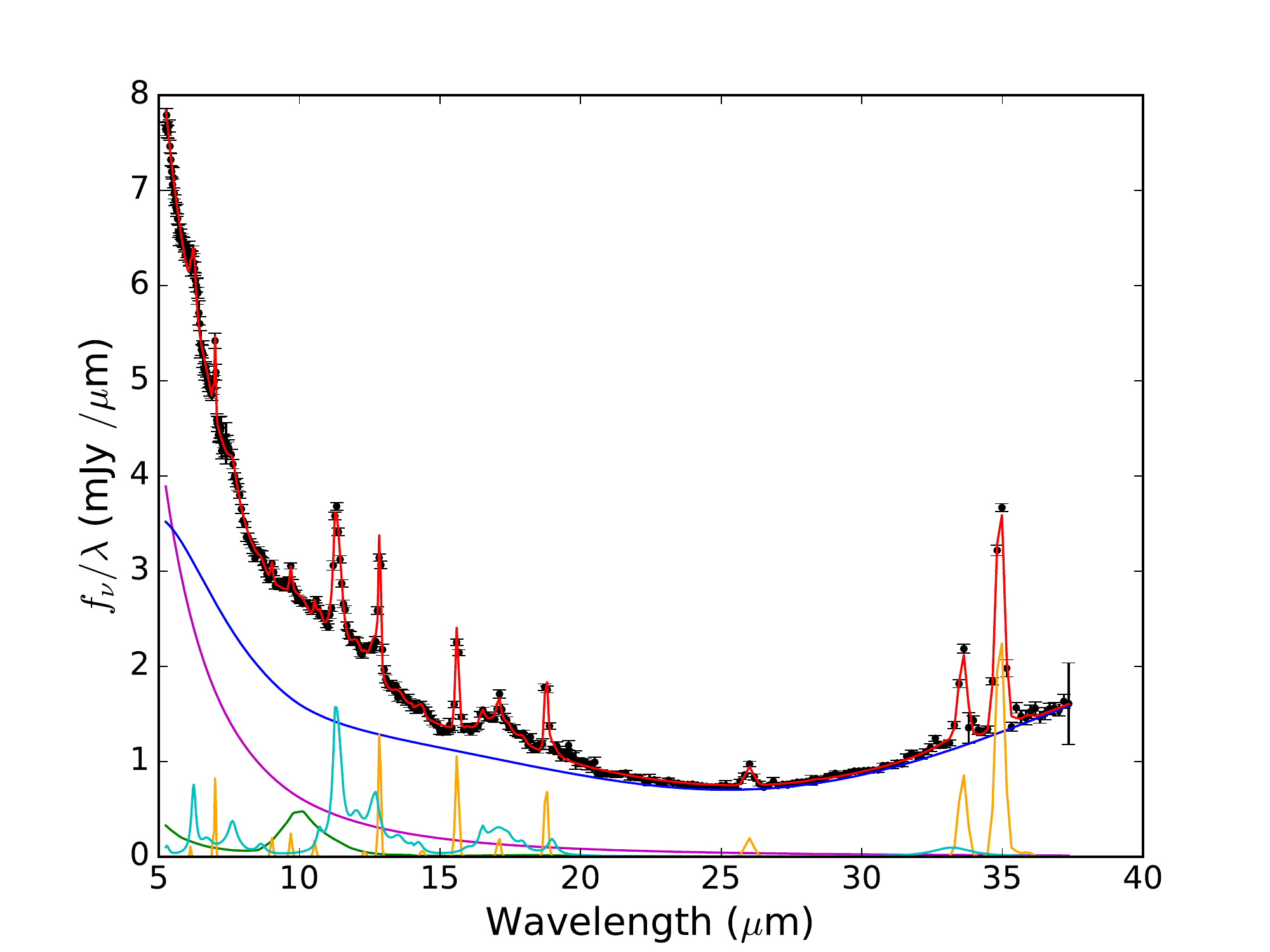}
  \captionof{figure}{M 84}
  \label{fig:PAH-M84}
\end{subfigure}

\begin{subfigure}{.5\textwidth}
  \centering
  \includegraphics[width=.97\linewidth]{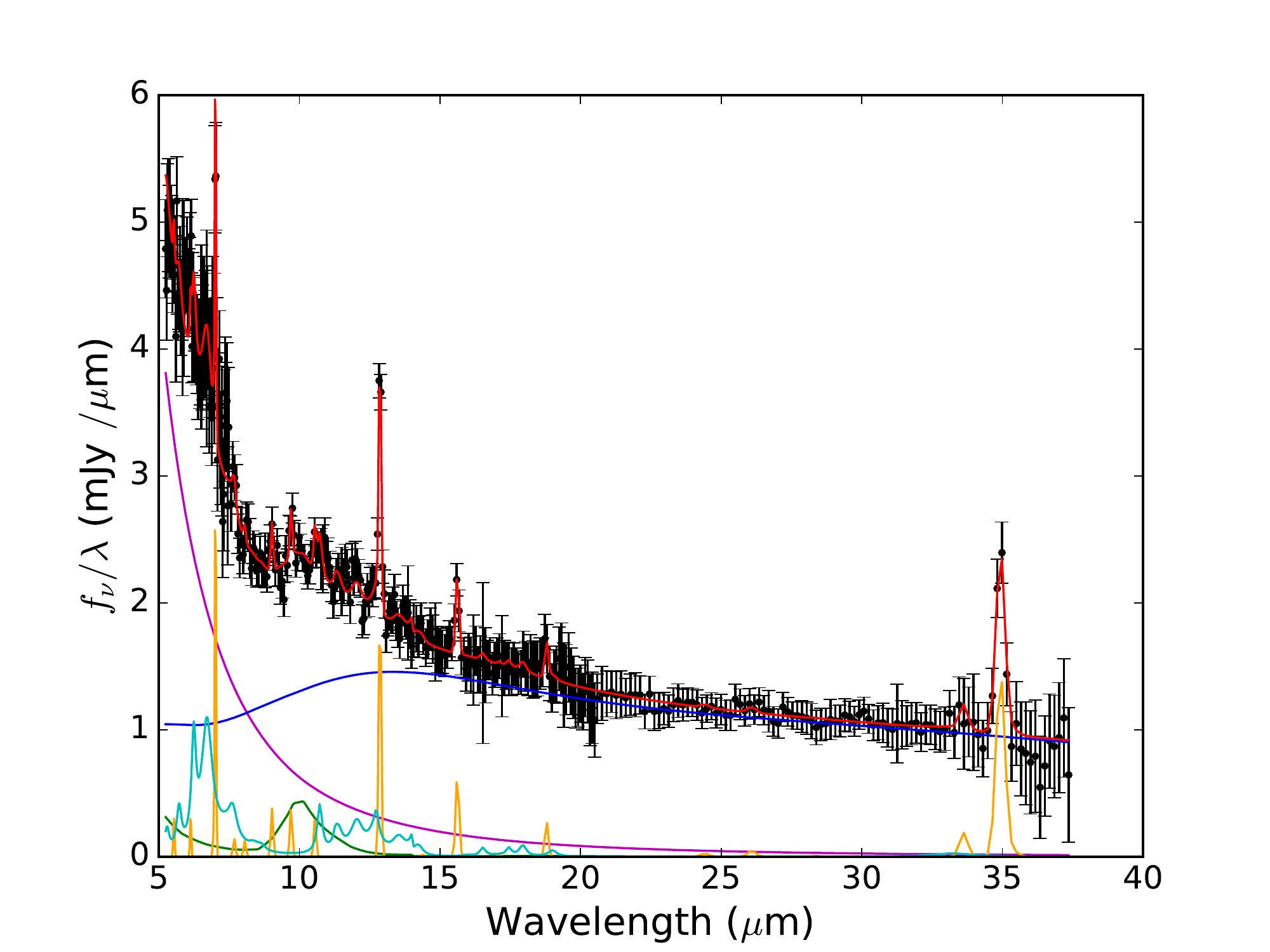}
  \captionof{figure}{M 87}
  \label{fig:PAH-M87}
\end{subfigure}%
\begin{subfigure}{.5\textwidth}
  \centering
  \includegraphics[width=.97\linewidth]{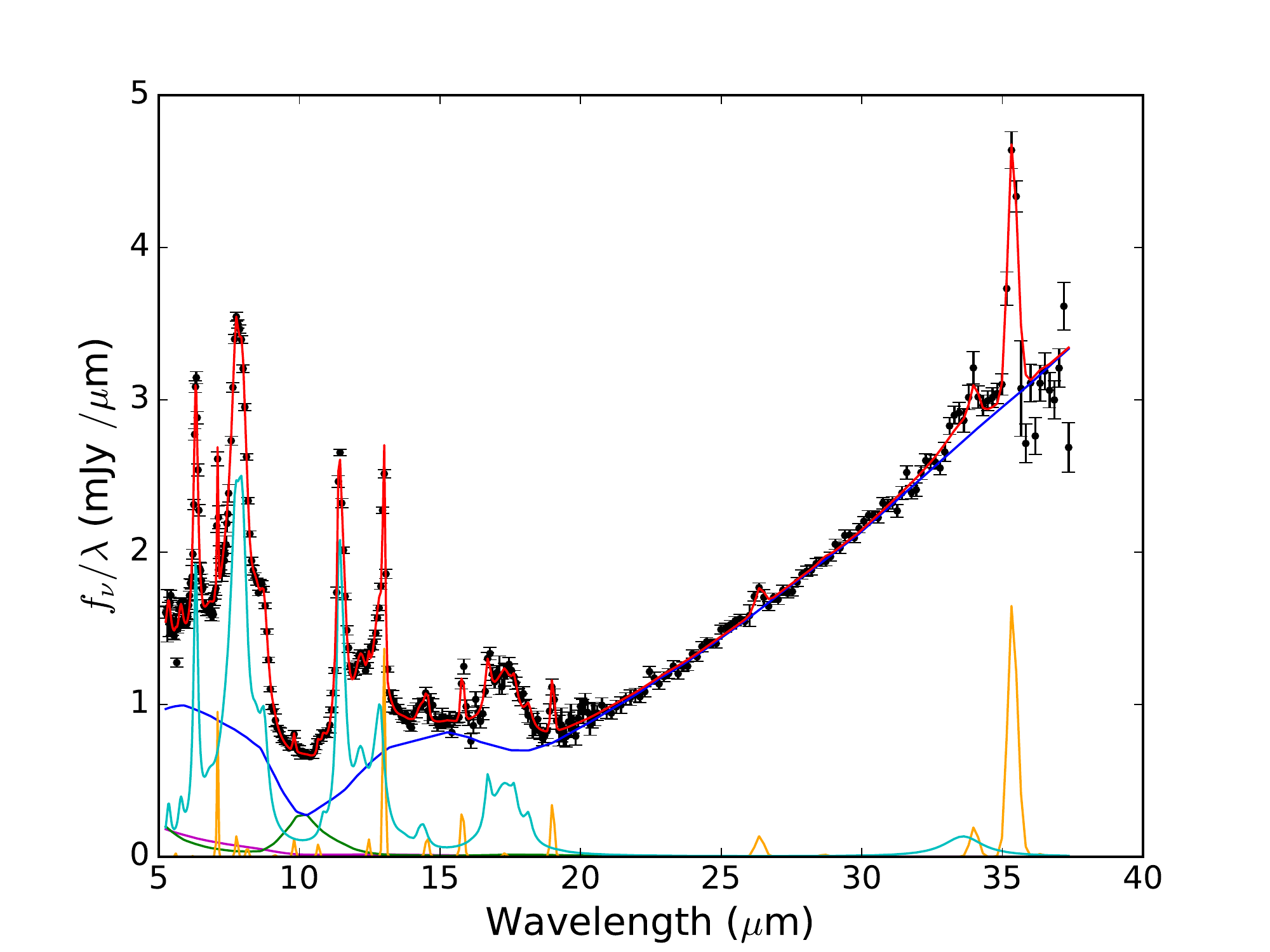}
  \captionof{figure}{NGC 7052}
  \label{fig:PAH-NGC7052}
\end{subfigure}%

\caption{Spectral line fits for six of the sample galaxies. In each fit the red line is the total fit, the dark blue line is the line-subtracted spectral energy distribution minus any power law component, the purple line is a power law component (likely associated with the Rayleigh-Jeans tail of starlight), the green line represents the silicate features, light blue represents polycyclic aromatic hydrocarbon (PAH) lines, and yellow represents the narrow ionization lines. Shown here in radiant flux per unit frequency $f_\nu$ per unit wavelength $\lambda$ because photon energy is linear with frequency and infrared spectroscopy is done in wavelength space. To convert to the more standard $\nu f_\nu$ form in $\mathrm{W}\,\mathrm{m}^{-2}$ multiply by $10^{-26}\,\mathrm{c}=2.998\times 10^{-12}\,\mu\mathrm{m}\,\mathrm{s}^{-1}$.\label{fig:pahFits}}
\end{figure}

\begin{figure}
  \centering
  \includegraphics[width=.9\linewidth]{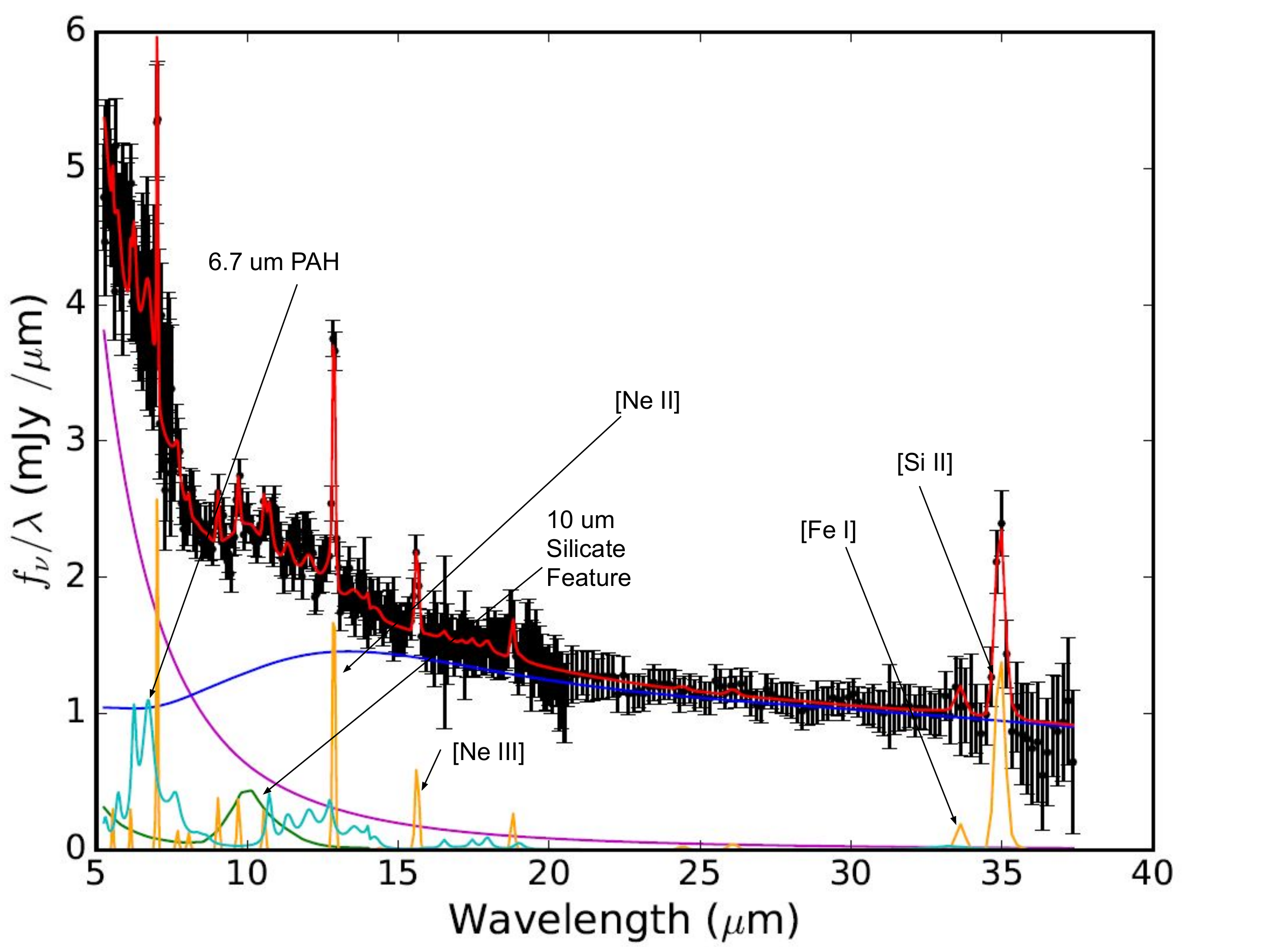}
\caption{Spectral line fit for M~87 from Figure~\ref{fig:pahFits} with some large or important spectral features labelled. As before, the red line is the total fit, the dark blue line is the line-subtracted spectral energy distribution minus any power law component, the purple line is a power law component, the green line represents the silicate features, light blue represents polycyclic aromatic hydrocarbon (PAH) lines, and yellow represents the narrow ionization lines. \label{fig:labelpahFits}}
\end{figure}

\subsection{Star formation rate estimates}
\label{sec:SFRs}
\FloatBarrier

We use $L_\text{[Ne\,II]}$ and $L_\text{[Ne\,III]}$, the luminosities [Ne II] and [Ne III] lines respectively, to estimate the star formation rate (SFR) in each sample galaxy. This estimate by \cite{HoKeto07}, $\mathrm{SFR}_\text{Ne}$, is given as 
\begin{equation}
\mathrm{SFR}_\text{Ne} =4.73 \times {10}^{-41} \,M_{\astrosun}\, {\text{yr}}^{-1} \text{s}~{\text{erg}}^{-1} ~\big[ L_\text{[Ne\,II]} + L_\text{[Ne\,III]} \big].
\label{eq:Neon-SFR}
\end{equation}
Similarly, we use the combined luminosity of the 6.2 and 11.3 $\mu$m PAH lines, $L_\text{PAH}$ to get a second independent estimate of the SFR in each source. This estimate by \cite{Farrah07}, $\text{SFR}_\text{PAH}$, is given by
\begin{equation}
\text{SFR}_\text{PAH} =1.18\times {10}^{-41} \,M_{\astrosun}\, {\text{yr}}^{-1} \text{s}~{\text{erg}}^{-1} ~\big[ L_\text{PAH} \big].
\label{eq:PAH-SFR}
\end{equation}
Finally, we also show SFR estimates using scalings by \cite{Roussel01} using $7\,\mu\mathrm{m}$ ($L_{7\,\mu \text{m}}$) and $15\,\mu\mathrm{m}$ ($L_{15\,\mu \text{m}}$) ``photometry" measurements created by integrating the \emph{Spitzer}/IRS spectra over 16.18~THz and 6.75~THz bandpasses respectively to approximate the bandpasses of the relevant \emph{Infrared Space Observatory} filters. These scalings are given by  
\begin{equation}
\mathrm{SFR}_{7\,\mu \text{m}} =6.3\times {10}^{-43} \,M_{\astrosun}\, {\text{yr}}^{-1} \text{s}~{\text{erg}}^{-1} ~\big[ L_{7\,\mu \text{m}} \big],
\label{eq:SFR7}
\end{equation}
and
\begin{equation}
\text{SFR}_{15\,\mu \text{m}} =1.7\times {10}^{-42} \,M_{\astrosun}\, {\text{yr}}^{-1} \text{s}~{\text{erg}}^{-1} ~\big[ L_{15\,\mu \text{m}} \big], 
\label{SFR15}
\end{equation}
with approximately $\sim 50\%$ calibration uncertainties. \cite{Farrah07} showed that Equations \ref{eq:Neon-SFR} and \ref{eq:PAH-SFR}  give equivalent results in their Figure~15; this is also supported by \cite{Willett10}. However, \cite{nopahSFR} found that AGN activity can dominate PAH emission in Seyfert galaxies and it is currently unknown whether this finding will generalize to include low-luminosity radio galaxies. 
Despite the different estimates, they are consistently below a few solar masses per year so we do not expect star formation to contribute a large thermal component in the MIR which could be mistaken for warm nuclear obscuring material.

The high $\text{SFR}_\text{PAH}$ estimates in 3C~31 and NGC~3862 are consistent with the high PAH emission \cite{Leipski09} found in these objects. Our relatively low $\text{SFR}_\text{PAH}$ estimates in 3C~66B and 3C~270 are also consistent with their findings of no PAH features in 3C~66B and weak PAH features in 3C~270. However, \cite{Leipski09} found significant 11.3$\,\mu$m PAH emission in M~84 which is inconsistent with our results.

\begin{table}
\centering
\begin{tabular}{ccccc}
Source & $\mathrm{SFR}_\mathrm{PAH}$ & $\mathrm{SFR}_\mathrm{Ne}$ & $\mathrm{SFR}_{7\,\mu \text{m}}$ & $\mathrm{SFR}_{15\,\mu \text{m}}$ \\
 & ($\mathrm{M}_\odot \, \mathrm{yr}^{-1}$) & ($\mathrm{M}_\odot \, \mathrm{yr}^{-1}$) & ($\mathrm{M}_\odot \, \mathrm{yr}^{-1}$) & ($\mathrm{M}_\odot \, \mathrm{yr}^{-1}$) \\

\hhline{=====}
NGC 315 & ${1.6}\pm {0.9}$ & ${1.6}\pm{0.9}$ & $0.2 \pm 0.1$ & $0.4 \pm 0.2$\\

3C 31 & ${4.}\pm {2.}$ & ${1.0}\pm {0.5}$ &$0.9 \pm 0.4$ & $1.1 \pm 0.5$\\

NGC 541 & ${0.9}\pm {0.5}$ & ${0.2}\pm {0.1}$ & $1.2 \pm 0.6$ & $1.4 \pm 0.7$\\

3C 66B & ${0.3}\pm {0.2}$ & ${0.7}\pm {0.4}$ & $0.9 \pm 0.5$ & $3. \pm 2.$ \\

NGC 3801 & ${2.}\pm {1.}$ & ${0.4}\pm {0.2}$ & $0.4 \pm 0.2$ & $0.5 \pm 0.2$\\

NGC 3862 & ${1.5}\pm {0.8}$ & ${0.8}\pm {0.4}$ & $0.4 \pm 0.2$ & $0.6 \pm 0.$\\

3C 270 & ${0.2}\pm {0.1}$ & ${0.3}\pm {0.2}$ & $3. \pm 1.$ & $8. \pm 4.$\\

M 84 & ${0.09}\pm {0.05}$ & ${0.10}\pm {0.05}$ & $0.5 \pm 0.2$ & $0.5 \pm 0.3$ \\

M 87* & ${1.}\pm {1.}$ & ${6.}\pm {3.}$ & $1.1 \pm 0.5$ & $1.3 \pm 0.7$\\

NGC 7052 & ${3.}\pm {2.}$ & ${1.2}\pm {0.6}$ &$0.09 \pm 0.04$ & $0.09 \pm 0.05$\\

\hline
\end{tabular}
\caption{SFR estimates for each source based on narrow emission features. Estimated uncertainties on \cite{Roussel01} 7~and~15~$\mu$m estimates are dominated by calibration uncertainties inferred from the \cite{Roussel01} 7~and~15~$\mu$m-H$\alpha$ relation and \cite{Kennicutt98} H$\alpha$-SFR relation for a total uncertainty of $\sim 50\%$. Note how even the highest of these estimates is on the order of a few solar masses per year which suggests that star formation is not likely to be a significant contributor to the overall spectral energy distribution of any galaxy in our sample. *\cite{Donahue11} found that some brightest cluster galaxies show elevated [Ne~II] due to contributions from a cooling intracluster medium. This may explain the discrepant estimates of SFR in M~87.\label{tab:SFR}}
\end{table}

\FloatBarrier
\subsection{High ionization lines}
\label{sec:hiIon}
\FloatBarrier

Significant emission lines from ionization states with a high ionization potential require a large source of high energy photons. These photons can come from massive stars but the central engine itself is the obvious photon source candidate. Indeed, \cite{Spinoglio92} showed that you can distinguish star formation from accretion based on which lines are excited; AGN produce photons that can ionize atoms with a higher ionization potential.  We list the luminosities of 4 high-ionization lines which are accretion tracers in Table~\ref{tab:highIon}: $[\mathrm{NeVI}]$~8~$\mu$m,  $[\mathrm{NeV}]$~14~$\mu$m, $[\mathrm{NeV}]$~24~$\mu$m, and  $[\mathrm{OIV}]$~26~$\mu$m. High-energy ionizing photons from an accretion disk should be unable to penetrate a dusty obscuring torus, so the detection  of infrared spectral lines from highly-ionized species suggests that we are seeing unobscured or minimally-obscured (in infrared) emission from clouds near the central engine. Note that the presence of these lines does not completely prohibit an obscuring torus, since these lines are infrared radiation which can pass through dust clouds better than UV or optical photons from the central engine. Still, they suggest some other explanation may be preferable.

\begin{table}
\centering
\begin{tabular}{ccccc}
Source & $L_{[\mathrm{NeVI}]}$ 8 $\mu$m & $L_{[\mathrm{NeV}]}$ 14 $\mu$m & $L_{[\mathrm{NeV}]}$ 24 $\mu$m &  $L_{[\mathrm{OIV}]}$ 26 $\mu$m \\
 & ($10^{39}\,\mathrm{erg}\,\mathrm{s}^{-1}$) 
 & ($10^{39}\,\mathrm{erg}\,\mathrm{s}^{-1}$)
 & ($10^{39}\,\mathrm{erg}\,\mathrm{s}^{-1}$) 
 & ($10^{39}\,\mathrm{erg}\,\mathrm{s}^{-1}$) \\
\hhline{=====}
\vspace{-0.5cm}\\
NGC 315 & $<5.9$  & $<4.8$ & $<3.0$ & $<5.0$ \\
3C 31 & $<5.2$ & $<3.3$ & $<0.82$ & $3.5\pm {0.7}$ \\
NGC 541 & $<5.8$ & $<7.9$ & $<4.7$ & $<6.3$ \\
3C 66B & $< 2.3$ & $<3.6$ & $<2.0$ & $<3.1$ \\
NGC 3801 & $<11.$ & $<2.$ & $<3.8$ & $<3.2$ \\
NGC 3862 & $<5.4$ & $<2.2$ & $<2.9$ & $<4.1$ \\
3C 270 & $0.35\pm {0.08}$ & $<0.49$ & $<0.27$ & $<0.92$ \\
M 84 & $<0.13$ & $<0.21$ & $<0.033$ & $0.32_{-0.06}^{+0.04}$ \\
M 87  & $<1.4$ & $<0.70$ & $<0.54$ & $<0.75$ \\
NGC 7052 & $4.\pm {1.}$ & $<12.$ & $<2.0$ & $<7.6$\\
\hline
\end{tabular}
\caption{High-ionization line luminosities. The majority of entries are upper limits on non-detections. Note that the higher upper limits here  are not restrictive; they would miss emission at the same luminosity as detected lines in other sources.\label{tab:highIon}}
\end{table}

\FloatBarrier
\subsection{Silicate constraints on torus models}
\label{sec:spectConstraint}
\FloatBarrier

\begin{figure}[ht]
\centering
\begin{subfigure}{.5\textwidth}
  \centering
  \includegraphics[width=.9\linewidth]{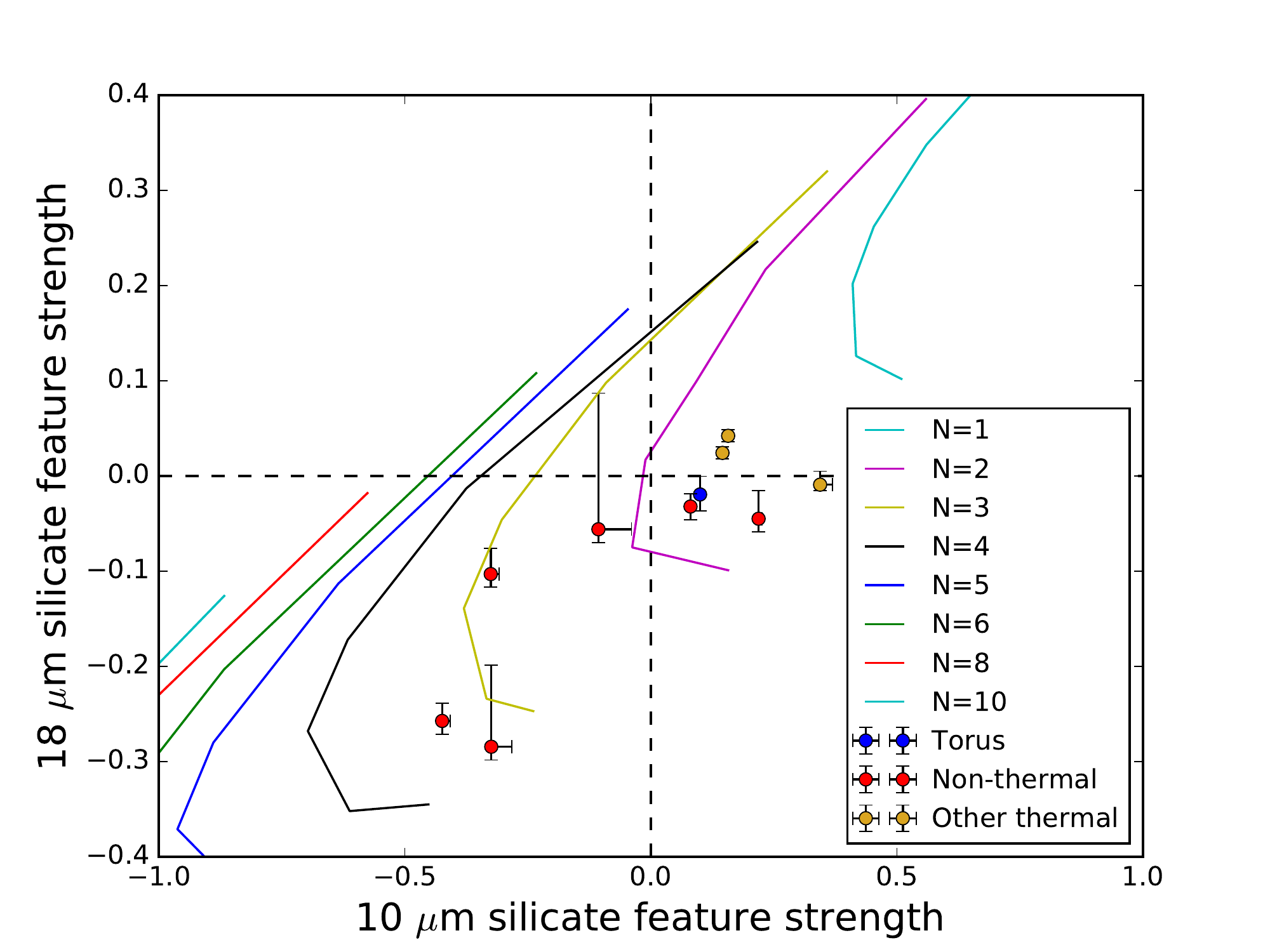}
  \captionof{figure}{Clumpy torus model, $q=0$ }
  \label{fig:clumpyQ0}
\end{subfigure}%
\begin{subfigure}{.5\textwidth}
  \centering
  \includegraphics[width=.9\linewidth]{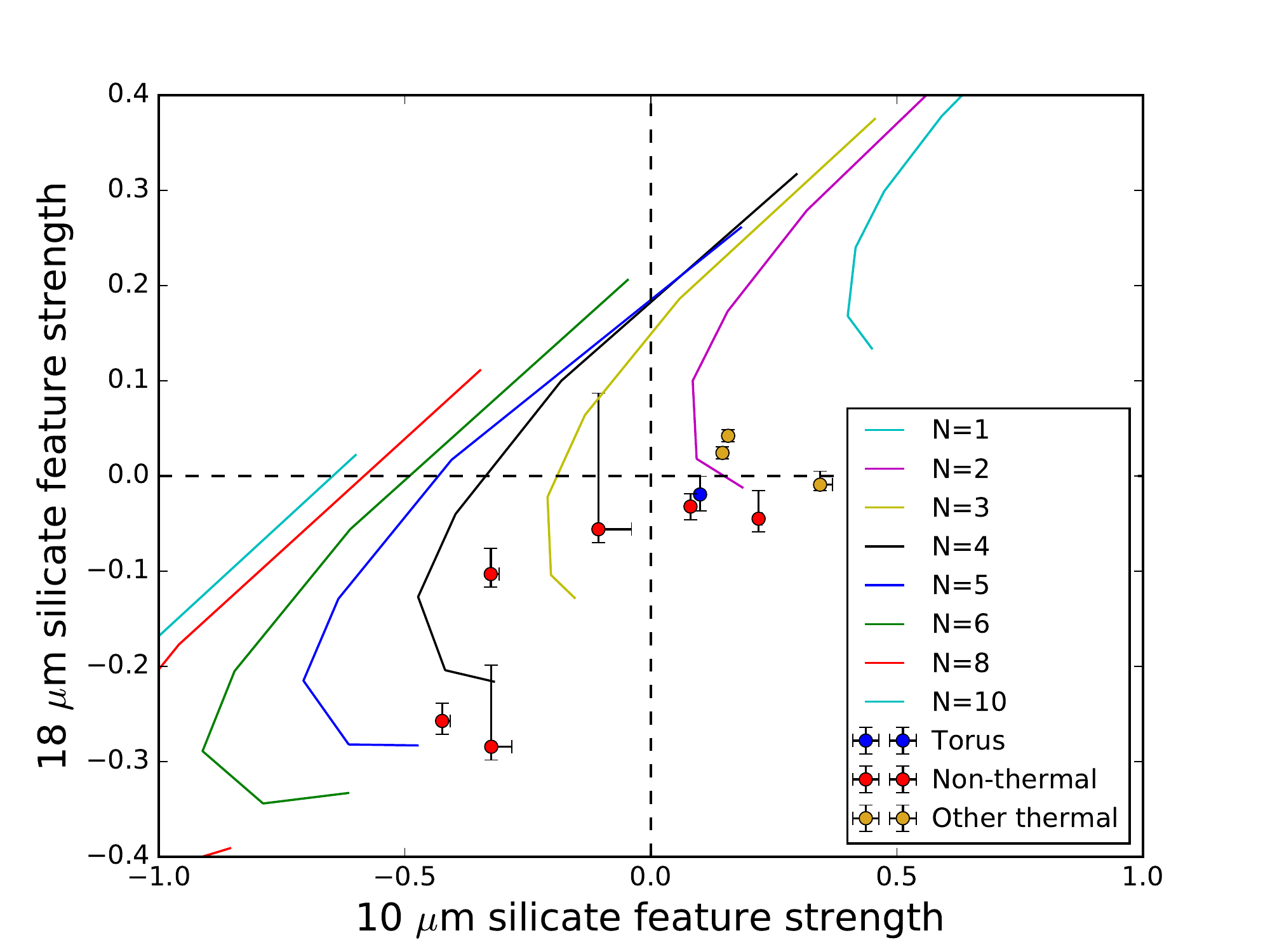}
  \captionof{figure}{Clumpy torus model, $q=1$}
  \label{fig:clumpyQ1}
\end{subfigure}%

\begin{subfigure}{.5\textwidth}
  \centering
  \includegraphics[width=.9\linewidth]{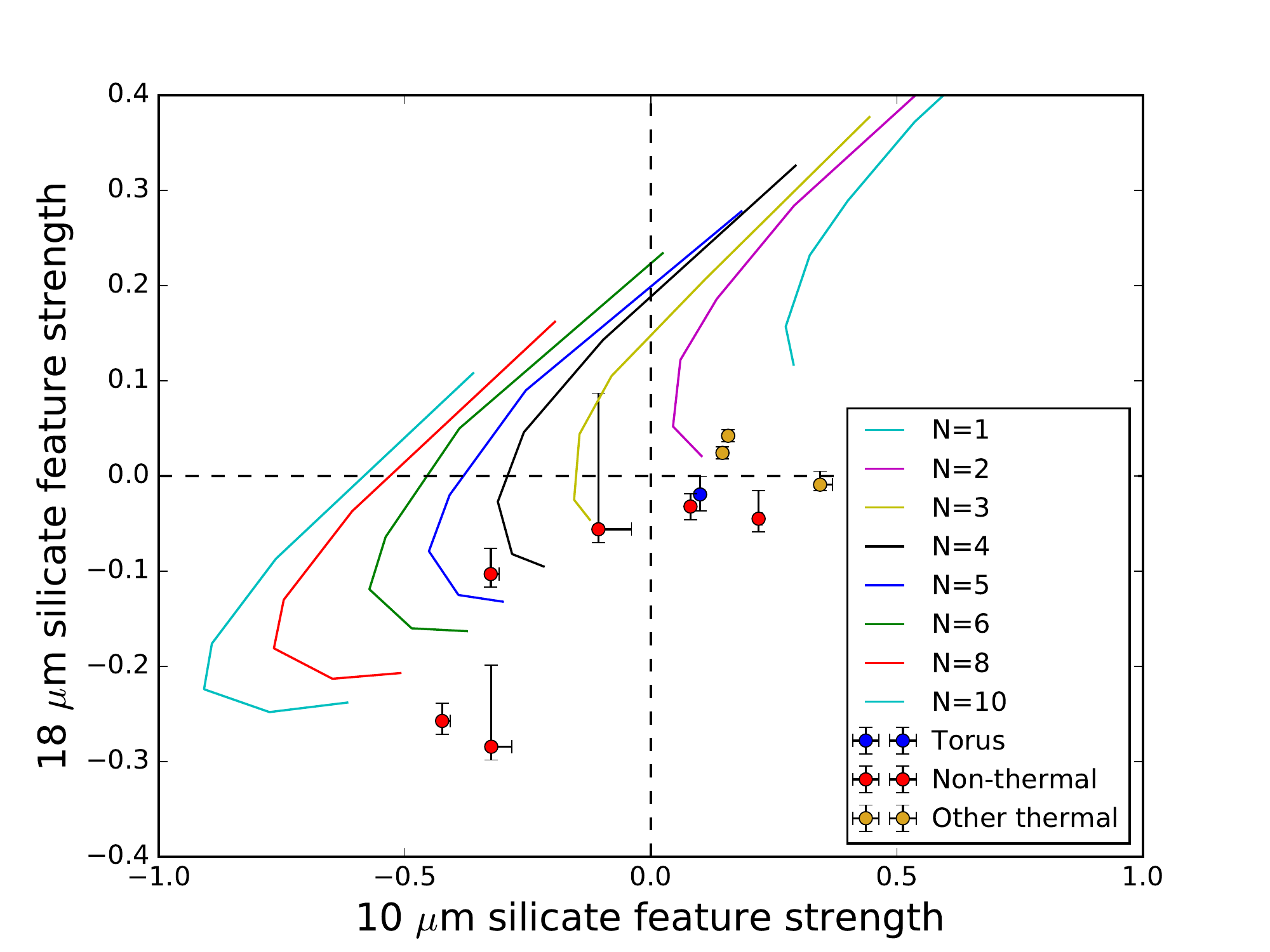}
  \captionof{figure}{Clumpy torus model, $q=2$ }
  \label{fig:clumpyQ2}
\end{subfigure}%
\begin{subfigure}{.5\textwidth}
  \centering
  \includegraphics[width=.9\linewidth]{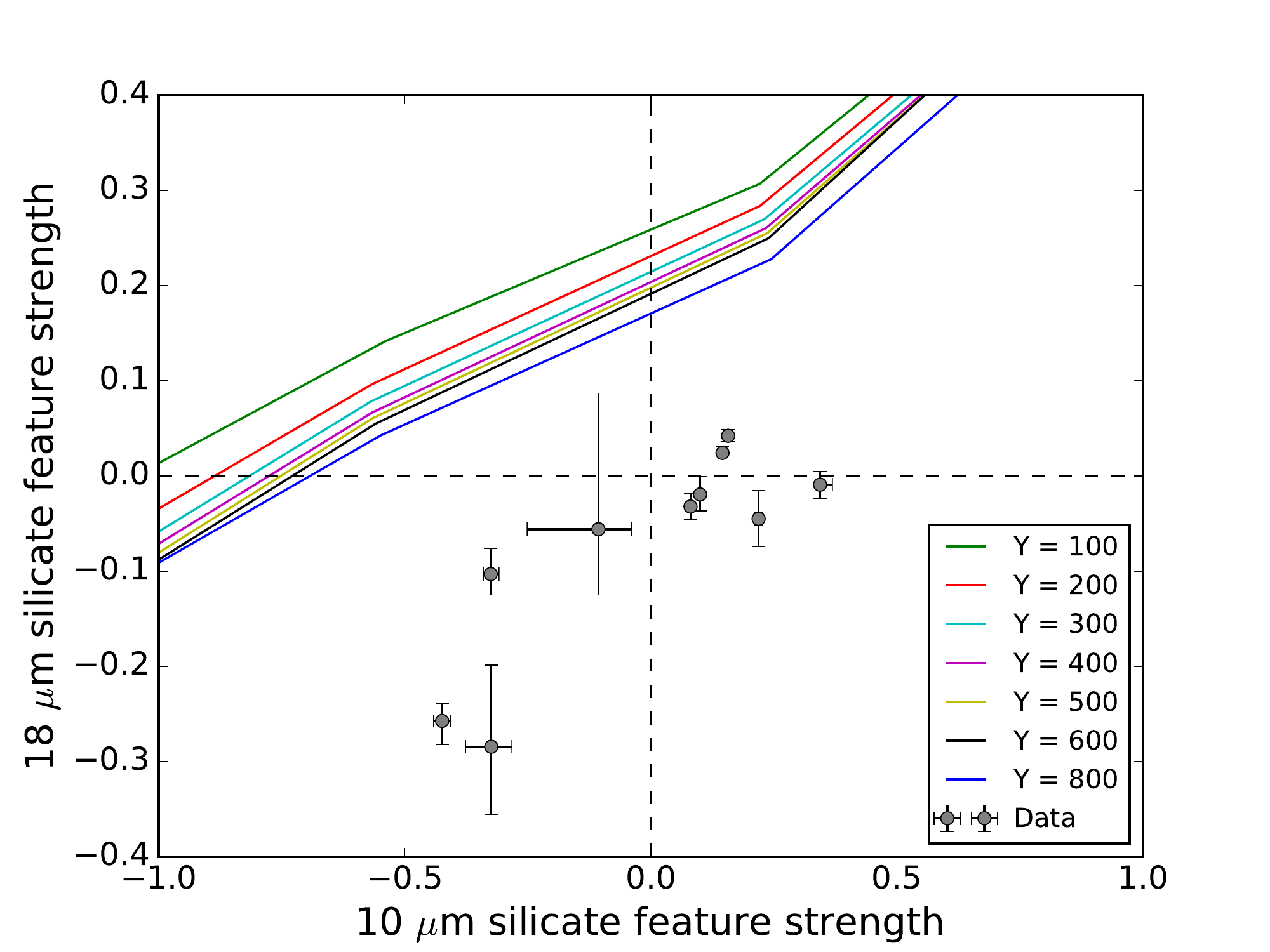}
  \captionof{figure}{Smooth shell model, $q=0$}
  \label{fig:shellQ0}
\end{subfigure}%

\centering
\begin{subfigure}{.5\textwidth}
  \centering
  \includegraphics[width=.9\linewidth]{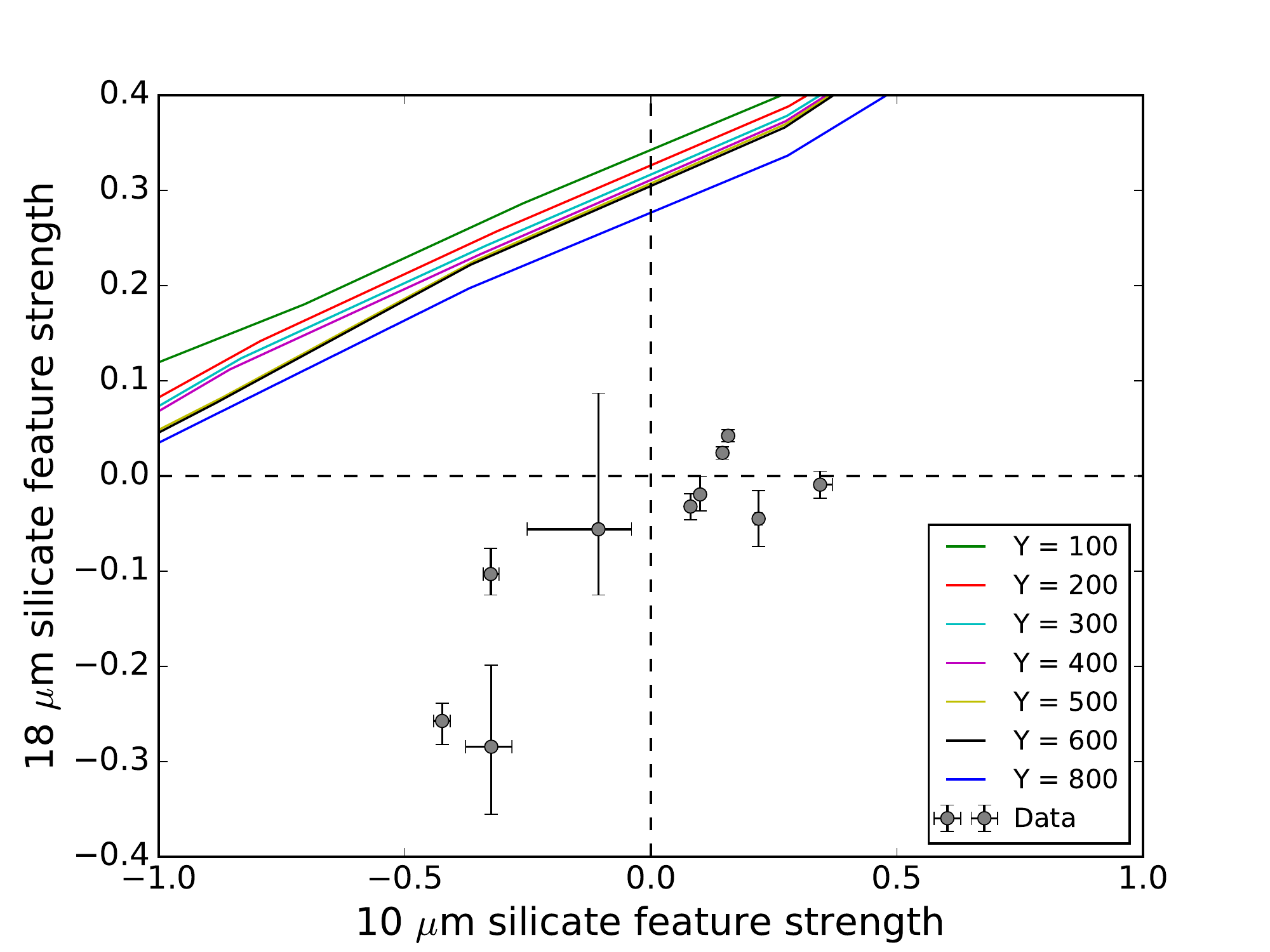}
  \captionof{figure}{Smooth shell model, $q=1$}
  \label{fig:shellQ1}
\end{subfigure}%
\begin{subfigure}{.5\textwidth}
  \centering
  \includegraphics[width=.9\linewidth]{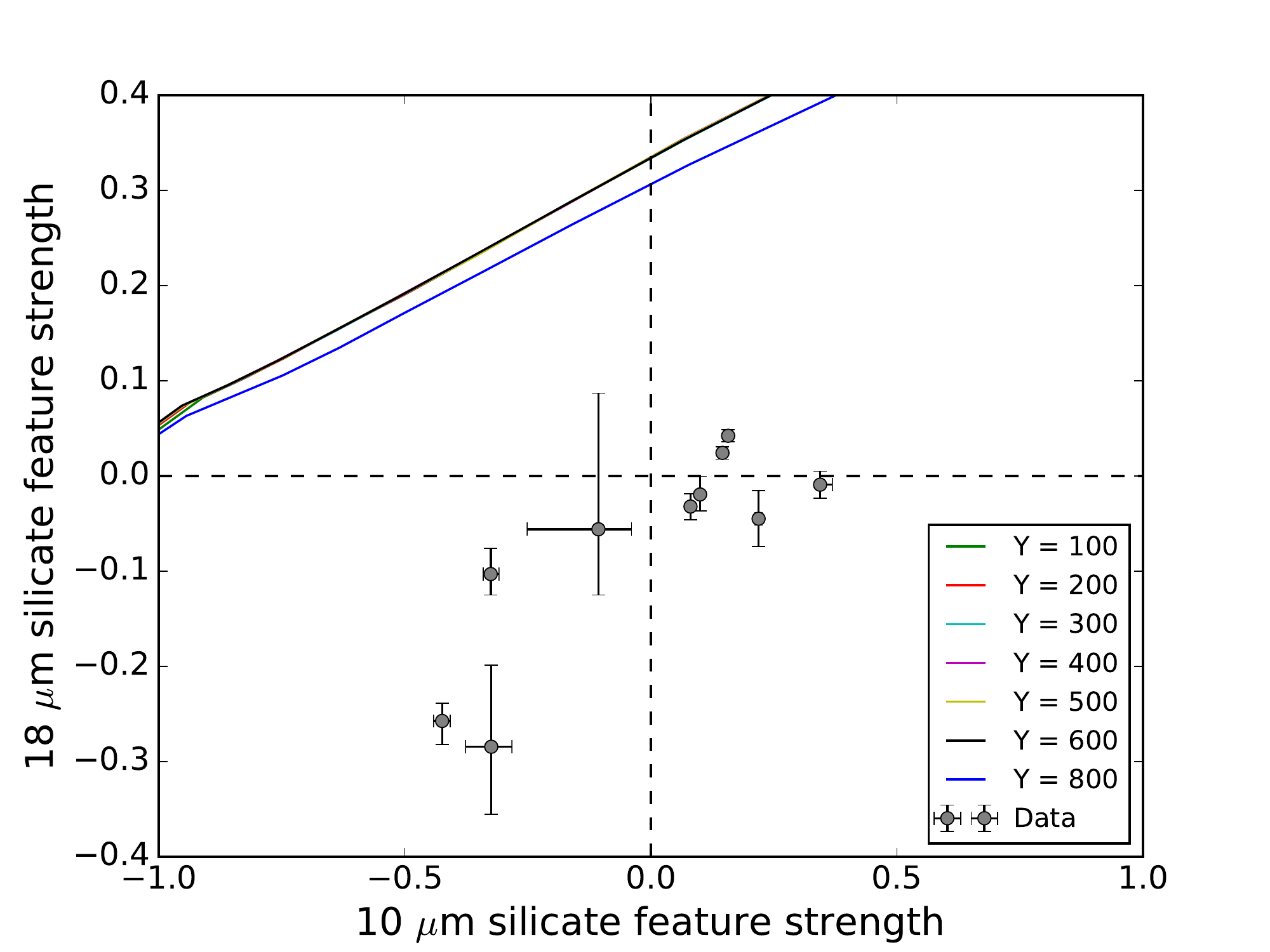}
  \captionof{figure}{Smooth shell model, $q=2$}
  \label{fig:shellQ2}
\end{subfigure}
\caption{Comparison of silicate absorption/emission strengths to predictions of various smooth and clumpy torus models. $Y$ is given in units of the dust sublimation radius. Note that there is significant overlap between the region occupied by the clumpy torus models with $q=0$ and the region occupied by the data. There is less overlap between the data and the $q=1$ clumpy torus models and no significant overlap with any other model sets. Galaxies with thermal components (torus, NLR, and/or hot dust) in our continuum fits (see Section \ref{sec:clumpy}) are flagged with different colors in the clumpy torus model plots.  \label{fig:silPlots}}
\end{figure}

We compare our 10 and 18 $\mu$m silicate feature observations to model sets by \cite{Sirocky08} to test whether our results are consistent with clumpy torus models, as this is an important diagnostic of the clumpiness of an obscuring torus. We show this comparison in Figure \ref{fig:silPlots}, where we compare our best-fit silicate line strengths $S_\text{sil}$, defined as 
\begin{equation}
S_\text{sil} = \log{\bigg(\frac{f_\text{obs}}{f_\text{cont}}\bigg)},
\label{eq:SilStrength}
\end{equation}
in terms of the observed flux of the feature $f_\text{obs}$ and the interpolated continuum flux $f_\text{cont}$ at the peak wavelength) that we found in our line fits to the  various \cite{Sirocky08} clumpy torus and smooth shell models. 

\cite{nenkova08} model the radial distribution of the clumpy torus $N_r$ as the power law distribution 
\begin{equation}
N_r \propto r^{-q}.
\label{eq:plDistClumpy}
\end{equation}
Figures \ref{fig:clumpyQ0}-\ref{fig:clumpyQ2} show the comparison to model sets, each with various numbers $N$ of clouds along the line of sight, by \cite{Sirocky08} with power law index $q$ of 0-2 respectively.

Similarly, the mass distribution of the single continuous cloud in the smooth shell model also follows a power law, 
\begin{equation}
\rho_r \propto r^{-q},
\label{eq:plDistShell}
\end{equation}
still described by the power law index, $q$, and dependent on the radial extent of the cloud, $Y$ (given in units of the dust sublimation radius).

We find that the silicate line strengths are inconsistent with smooth shell models of AGN obscuration and with clumpy models with power law indices $q>1$ and are consistent with clumpy models with $q\leq 1$ and $N \leq 5$ although a few are also consistent with the complete absence of any $18\,\mu \mathrm{m}$ silicate feature. Note that in Section \ref{sec:clumpy} we find that only four of our ten  sources show a significant thermal component in their continua, which we denote with blue and yellow markers in Figures \ref{fig:clumpyQ0}-\ref{fig:clumpyQ2}.

\FloatBarrier
\section{Continuum fitting}
\label{sec:clumpy}



The infrared spectra, including low-resolution, narrow-line-subtracted IRS spectra, were decomposed using clumpyDREAM \citep{clumpyDREAM}.  This software assumes that the infrared spectra of active galaxies include contributions from starlight, dusty star-forming regions, obscuring molecular clouds near the AGN, etc. (see, e.g., \citealt{Genzel00}). To decompose the observed infrared spectrum, clumpyDREAM simultaneously fits models for these different infrared sources.  During the fit, gridded model parameters are allowed to vary smoothly over the range allowed by the model grid, and a model spectrum is extracted from the model grid by $n_p$-dimensional spline interpolation, where $n_p$ is the number of parameters defining the model grid. 

clumpyDREAM uses a custom implementation of the DREAM(ZS) MCMC stepper algorithm \citep{dreamzs} to fit models to the observed spectra. The DREAM(ZS) stepper uses differential evolution, as described by \cite{Storn97}, to determine most of its steps in parameter space but it uses the snooker updater described by \cite{dreamzs} for every tenth iteration. It uses $\sim 100$ walkers randomly initialized with uniform priors \citep{clumpyDREAM}. For further details on clumpyDREAM and futher overview of DREAM(ZS) see the appendix of \cite{clumpyDREAM}.

Fitting a spectrum was performed in two stages. First, we used a modified version of clumpyDREAM, clumpyDREAMBIC, to determine which model components are likely needed to best fit the observed spectrum. To this end, clumpyDREAMBIC searches for best-fit model combinations with different model components (e.g., starlight, star-forming regions, etc.) toggled off or on. A given model combination is evaluated for goodness-of-fit using the Bayes Information Criterion (BIC; \citealt{Schwarz78}), 
\begin{equation}
\mathrm{BIC} = -2\ln \mathcal{L} + k \ln n,
\end{equation}
where $\mathcal{L}$ is the likelihood function, $k$ is the number of parameters in the trial set of model components, and $n$ is the number of data points in the observed spectrum. By searching for model combinations that minimize the BIC, the result is a model that maximizes the likelihood function while penalizing model combinations that over-fit the observed spectrum (i.e., too large $k$). 
After clumpyDREAMBIC estimates the BIC for different combinations of model components, the fit is repeated with for fixed sets of model components in clumpyDREAM. A typical run for a single spectrum involves first running clumpyDREAMBIC with three parallel chains for $10^6$ iterations with chains thinned by a factor of ten (i.e., parameter values are stored to the Markov chains every ten iterations). Then, based on the output of clumpyDREAMBIC, clumpyDREAM is run on fixed model combinations, again with three parallel chains, $10^6$ iterations, and factor of ten thinning.

%

In this Section we describe each of the model components we try to fit with clumpyDREAM except the clumpy torus model itself which we describe in Section~\ref{sec:AGNuni}. For quick reference we list all of these and their source papers in Table \ref{tab:components}. 

These targets are galaxies so we force the model to include a simple stellar population for every source as well as a diffuse interstellar medium (ISM). We use the GRASIL stellar population model for elliptical galaxies developed by \cite{GRASIL} and the diffuse ISM model described by \cite{Draine07}. The GRASIL models simulate a 1~Gyr burst of star formation followed by passive evolution and so are described solely by the age and scaling of the stellar population. The stellar population models are not expected to contribute significantly in the IRS band but that they should dominate the NIR and so should be well constrained by our 2MASS and IRAC photometry. Our results in some cases favor models in which stars contribute significantly to the IRS band, particularly at the short-wavelength end, but this is not surprising due to the dim nucleus. In this section, however, there is significant contribution from the stellar population model through the short-wave end of the IRS band in several of our sources which is likely due to the low overall IRS band luminosity of those sources. 

The diffuse ISM model by \cite{Draine07} has three main parameters: the mass fraction of polycyclic aromatic hydrocarbons $q_\text{PAH}$ relative to the total dust mass, the lower limit of the interstellar radiation field as a scaling of that in the solar neighborhood $U_\text{min}$, and photo-dissociation regions exposed to $U_\text{max} = 10^6\, U_\text{min}$ which are characterized by their relative contribution $\gamma$ to the spectrum.


We use a narrow-line region (NLR) model, based on the \emph{DUSTY} continuum radiative transfer code by \cite{DUSTY}, together with a $800\,\text{K} \leq T\leq 1600\,\text{K}$ blackbody hot dust model. This approach is similar to the one taken by \cite{Mor09} 
in order to fit IRS spectra of quasi-stellar objects (QSOs). \cite{Mor09} also attempted torus models and models with both a torus and an NLR and found that neither were adequate. Since we are extending our dataset beyond the IRS band into the far-infrared/sub-mm we consider a cold dust component as well. All three components are described in our models by two parameters each --- their temperature and their luminosity. Note that the NLR clouds and hot dust are optically thin.

We also try a foreground extinction component, as described by \cite{Fischera03}, on the torus in 3C 31, NGC 3801, and M 84. These sources appear to have regions with decreased brightness compared to adjacent regions at the same radius which could be foreground dust in SDSS images by \cite{SDSS4} although it is only obvious in NGC 3801. On a similar note we only tested power-law synchrotron spectra in sources with known optical/IR jets. These are 3C 31 \citep{3C31jet}, 3C 66B \citep{3C66Bjet}, NGC 3862 \citep{3862jet}, and M 87 \citep{M87jet}. We constrain the power law indices to the widest ranges given in the cited papers.
We list these model components in Table \ref{tab:components} for reference. 

\begin{table}[htp]
\centering
\begin{tabular}{ccc} 
Component & Plotting colour & Reference\\
\hhline{===}
\vspace{-0.4cm}\\
Clumpy torus & Green & \cite{nenkova08}\\

Hot dust & Pink &  - \\

Cold dust & - &  - \\

Diffuse ISM & Blue &  \cite{Draine07}\\

Simple stellar population & Yellow &  \cite{GRASIL} \\

Foreground extinction on torus & - &  \cite{Fischera03} \\

Power law & Purple &  - \\

Narrow line region & Cyan &  \cite{DUSTY}\\

Aperture correction & - &  - \\

\hline
\end{tabular}
\vspace{-0.2cm} 
\caption{All spectrum model components we considered. Extinction and aperture correction aren't plotted as model components; they alter other model components. No sources require a cold dust component however our far-infrared data  are only upper limits. The ``hot dust" and ``cold dust" components are blackbodies. \label{tab:components}} 
\end{table}

In this section we present the best fit models for the continuum emission from each of our sample galaxies. Additionally, we present the results of forcing a torus component and suppressing all other IR-peaked thermal components (i.e., requiring that the software include a clumpy torus component and not include any other thermal components which peak in IR) in order to establish a firm upper limit on the contribution to the total flux from a possible torus component. Low contributions from a compact clumpy torus may otherwise be hidden by fitting degeneracies between thermal model components; removing the possibility of such fitting degeneracies allows us to determine the highest contribution from a torus component to the overall spectrum which would still be consistent with observations. Finally, we present an overview of the results for the entire sample at the end of this section.

\subsection{NGC~315}
\label{sec:NGC315cont}
\FloatBarrier

This source favors a torus component and that torus component is a significant part of the best fit, contributing $\sim 70\%$ of the flux at 15 and 30 $\mu$m and $\sim 50\%$ at 60 $\mu$m. There is possible degeneracy between the torus component and NLR component but we doubt that this is a significant concern in this case due to the fact that the BIC will penalize the torus model more harshly due to its higher number of parameters. We therefore see no particular reason to doubt the presence of an obscuring warm torus-like structure in this AGN. 

The presence of a torus component in the spectral energy distribution of NGC 315 is consistent with the discovery of broad polarized H$\alpha$ emission in this source by \cite{Barth99}. Which may indicate the presence of a Thomson-scattered hidden broad line region.

Similarly, \cite{Gu07} detected a low-luminosity compact AGN in IRAC and MIPS images of NGC 315 after removal of stellar IR emission and accounting for emission from a central dusty disk. They make no claims about the nature of this AGN however its presence is consistent with our result that the IR spectrum of NGC 315 is not entirely dominated by emission from the host galaxy.

\begin{center}
\begin{figure}[ht]
\centering
\includegraphics[width=0.8\linewidth]{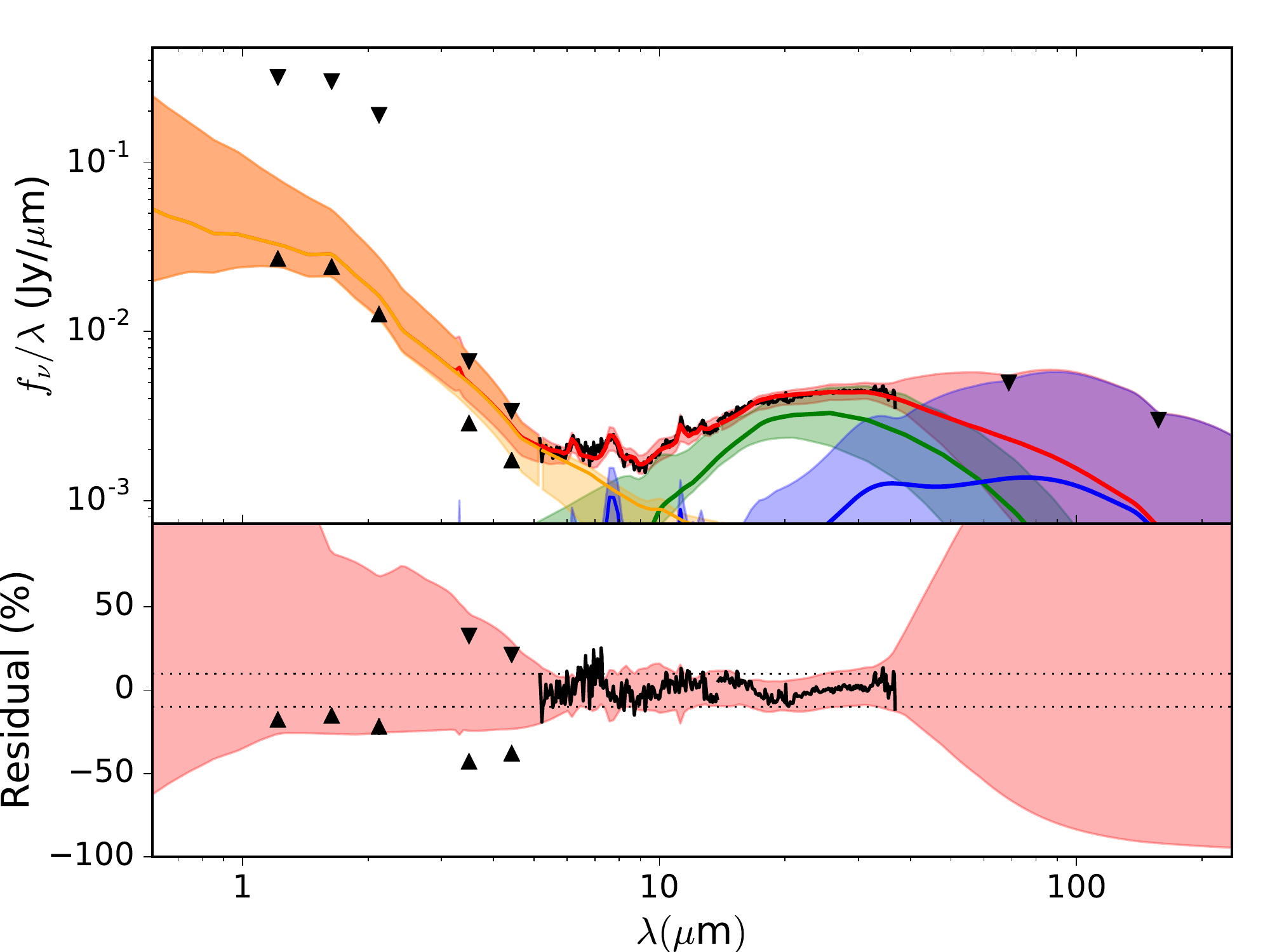} 
\caption{Best continuum fit for NGC 315. Model consists of a simple stellar population (yellow), diffuse ISM (blue) and clumpy torus (green). The total fit is shown in red although in the NIR it appears as orange due to overlap with the stellar population.}
\label{fig:SED-NGC0315-cdo}
\end{figure}
\end{center}

\begin{table}
\centering
\begin{tabular}{ccccc}
Component & Wavelength & Best fit & Minimum & Maximum\\
 & ($\mu$m) & (\%) & (\%) & (\%)\\
\hhline{=====}
\multirow{4}{1.5cm}{Torus} & 5.0 & 1.4 & 0.8 & 33.5\\
 & 15.0 & 69.0 & 59.4 & 97.8\\
 & 30.0 & 69.9 & 40.8 & 100\\
 & 60.0 & 47.8 & 13.5 & 91.9\\
\hline
\multirow{4}{1.5cm}{Diffuse ISM} & 5.0 & 1.8 & 0.2 & 2.9\\
 & 15.0 & 10.4 & 1.7 & 20.1\\
 & 30.0 & 24.7 & 1.7 & 64.9\\
 & 60.0 & 50.1 & 4.2 & 100\\
\hline
\multirow{4}{1.5cm}{Stars} & 5.0 & 96.8 & 58.8 & 100\\
 & 15.0 & 20.6 & 2.1 & 22.8\\
 & 30.0 & 5.4 & 0.2 & 6.0\\
 & 60.0 & 2.0 & 0.0 & 2.2\\
\hline
\end{tabular}
\caption{Fractional contributions of the various model components to the overall continuum in NGC 315. \label{tab:NGC0315_frac_cdo}}
\end{table}

\subsection{3C~31}
\label{sec:3C31cont}
\FloatBarrier

This source appears to be dominated by emission from the host galaxy in the IRS band. This source contains an optical jet, first noticed by \cite{Butcher80}, and a radio jet which \cite{3C31jetopt} identifies as the same structure as in the optical. The optical jet does not contribute substantially to the IR core flux. It is worth noting, however, that \cite{3C31jet} detect extended NIR emission from the jet at kpc scales in \emph{Spitzer}/IRAC images. As shown in Figure \ref{fig:SED-3C31-cdo} and in Table~\ref{tab:3C31_frac_cdo}, when a torus component is forced it contributes a significant amount of flux at $15\,\mu\mathrm{m}$ and $30\,\mu\mathrm{m}$ with best fit fractional contributions of 34.2\% and 18.7\% respectively.  However, as shown in Table~\ref{tab:BICs}, the BIC difference between galaxy models with a forced torus and galaxy-only models is non-negligable (the BIC of the forced-torus model is $22.2$, or $\sim 6\%$, higher) and we notice no obviously unphysical best fit model parameters, therefore we conclude that we do not find evidence for a warm obscuring torus in the infrared spectra of 3C 31. We note, however, that \cite{541OIII} detected broad H$\alpha$ emission with a FWHM of $2710\,\mathrm{km}\,\mathrm{s}^{-1}$, contributing 86\% of the combined narrow+broad H$\alpha$ flux. This broad H$\alpha$ is consistent with the presence of a broad line region and therefore possibly an accretion disk.

\begin{center}
\begin{figure}[ht]
\centering
\includegraphics[width=0.8\linewidth]{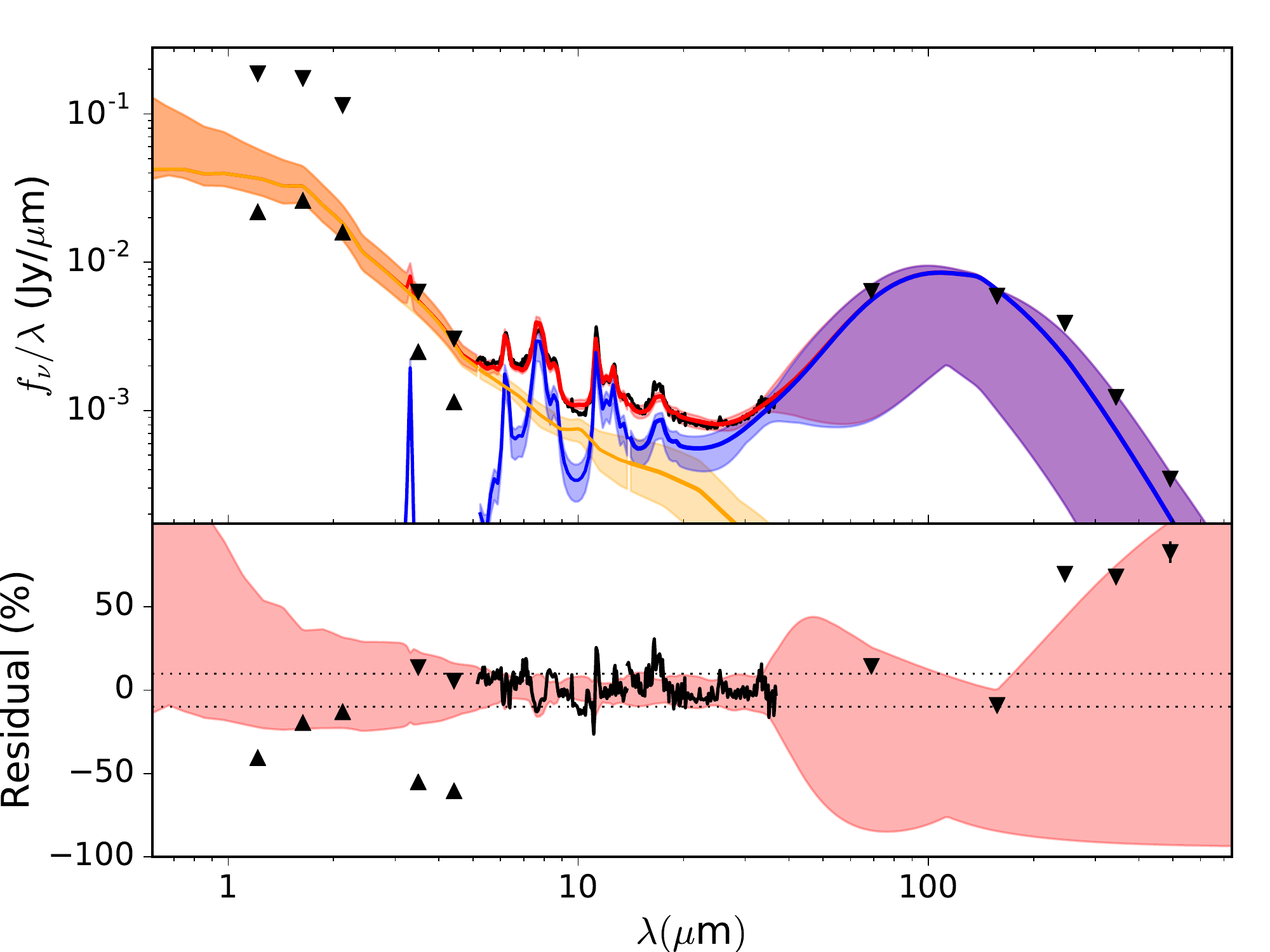} 
\caption{Best continuum fit for 3C 31. Model consists of a simple stellar population (yellow) and diffuse ISM (blue). The total fit is shown in red although in the NIR it appears as orange due to overlap with the stellar population. This is the only source with a complete data set.}
\label{fig:SED-3C31-do}
\end{figure}
\end{center}

\begin{table}
\centering
\begin{tabular}{ccccc}
Component & Wavelength & Best fit & Minimum & Maximum\\
 & ($\mu$m) & (\%) & (\%) & (\%)\\
\hhline{=====}
\multirow{4}{1.5cm}{Diffuse ISM} & 5.0 & 4.2 & 3.0 & 5.0\\
 & 15.0 & 57.0 & 43.2 & 64.4\\
 & 30.0 & 83.6 & 65.7 & 100\\
 & 60.0 & 99.1 & 18.7 & 100\\
\hline
\multirow{4}{1.5cm}{Stars} & 5.0 & 95.8 & 83.0 & 100\\
 & 15.0 & 43.0 & 27.2 & 64.6\\
 & 30.0 & 16.4 & 5.6 & 25.0\\
 & 60.0 & 0.9 & 0.2 & 1.2\\
\hline
\end{tabular}
\caption{Fractional contributions of the various best fit model components to the overall continuum in 3C~31. \label{tab:3C31_frac_do}}
\end{table}

\begin{center}
\begin{figure}[ht]
\centering
\includegraphics[width=0.8\linewidth]{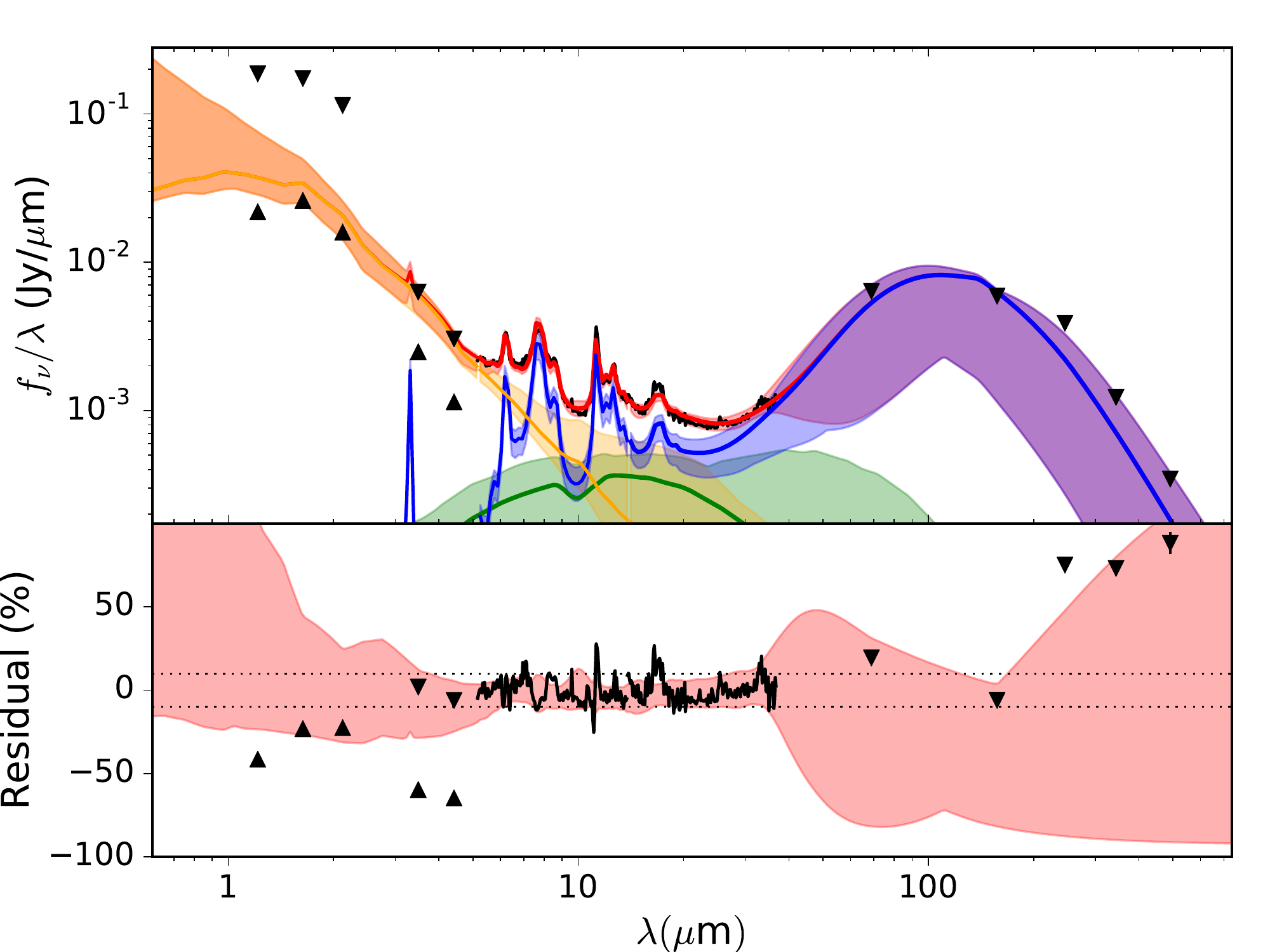} 
\caption{Continuum fit for 3C 31 forcing a torus component (green). Model is otherwise the same as in Figure \ref{fig:SED-3C31-do}.}
\label{fig:SED-3C31-cdo}
\end{figure}
\end{center}

\begin{table}
\centering
\begin{tabular}{ccccc}
Component & Wavelength & Best fit & Minimum & Maximum\\
 & ($\mu$m) & (\%) & (\%) & (\%)\\
\hhline{=====}
\multirow{4}{1.5cm}{Torus} & 5.0 & 8.0 & 0.0 & 13.6\\
 & 15.0 & 34.2 & 0.0 & 48.7\\
 & 30.0 & 18.7 & 0.0 & 54.2\\
 & 60.0 & 0.4 & 0.0 & 11.2\\
\hline
\multirow{4}{1.5cm}{Diffuse ISM} & 5.0 & 3.6 & 2.8 & 4.4\\
 & 15.0 & 51.1 & 39.6 & 59.3\\
 & 30.0 & 77.9 & 45.2 & 98.4\\
 & 60.0 & 99.4 & 19.9 & 100\\
\hline
\multirow{4}{1.5cm}{Stars} & 5.0 & 88.4 & 73.5 & 98.9\\
 & 15.0 & 14.8 & 8.0 & 59.3\\
 & 30.0 & 3.4 & 1.2 & 24.7\\
 & 60.0 & 0.2 & 0.0 & 1.3\\
\hline
\end{tabular}
\caption{Fractional contributions of the various model components to the overall continuum in 3C 31 with a forced torus component.\label{tab:3C31_frac_cdo}}
\end{table}

\subsection{NGC~541}
\label{sec:NGC541cont}
\FloatBarrier

Not only does the best fit to the continuum emission in this source  not require a torus component but the best fit contribution from a forced torus component is not significant, contributing only $\sim 20\%$ at its brightest. The largest difference between the best fit model and the data is the 70$\,\mu$m but that point is an upper limit and the best fit model is within the uncertainty of the data point. We therefore conclude that we can explain all the emission with only contributions from the host galaxy and none from the nucleus and so rule out any significant MIR-bright obscuration. 

\begin{center}
\begin{figure}[ht]
\centering
\includegraphics[width=0.8\linewidth]{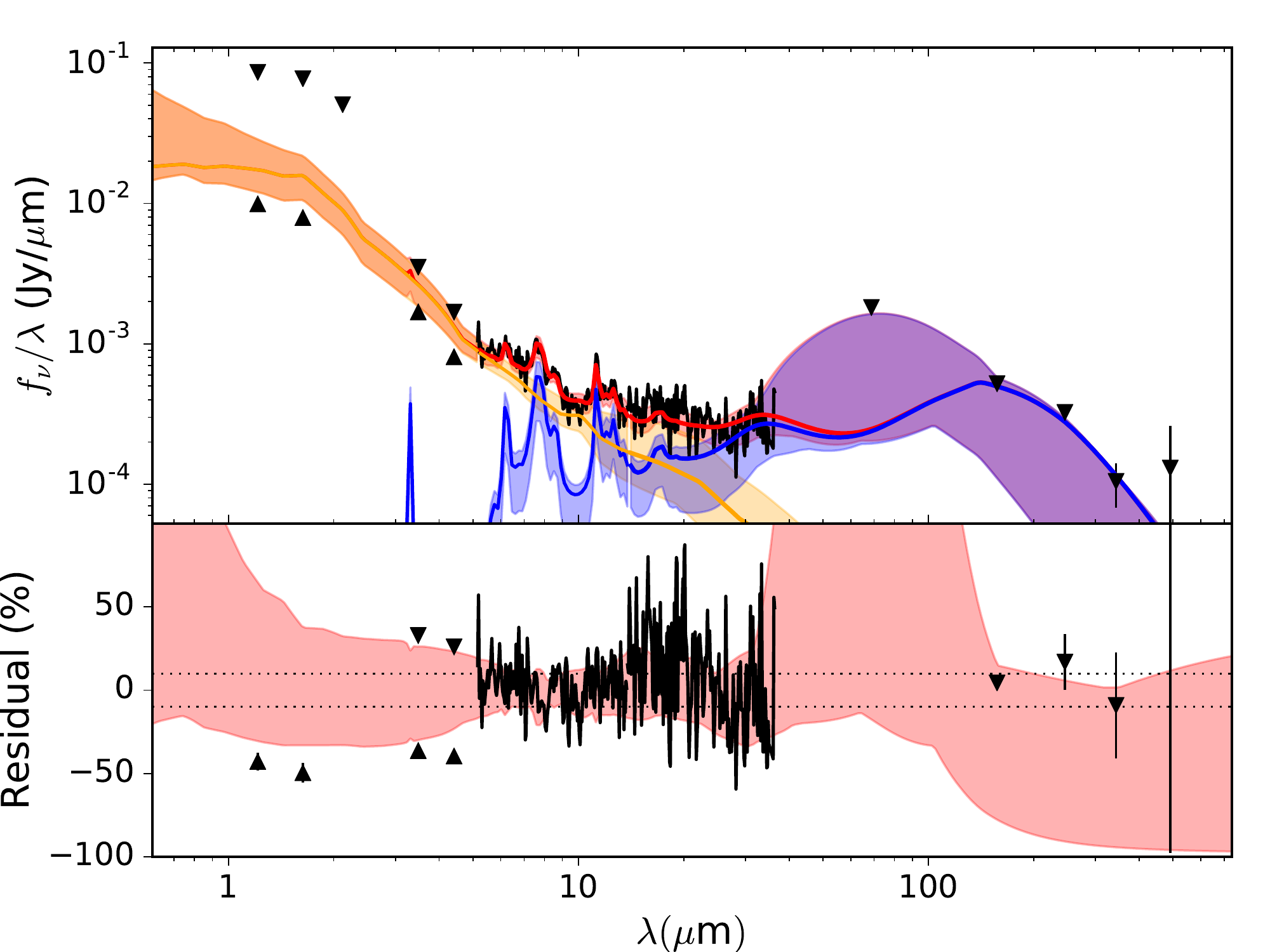} 
\caption{Best continuum fit for NGC 541. Model consists of a simple stellar population (yellow) and diffuse ISM (blue). The total fit is shown in red although in the NIR it appears as orange due to overlap with the stellar population. Upright triangles represent lower limits from photometry and inverted triangles represent upper limits.}
\label{fig:SED-NGC0541-do}
\end{figure}
\end{center}

\begin{table}
\centering
\begin{tabular}{ccccc}
Component & Wavelength & Best fit & Minimum & Maximum\\
 & ($\mu$m) & (\%) & (\%) & (\%)\\
\hhline{=====}
\multirow{4}{1.5cm}{Diffuse ISM} & 5.0 & 2.0 & 0.9 & 2.5\\
 & 15.0 & 43.8 & 21.1 & 52.1\\
 & 30.0 & 82.1 & 36.3 & 100\\
 & 60.0 & 94.7 & 79.0 & 100\\
\hline
\multirow{4}{1.5cm}{Stars} & 5.0 & 98.0 & 80.3 & 100\\
 & 15.0 & 56.2 & 33.0 & 100\\
 & 30.0 & 17.9 & 6.2 & 35.4\\
 & 60.0 & 5.3 & 1.3 & 10.1\\
\hline
\end{tabular}
\caption{Fractional contributions of the various model components to the overall continuum in the best fit model of NGC 541.\label{tab:NGC0541_frac_do}}
\end{table}

\begin{center}
\begin{figure}[ht]
\centering
\includegraphics[width=0.8\linewidth]{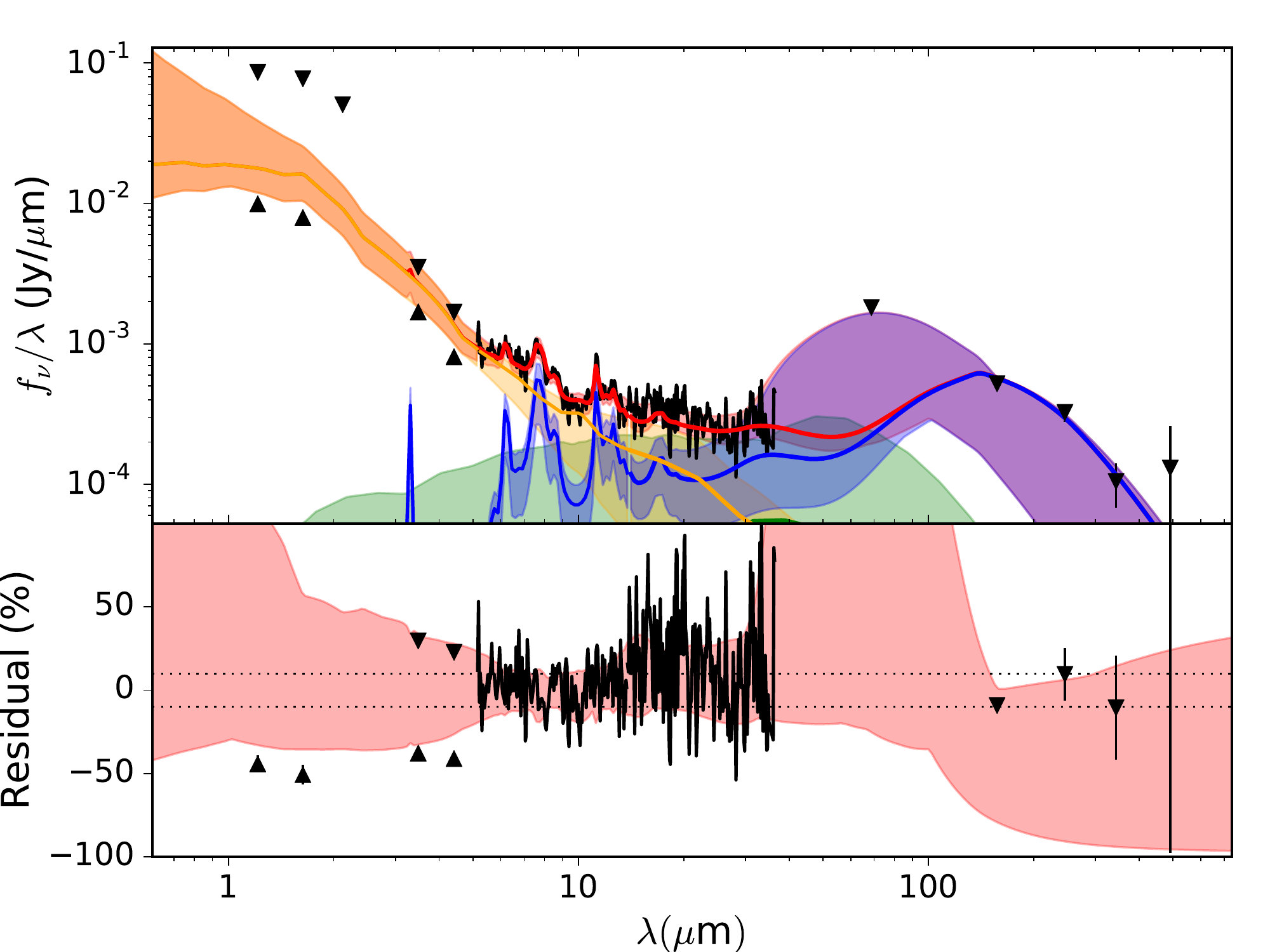} 
\caption{Continuum fit for NGC 541 forcing a torus component (green). Model is otherwise the same as in Figure \ref{fig:SED-NGC0541-do}.}
\label{fig:SED-NGC0541-cdo}
\end{figure}
\end{center}

\begin{table}
\centering
\begin{tabular}{ccccc}
Component & Wavelength & Best fit & Minimum & Maximum\\
 & ($\mu$m) & (\%) & (\%) & (\%)\\
\hhline{=====}
\multirow{4}{1.5cm}{Torus} & 5.0 & 0.0 & 0.0 & 13.9\\
 & 15.0 & 4.3 & 0.0 & 78.8\\
 & 30.0 & 20.7 & 0.0 & 81.4\\
 & 60.0 & 15.7 & 0.0 & 100\\
\hline
\multirow{4}{1.5cm}{Diffuse ISM} & 5.0 & 1.8 & 0.8 & 2.4\\
 & 15.0 & 36.8 & 20.2 & 51.4\\
 & 30.0 & 57.4 & 23.3 & 100\\
 & 60.0 & 78.6 & 39.9 & 100\\
\hline
\multirow{4}{1.5cm}{Stars} & 5.0 & 98.1 & 74.6 & 100\\
 & 15.0 & 58.9 & 11.9 & 99.6\\
 & 30.0 & 21.9 & 1.7 & 40.1\\
 & 60.0 & 5.7 & 0.2 & 10.1\\
\hline
\end{tabular}
\caption{Fractional contributions of the various model components to the overall continuum in NGC 541 with a forced torus component.\label{tab:NGC0541_frac_cdo}}
\end{table}

\subsection{3C~66B}
\label{sec:3C66Bcont}
\FloatBarrier

The best fit for this source requires a prominent hot dust component and a significant NLR model component. The best fit to the diffuse ISM component is unusually dim but the uncertainties are wide.

The upper limit provided by the fit with a forced torus component is a significant fraction of the total flux but the forced torus is unusually cold, suppresses the diffuse ISM component, and those two fits have a large BIC difference (best: 472.3, forced torus: 506.8) of 34.5. We note that the BIC difference may be primarily due to the removal of the hot dust component since we expect the most degeneracy between the NLR and torus components as they are both MIR thermal components. Overall, we cannot rule out the possibility of a torus, especially given the presence of another MIR thermal component, but we also do not favor the presence of a torus so we consider our results inconclusive with regard to the presence of a clumpy torus but our results quite strongly indicate the presence of some MIR thermal component.

\begin{center}
\begin{figure}[htb]
\centering
\includegraphics[width=0.8\linewidth]{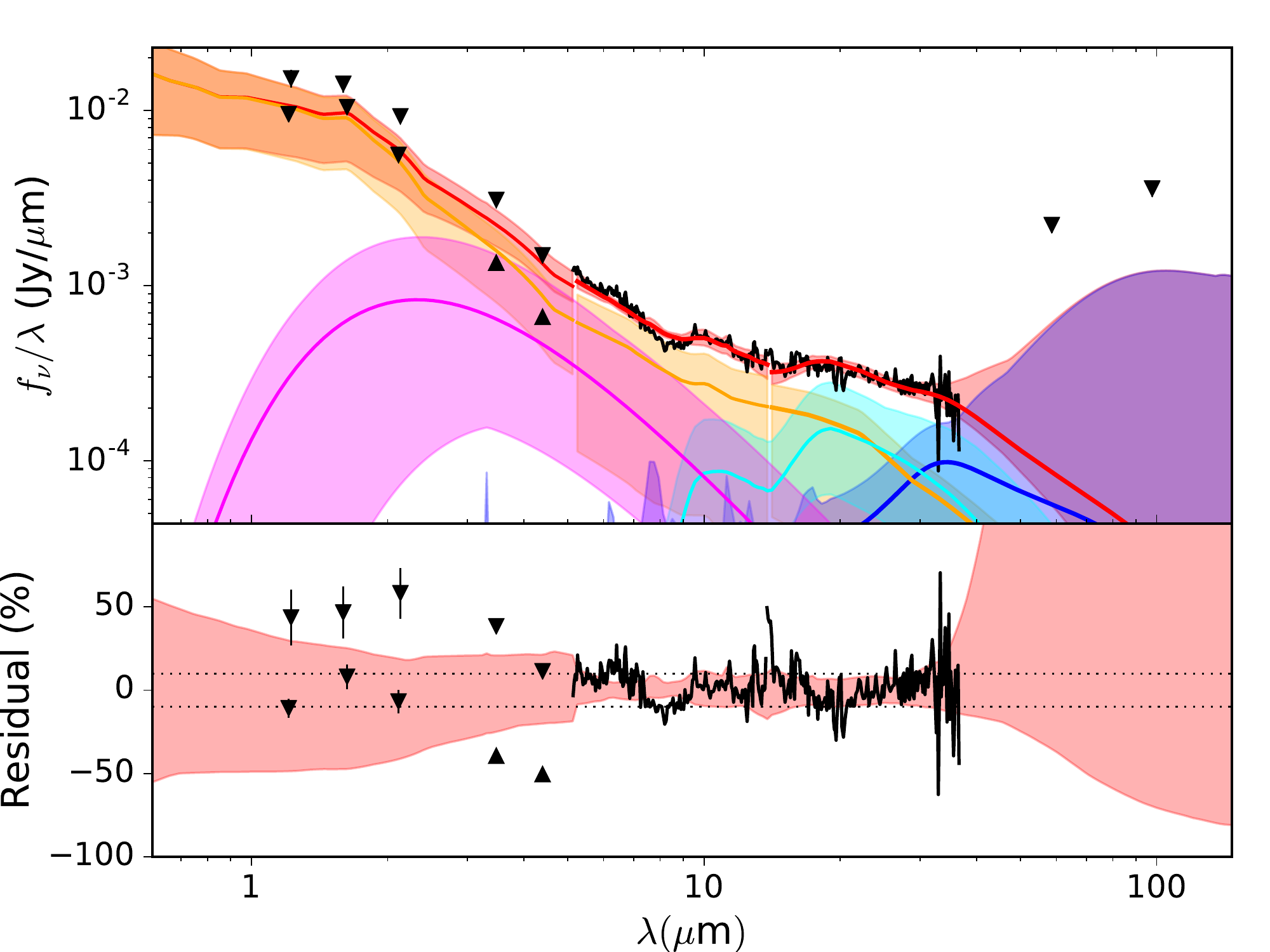} 
\caption{Best continuum fit for 3C 66B. Model consists of a simple stellar population (yellow), hot dust (pink), NLR (cyan), and diffuse ISM (blue). The total fit is shown in red although in the NIR it appears as orange due to overlap with the stellar population. We have applied aperture correction between the SL and LL regions. Upright triangles represent lower limits from photometry and inverted triangles represent upper limits.}
\label{fig:SED-UGC01841-nbdoa}
\end{figure}
\end{center}

\begin{table}
\centering
\begin{tabular}{ccccc}
Component & Wavelength & Best fit & Minimum & Maximum\\
 & ($\mu$m) & (\%) & (\%) & (\%)\\
\hhline{=====}
\multirow{4}{1.5cm}{Diffuse ISM} & 5.0 & 0.1 & 0.0 & 0.5\\
 & 15.0 & 5.9 & 0.1 & 11.8\\
 & 30.0 & 35.1 & 0.5 & 58.9\\
 & 60.0 & 66.7 & 1.1 & 100\\
\hline
\multirow{4}{1.5cm}{Stars} & 5.0 & 64.0 & 31.4 & 83.2\\
 & 15.0 & 59.5 & 13.2 & 80.3\\
 & 30.0 & 28.7 & 3.1 & 38.6\\
 & 60.0 & 19.6 & 1.6 & 26.1\\
\hline
\multirow{4}{1.5cm}{NLR} & 5.0 & 0.0 & 0.0 & 0.1\\
 & 15.0 & 26.1 & 11.6 & 49.2\\
 & 30.0 & 34.6 & 13.0 & 70.6\\
 & 60.0 & 12.9 & 4.4 & 48.2\\
\hline
\multirow{4}{1.5cm}{Hot dust} & 5.0 & 35.9 & 8.8 & 84.5\\
 & 15.0 & 8.5 & 2.3 & 24.1\\
 & 30.0 & 1.7 & 0.4 & 5.1\\
 & 60.0 & 0.7 & 0.2 & 2.2\\
\hline
\end{tabular}
\caption{Fractional contributions of the various model components to the overall continuum in the best fit model for 3C 66B.\label{tab:UGC01841_frac_nbdoa}}
\end{table}

\begin{center}
\begin{figure}[htb]
\centering
\includegraphics[width=0.8\linewidth]{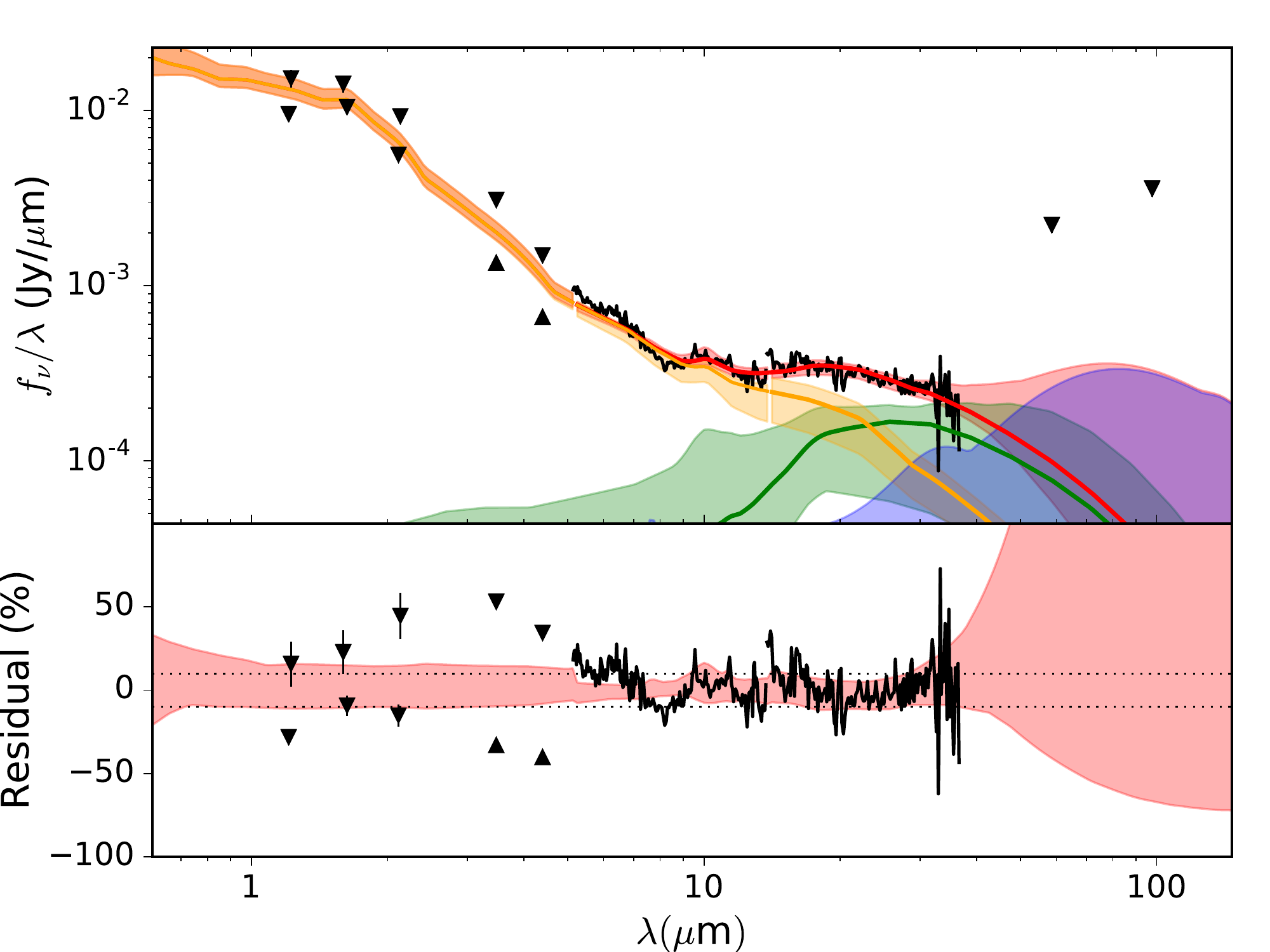} 
\caption{Continuum fit for 3C 66B forcing a torus component (green). To ensure maximal torus component we excluded the NLR and hot dust components seen in Figure \ref{fig:SED-UGC01841-nbdoa}.}
\label{fig:SED-UGC01841-cdoa}
\end{figure}
\end{center}

\begin{table}
\centering
\begin{tabular}{ccccc}
Component & Wavelength & Best fit & Minimum & Maximum\\
 & ($\mu$m) & (\%) & (\%) & (\%)\\
\hhline{=====}
\multirow{4}{1.5cm}{Torus} & 5.0 & 0.2 & 0.0 & 7.2\\
 & 15.0 & 26.7 & 11.4 & 50.0\\
 & 30.0 & 65.1 & 20.7 & 81.9\\
 & 60.0 & 78.7 & 17.2 & 100\\
\hline
\multirow{4}{1.5cm}{Diffuse ISM} & 5.0 & 0.0 & 0.0 & 0.3\\
 & 15.0 & 0.0 & 0.0 & 7.7\\
 & 30.0 & 0.0 & 0.0 & 42.6\\
 & 60.0 & 0.1 & 0.0 & 100\\
\hline
\multirow{4}{1.5cm}{Stars} & 5.0 & 99.8 & 90.9 & 100\\
 & 15.0 & 73.2 & 48.2 & 88.3\\
 & 30.0 & 34.9 & 22.5 & 41.8\\
 & 60.0 & 21.3 & 14.3 & 25.1\\
\hline
\end{tabular}
\caption{Fractional contributions of the various model components to the overall continuum in 3C 66B with a forced torus component.\label{tab:UGC01841_frac_cdoa}}
\end{table}

\subsection{NGC~3801}
\label{sec:NGC3801cont}
\FloatBarrier

The best fit for NGC 3801 requires a NLR component, however it is very dim so we suspect it is likely either a fitting artifact or possibly associated with a large-scale structure in the host galaxy. A likely candidate for this component from the host is the large dust lane visible near the AGN in \emph{HST} images by \cite{VerdoesKleijn99}. The forced-torus fit is significantly worse, with a BIC of 453.7 compared to the best-fit BIC of 423.5. We also show an extincted galaxy-only fit with a BIC of 437.3 in Figure \ref{fig:SED-NGC3801-doae} as a comparison since the dust lane which is a candidate thermal source would also provide foreground extinction on the core. Overall, we do not find evidence for nuclear obscuration in the form of a clumpy torus and our results are ambiguous for any other warm thermal component in the nucleus itself due to likely emission from the host galaxy.

\cite{Das05} detected a flat-spectrum non-thermal core in NGC~3801 in $3\,\mathrm{mm}$ images from the Berkeley-Illinois-Maryland Association millimeter-wave array. This does not appear to contribute significantly to the IRS band and we have too little far-infrared photometry to properly compare. They also detected carbon monoxide CO(1-0) emission from the dust disk seen in \emph{HST} images \citep{VerdoesKleijn99} with a velocity gradient indicating a $\sim 2\,\mathrm{kpc}$ rotating disk of molecular gas and dust with an inferred $3\times 10^8\,\mathrm{M}_\odot$ of molecular hydrogen. They also detected a $\sim 10^8\,\mathrm{M}_\odot$ infalling molecular gas clump which they attribute to a recent merger. 

\cite{Hota12} found evidence in \emph{GALEX} images of a kinematically decoupled core in NGC~3801 in which star formation has recently declined following a post-merger burst but the jet-driven shock has not yet triggered a burst of star formation in the outer regions to the galaxy. In comparison, our SFR estimates based on PAH fit results in Section \ref{sec:spectFit} are higher than average for our sample and our SFR estimates based on narrow Neon lines are typical for the sample. Our continuum models have a bright diffuse ISM which may be associated with the large-scale gas \cite{Hota12} predict will begin to form stars within the next 10~Myr. Our stellar population models show a high stellar age of 9.4~Gyr but as mentioned earlier our model only accounts for one burst of star formation followed by passive evolution so our results only suggest that the population of low-mass stars is still dominated by the older sub-population.

\begin{center}
\begin{figure}[ht]
\centering
\includegraphics[width=0.8\linewidth]{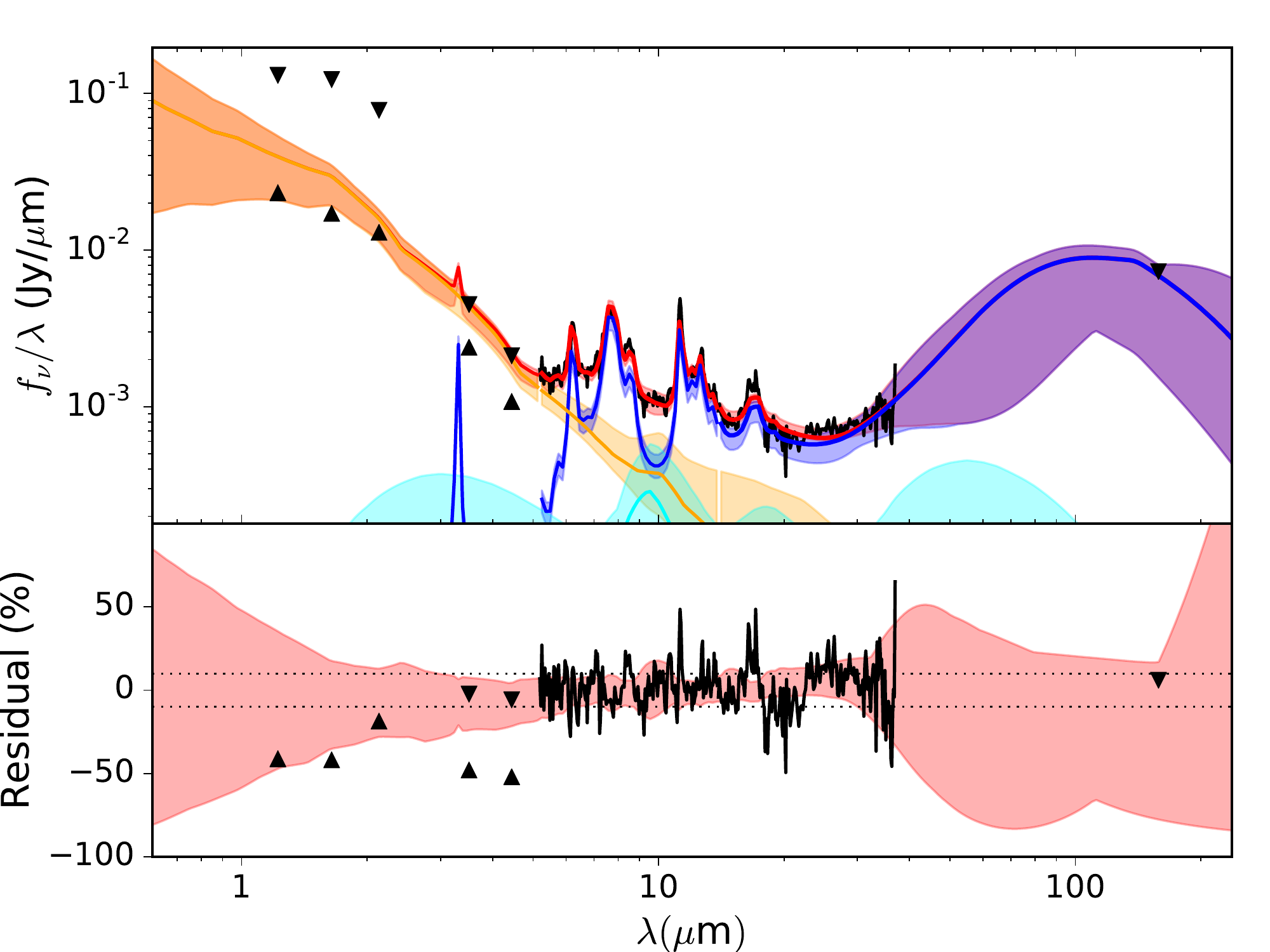} 
\caption{Best continuum fit for NGC 3801. Model consists of a simple stellar population (yellow), NLR (cyan) and diffuse ISM (blue). The total fit is shown in red although in the NIR it appears as orange due to overlap with the stellar population.Upright triangles represent lower limits from photometry and inverted triangles represent upper limits.}
\label{fig:SED-NGC3801-ndo}
\end{figure}
\end{center}

\begin{center}
\begin{figure}[ht]
\centering
\includegraphics[width=0.8\linewidth]{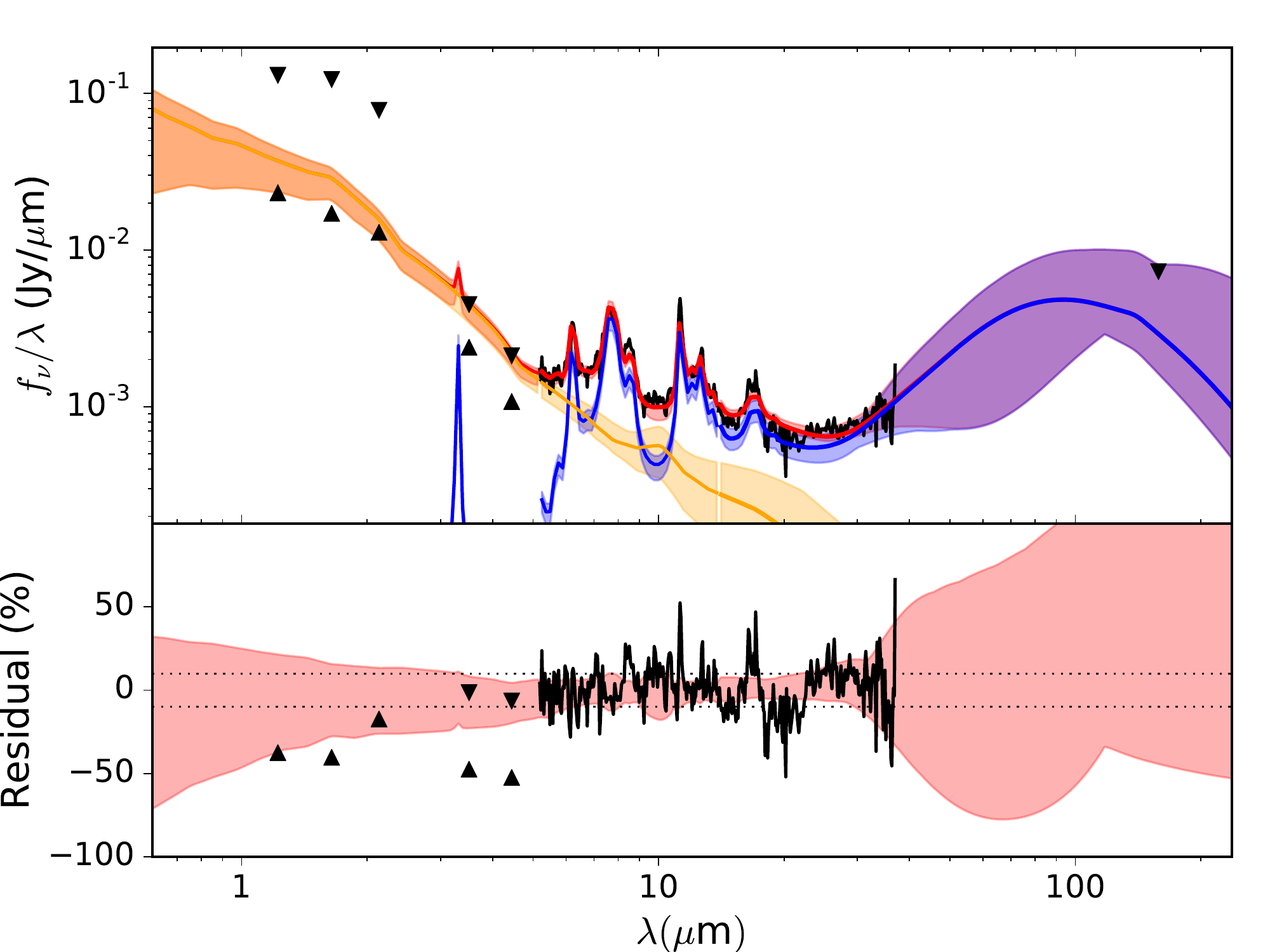} 
\caption{Galaxy-only continuum fit for NGC 3801. Model consists of a simple stellar population (yellow) and diffuse ISM (blue). The total fit is shown in red although in the NIR it appears as orange due to overlap with the stellar population. This is a somewhat poorer fit than that in Figure \ref{fig:SED-NGC3801-ndo} and still better than the forced-torus fit in Figure \ref{fig:SED-NGC3801-cdo}. This model also requires a foreground extinction component. Upright triangles represent lower limits from photometry and inverted triangles represent upper limits.}
\label{fig:SED-NGC3801-doae}
\end{figure}
\end{center}

\begin{table}
\centering
\begin{tabular}{ccccc}
Component & Wavelength & Best fit & Minimum & Maximum\\
 & ($\mu$m) & (\%) & (\%) & (\%)\\
\hhline{=====}
\multirow{4}{1.5cm}{Diffuse ISM} & 5.0 & 7.0 & 5.7 & 8.0\\
 & 15.0 & 79.6 & 64.2 & 88.2\\
 & 30.0 & 96.4 & 74.9 & 100\\
 & 60.0 & 99.9 & 19.7 & 100\\
\hline
\multirow{4}{1.5cm}{Stars} & 5.0 & 85.9 & 69.5 & 99.6\\
 & 15.0 & 16.1 & 6.7 & 44.6\\
 & 30.0 & 2.8 & 1.0 & 17.8\\
 & 60.0 & 0.1 & 0.0 & 0.7\\
\hline
\multirow{4}{1.5cm}{NLR} & 5.0 & 7.1 & 0.0 & 14.3\\
 & 15.0 & 4.3 & 0.0 & 21.3\\
 & 30.0 & 0.8 & 0.0 & 14.8\\
 & 60.0 & 0.0 & 0.0 & 10.5\\
\hline
\end{tabular}
\caption{Fractional contributions of the various model components to the overall continuum in the best fit model for NGC 3801.\label{tab:NGC3801_frac_ndo}}
\end{table}

\begin{center}
\begin{figure}[ht]
\centering
\includegraphics[width=0.8\linewidth]{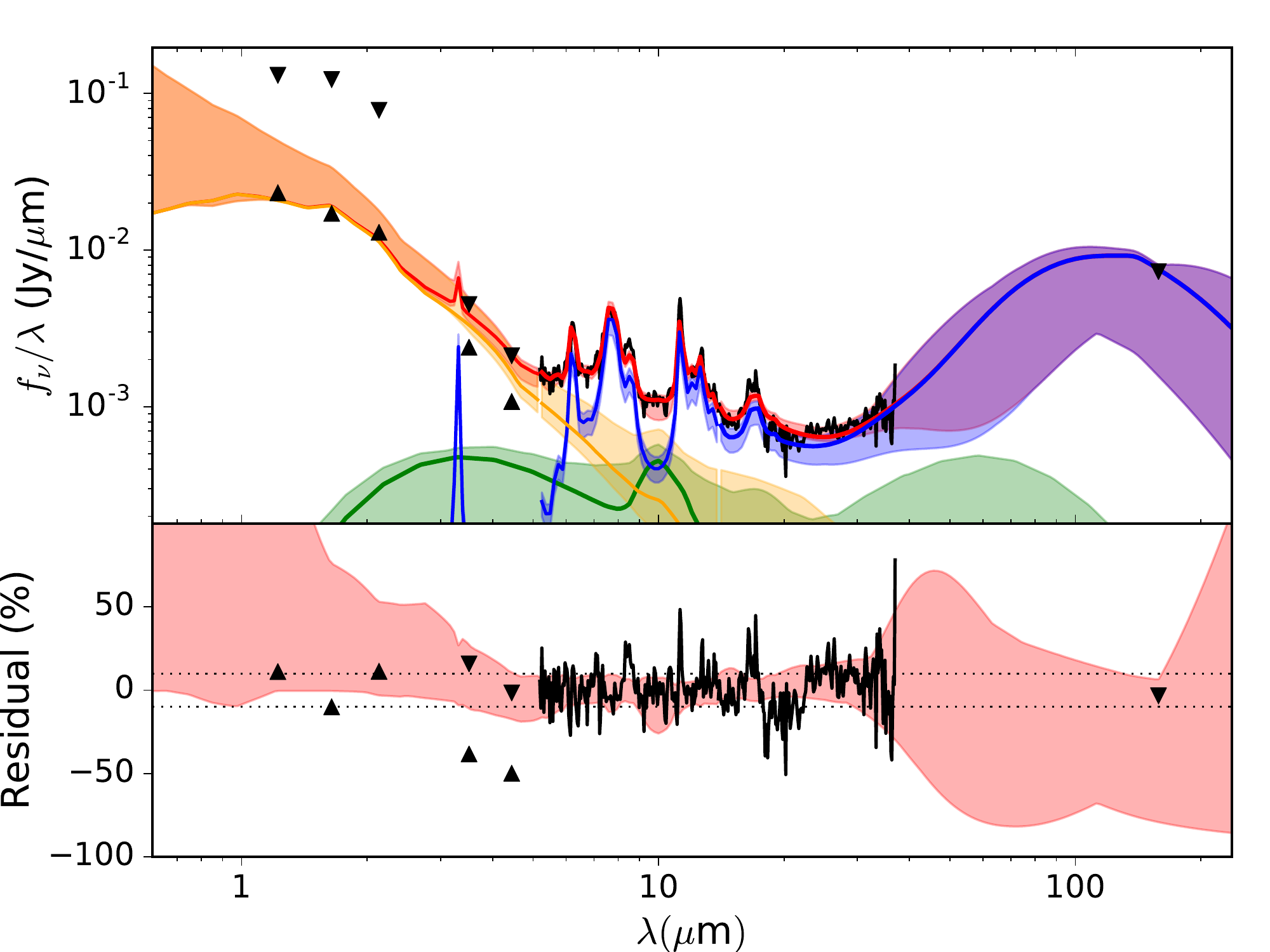} 
\caption{Continuum fit for NGC 3801 forcing a torus component (green). To ensure maximal torus component we excluded an NLR component seen in Figure \ref{fig:SED-NGC3801-ndo} from the fit.}
\label{fig:SED-NGC3801-cdo}
\end{figure}
\end{center}

\begin{table}
\centering
\begin{tabular}{ccccc}
Component & Wavelength & Best fit & Minimum & Maximum\\
 & ($\mu$m) & (\%) & (\%) & (\%)\\
\hhline{=====}
\multirow{4}{1.5cm}{Torus} & 5.0 & 23.0 & 0.0 & 30.0\\
 & 15.0 & 13.0 & 0.0 & 34.2\\
 & 30.0 & 3.8 & 0.0 & 34.3\\
 & 60.0 & 0.0 & 0.0 & 13.1\\
\hline
\multirow{4}{1.5cm}{Diffuse ISM} & 5.0 & 6.6 & 5.5 & 8.1\\
 & 15.0 & 76.7 & 61.2 & 87.2\\
 & 30.0 & 93.7 & 61.5 & 100\\
 & 60.0 & 99.9 & 18.2 & 100\\
\hline
\multirow{4}{1.5cm}{Stars} & 5.0 & 70.4 & 58.6 & 100\\
 & 15.0 & 10.3 & 5.3 & 45.3\\
 & 30.0 & 2.5 & 0.8 & 18.7\\
 & 60.0 & 0.1 & 0.0 & 0.9\\
\hline
\end{tabular}
\caption{Fractional contributions of the various model components to the overall continuum in NGC 3801 with a forced torus component.\label{tab:NGC3801_frac_cdoa}}
\end{table}

\subsection{NGC~3862}
\label{sec:NGC3862cont}
\FloatBarrier

The best fit model for this source requires only components from the host galaxy suggesting that the AGN does not contribute significantly to the IR flux. The largest difference between the best fit model and the data is the 70$\,\mu$m but that point is an upper limit and within the uncertainty on the best fit model spectrum.

The forced torus model in this source is quite reasonable albeit rather warm with a peak around $10\,\mu \mathrm{m}$, providing a firm upper limit on any possible torus component of $\sim 17\%$ of the flux at $5\,\mu \mathrm{m}$, $\sim 50\%$ of the flux at $15\,\mu \mathrm{m}$, and $\sim 13\%$ of the flux at $30\,\mu \mathrm{m}$. However, it is also unnecessary to explain the infrared spectrum as the best fit has a BIC of $391.9$ and the forced-torus fit has a BIC of $420.3$. Therefore, we see no evidence for a warm IR-bright obscuring structure in NGC~3862.

\begin{center}
\begin{figure}[ht]
\centering
\includegraphics[width=0.8\linewidth]{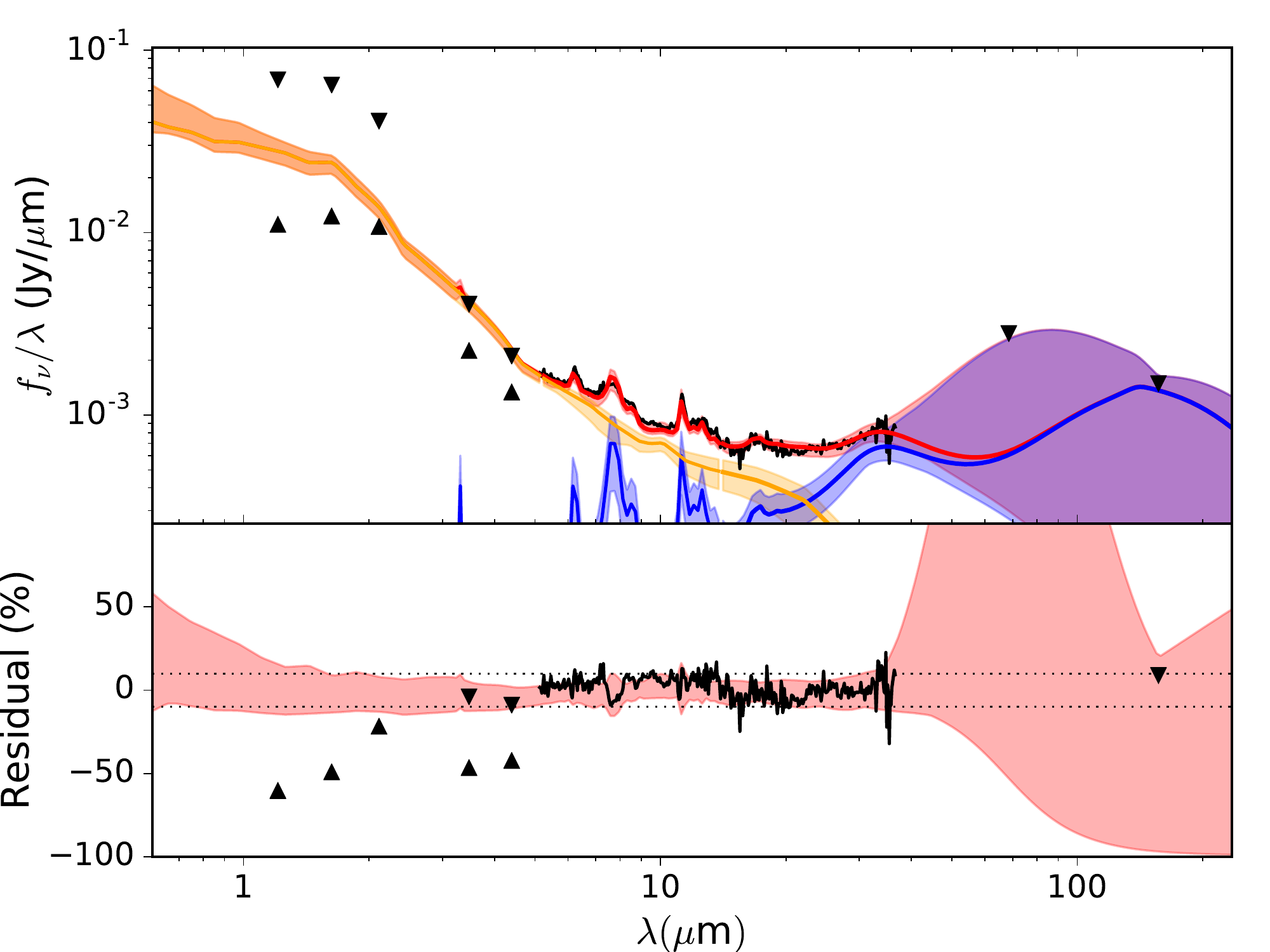} 
\caption{Best continuum fit for NGC~3862. Model consists of a simple stellar population (yellow) and diffuse ISM (blue). The total fit is shown in red although in the NIR it appears as orange due to overlap with the stellar population. Upright triangles represent lower limits from photometry and inverted triangles represent upper limits.}
\label{fig:SED-NGC3862-do}
\end{figure}
\end{center}

\begin{table}
\centering
\begin{tabular}{ccccc}
Component & Wavelength & Best fit & Minimum & Maximum\\
 & ($\mu$m) & (\%) & (\%) & (\%)\\
\hhline{=====}
\multirow{4}{1.5cm}{Diffuse ISM} & 5.0 & 1.5 & 0.8 & 2.0\\
 & 15.0 & 30.1 & 21.2 & 37.6\\
 & 30.0 & 77.4 & 63.5 & 95.7\\
 & 60.0 & 93.3 & 54.8 & 100\\
\hline
\multirow{4}{1.5cm}{Stars} & 5.0 & 98.5 & 89.2 & 100\\
 & 15.0 & 69.9 & 54.9 & 81.6\\
 & 30.0 & 22.6 & 10.2 & 26.1\\
 & 60.0 & 6.7 & 2.5 & 7.6\\
\hline
\end{tabular}
\caption{Fractional contributions of the various model components to the overall continuum in the best fit model for NGC 3862.\label{tab:NGC3862_frac_do}}
\end{table}

\begin{center}
\begin{figure}[ht]
\centering
\includegraphics[width=0.8\linewidth]{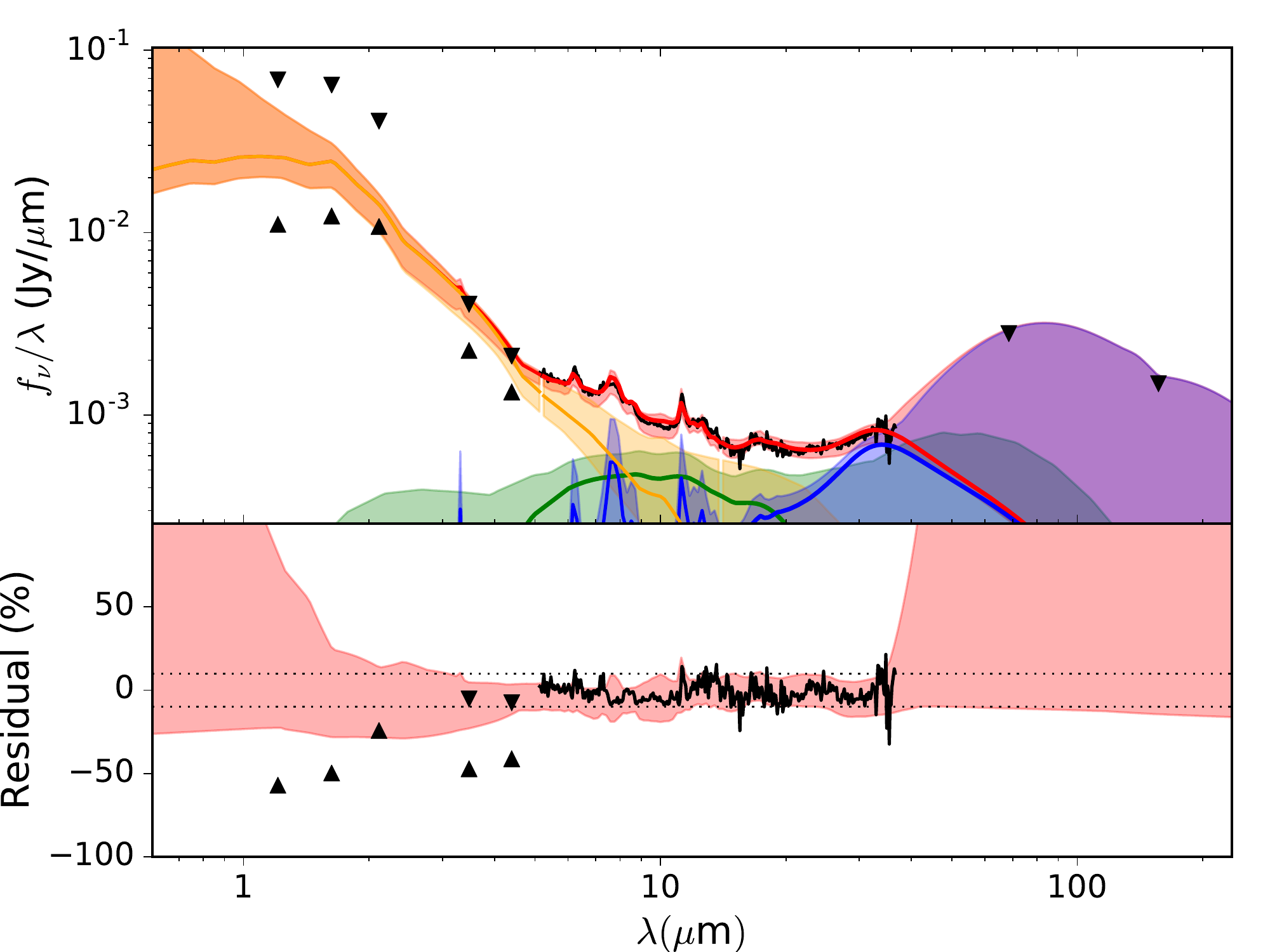} 
\caption{Continuum fit for NGC 3862 forcing a torus component (green). Model is otherwise the same as in Figure \ref{fig:SED-NGC3862-do}.}
\label{fig:SED-NGC3862-cdo}
\end{figure}
\end{center}

\begin{table}
\centering
\begin{tabular}{ccccc}
Component & Wavelength & Best fit & Minimum & Maximum\\
 & ($\mu$m) & (\%) & (\%) & (\%)\\
\hhline{=====}
\multirow{4}{1.5cm}{Torus} & 5.0 & 16.7 & 0.0 & 27.3\\
 & 15.0 & 49.7 & 0.0 & 69.6\\
 & 30.0 & 13.0 & 0.0 & 68.5\\
 & 60.0 & 1.9 & 0.0 & 100\\
\hline
\multirow{4}{1.5cm}{Diffuse ISM} & 5.0 & 1.3 & 0.5 & 2.0\\
 & 15.0 & 27.1 & 13.4 & 36.1\\
 & 30.0 & 78.1 & 12.4 & 87.6\\
 & 60.0 & 92.6 & 37.9 & 100\\
\hline
\multirow{4}{1.5cm}{Stars} & 5.0 & 82.0 & 63.6 & 100\\
 & 15.0 & 23.2 & 8.1 & 82.4\\
 & 30.0 & 9.0 & 0.9 & 25.1\\
 & 60.0 & 5.5 & 0.2 & 12.1\\
\hline
\end{tabular}
\caption{Fractional contributions of the various model components to the overall continuum in NGC 3862 with a forced torus component.\label{tab:NGC3862_frac_cdo}}
\end{table}

\subsection{3C~270}
\label{sec:3C270cont}
\FloatBarrier

\begin{center}
\begin{figure}[ht]
\centering
\includegraphics[width=0.8\linewidth]{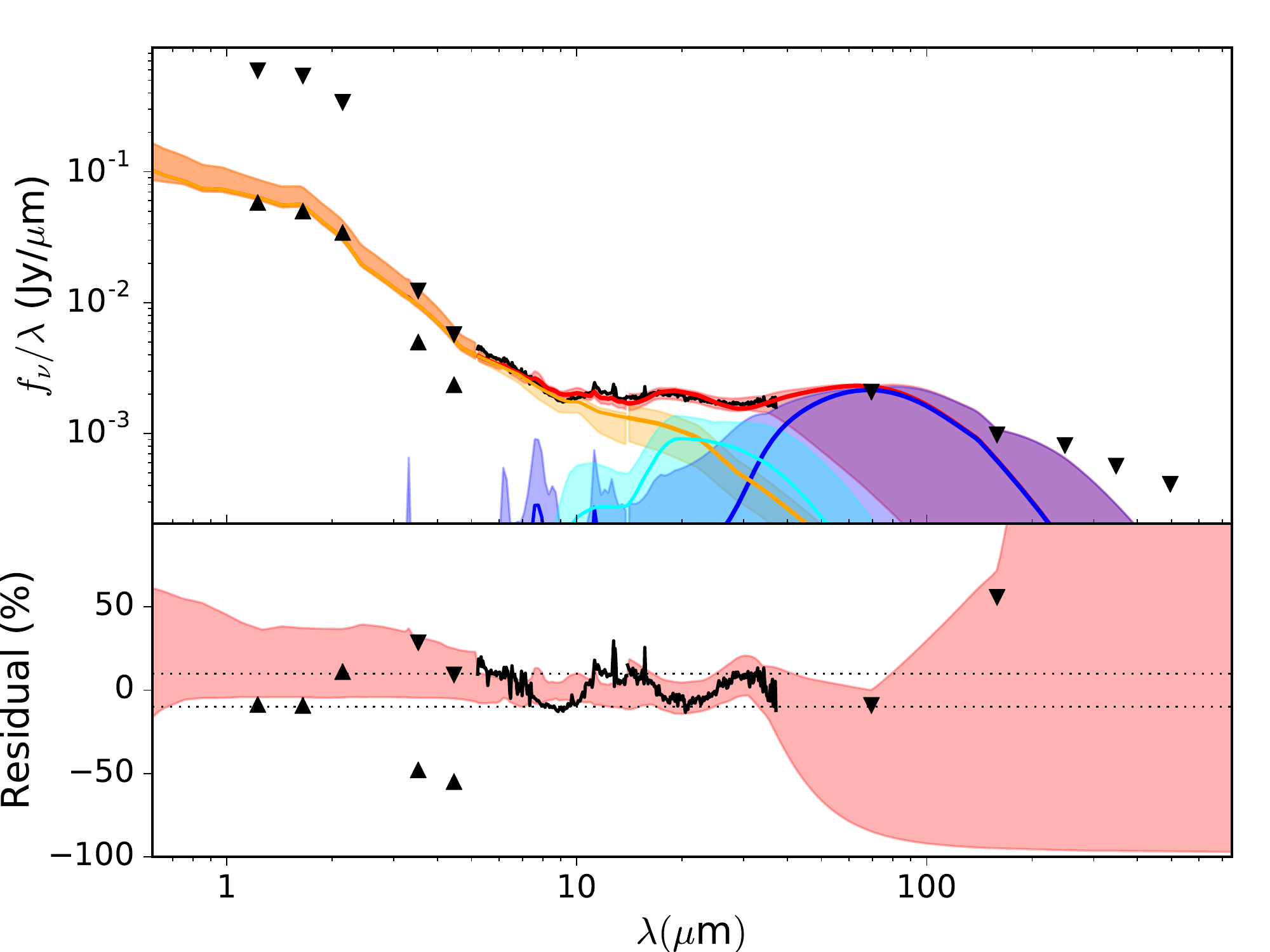} 
\caption{Best continuum fit for 3C 270. Model consists of a simple stellar population (yellow), NLR (cyan) and diffuse ISM (blue). The total fit is shown in red although in the NIR it appears as orange due to overlap with the stellar population. We have applied aperture correction between the SL and LL regions. Upright triangles represent lower limits from photometry and inverted triangles represent upper limits.}
\label{fig:SED-3C270-ndoa}
\end{figure}
\end{center}

\begin{center}
\begin{figure}[ht]
\centering
\includegraphics[width=0.8\linewidth]{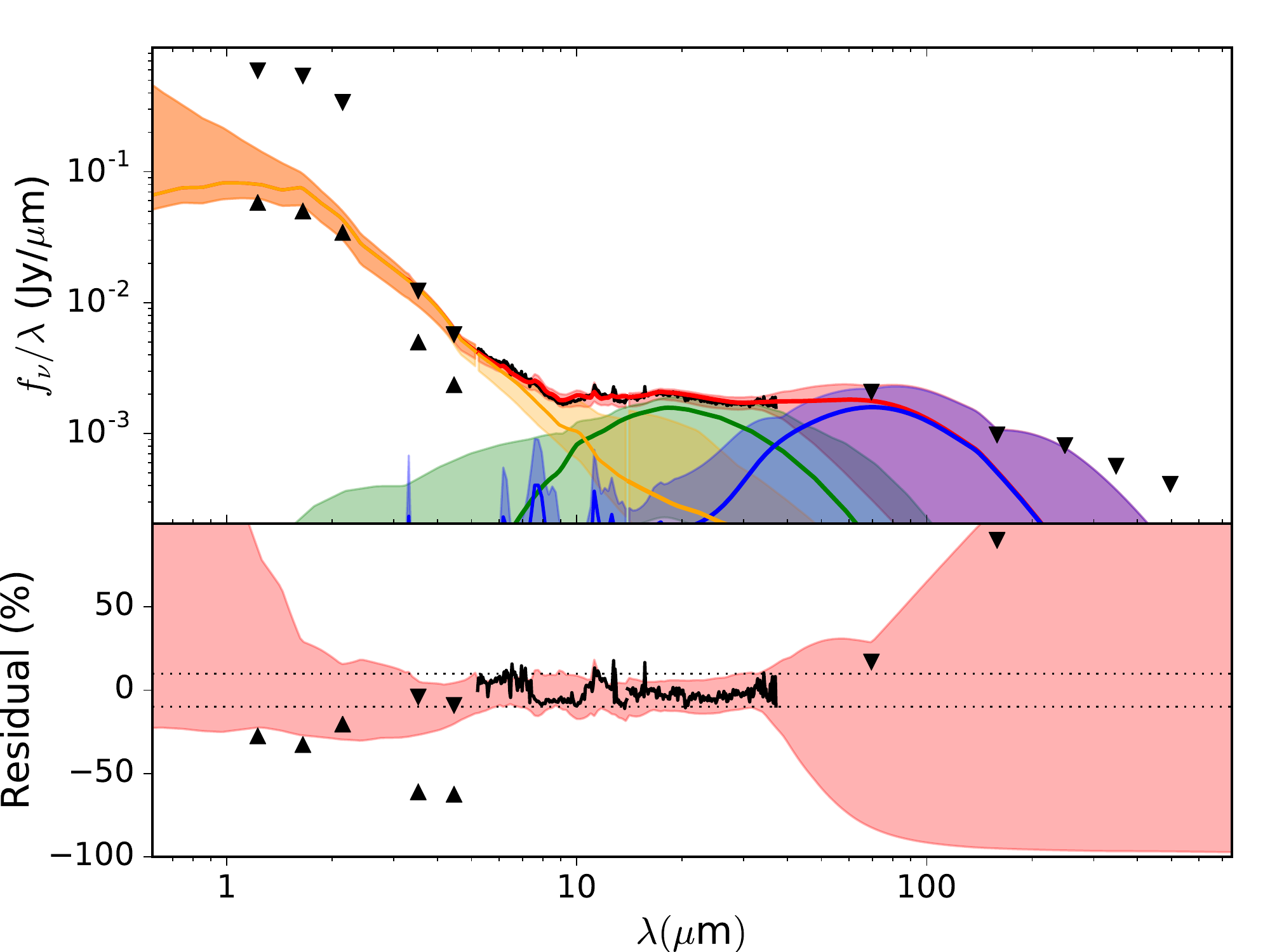} 
\caption{Continuum fit for 3C 270 forcing a torus component (green). To ensure maximal contribution from the torus component we excluded the NLR component from Figure \ref{fig:SED-3C270-ndoa}.}
\label{fig:SED-3C270-cdoa}
\end{figure}
\end{center}

We show our best fit for 3C 270 in Figure \ref{fig:SED-3C270-ndoa}, the best fit with a forced torus component in Figure \ref{fig:SED-3C270-cdoa}, and the fractional contributions to the flux in each of the two models in Tables \ref{tab:3C270_frac_ndoa} and \ref{tab:3C270_frac_cdoa} respectively. We suspect that the preference for a narrow-line region model in this source is due to a bias in our analysis since BIC favors models with fewer parameters and the NLR model has fewer parameters than a clumpy torus model. Therefore we cannot rule out a torus in this source. The presence of a torus in 3C 270 would be consistent with results by Antonucci (private communication) who detected broad polarized H$\alpha$ emission. This is potentially indicative of a Thomson-scattered hidden broad line region such as that expected in a high-excitation AGN, however Antonucci (private communication) caution that it could instead be due to narrow lines in the complex. Additionally,  although \cite{VanderWolk10} ultimately do not conclude the existence of a warm MIR thermal component in any of the eight FR-I radio galaxies in their sample they list 3C 270 as a possible exception due to a weak MIR consistent with 200~K dust. \cite{Jaffe93} note a $\sim 60\,\mathrm{pc}$ disk of cold dust in \emph{HST} images of 3C 270 surrounding its unresolved nucleus, the inner regions of which may be heated by the AGN (or star formation but our SFR estimates are too low for that) to produce the MIR excess detected.

\begin{table}
\centering
\begin{tabular}{ccccc}
Component & Wavelength & Best fit & Minimum & Maximum\\
 & ($\mu$m) & (\%) & (\%) & (\%)\\
\hhline{=====}
\multirow{4}{1.5cm}{Diffuse ISM} & 5.0 & 0.2 & 0.1 & 0.9\\
 & 15.0 & 5.0 & 2.9 & 17.2\\
 & 30.0 & 23.5 & 5.7 & 80.4\\
 & 60.0 & 90.3 & 7.7 & 94.5\\
\hline
\multirow{4}{1.5cm}{Stars} & 5.0 & 99.8 & 93.1 & 100\\
 & 15.0 & 72.4 & 46.7 & 90.2\\
 & 30.0 & 29.8 & 18.1 & 37.0\\
 & 60.0 & 4.5 & 2.5 & 5.6\\
\hline
\multirow{4}{1.5cm}{NLR} & 5.0 & 0.0 & 0.0 & 0.1\\
 & 15.0 & 22.6 & 3.1 & 37.4\\
 & 30.0 & 46.7 & 1.8 & 78.0\\
 & 60.0 & 5.2 & 0.1 & 15.6\\
\hline
\end{tabular}
\caption{Fractional contributions of the various model components to the overall continuum in the best fit model for 3C 270.\label{tab:3C270_frac_ndoa}}
\end{table}

\begin{table}
\centering
\begin{tabular}{ccccc}
Component & Wavelength & Best fit & Minimum & Maximum\\
 & ($\mu$m) & (\%) & (\%) & (\%)\\
\hhline{=====}
\multirow{4}{1.5cm}{Torus} & 5.0 & 1.5 & 0.1 & 15.6\\
 & 15.0 & 73.6 & 10.6 & 86.7\\
 & 30.0 & 63.8 & 8.6 & 100.0\\
 & 60.0 & 13.0 & 1.6 & 44.0\\
\hline
\multirow{4}{1.5cm}{Diffuse ISM} & 5.0 & 0.3 & 0.0 & 0.7\\
 & 15.0 & 6.5 & 0.9 & 13.9\\
 & 30.0 & 25.8 & 1.4 & 63.7\\
 & 60.0 & 84.0 & 4.5 & 100\\
\hline
\multirow{4}{1.5cm}{Stars} & 5.0 & 98.2 & 75.7 & 100\\
 & 15.0 & 20.0 & 9.2 & 74.7\\
 & 30.0 & 10.4 & 1.3 & 30.7\\
 & 60.0 & 3.0 & 0.2 & 6.6\\
\hline
\end{tabular}
\caption{Fractional contributions of the various model components to the overall continuum in 3C 270 with a forced torus component.\label{tab:3C270_frac_cdoa}}
\end{table}

\subsection{M~84}
\label{sec:M84cont}
\FloatBarrier

The best fit model for this source requires two AGN components: a hot dust component and an NLR. Unfortunately, we have no MIPS photometry and only one IRAC channel for M~84. 

The forced torus component in the forced-torus fit to M~84 is quite reasonable albeit unusually dim with a best-fit AGN luminosity of $1.2\times 10^{41}\,\mathrm{erg}\,\mathrm{s}^{-1}$ (see Table~\ref{tab:torusparams2} and discussion in Section~\ref{sec:discuss}). However, it has the largest difference in goodness-of-fit out of any of our pairs of models with the best fit having a BIC of $386.9$ compared to $444.4$ for the forced-torus fit. At least part of that is likely due to forced suppression of the hot dust component. Recall that the purpose of the forced-torus fits was to provide an upper limit on any potential torus contribution. Overall, we conclude that there is likely an optically thin thermal hot dust component and that the bumps in the spectrum toward the MIR are probably a thermal component although likely not in the form of a \cite{nenkova02} clumpy torus.

\begin{center}
\begin{figure}[ht]
\centering
\includegraphics[width=0.8\linewidth]{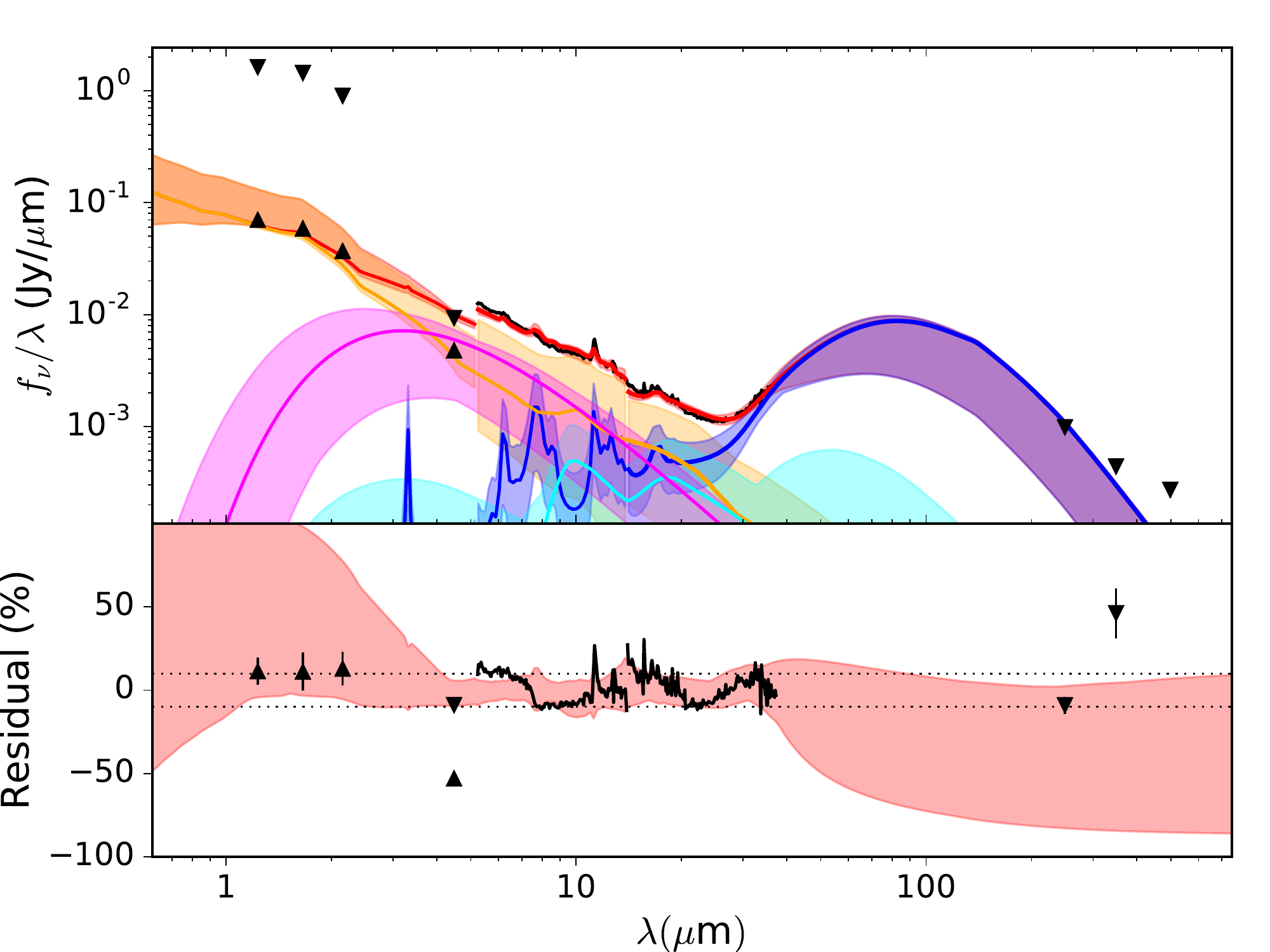} 
\caption{Best continuum fit for M 84. Model consists of a simple stellar population (yellow), hot dust (pink), NLR (cyan), and diffuse ISM (blue). The total fit is shown in red although in the NIR it appears as orange due to overlap with the stellar population. We have applied aperture correction between the SL and LL regions but this aperture correction was a fitting parameter and it seems to have not been fit well. Upright triangles represent lower limits from photometry and inverted triangles represent upper limits.}
\label{fig:SED-M84-nbdoa}
\end{figure}
\end{center}

\begin{table}
\centering
\begin{tabular}{ccccc}
Component & Wavelength & Best fit & Minimum & Maximum\\
 & ($\mu$m) & (\%) & (\%) & (\%)\\
\hhline{=====}
\multirow{4}{1.5cm}{Diffuse ISM} & 5.0 & 0.5 & 0.1 & 1.3\\
 & 15.0 & 19.9 & 8.8 & 35.7\\
 & 30.0 & 71.5 & 51.2 & 94.5\\
 & 60.0 & 99.3 & 40.3 & 100\\
\hline
\multirow{4}{1.5cm}{Stars} & 5.0 & 37.6 & 27.4 & 77.4\\
 & 15.0 & 37.7 & 9.3 & 86.9\\
 & 30.0 & 11.3 & 1.8 & 35.0\\
 & 60.0 & 0.4 & 0.0 & 1.4\\
\hline
\multirow{4}{1.5cm}{NLR} & 5.0 & 0.1 & 0.0 & 2.8\\
 & 15.0 & 13.2 & 0.0 & 27.6\\
 & 30.0 & 10.7 & 0.0 & 26.6\\
 & 60.0 & 0.2 & 0.0 & 8.4\\
\hline
\multirow{4}{1.5cm}{Hot dust} & 5.0 & 61.8 & 17.0 & 73.1\\
 & 15.0 & 29.2 & 5.8 & 43.5\\
 & 30.0 & 6.6 & 1.2 & 10.7\\
 & 60.0 & 0.2 & 0.0 & 0.3\\
\hline
\end{tabular}
\caption{Fractional contributions of the various model components to the overall continuum in the best fit model for M 84.\label{tab:M84_frac_nbdoa}}
\end{table}

\begin{center}
\begin{figure}[ht]
\centering
\includegraphics[width=0.8\linewidth]{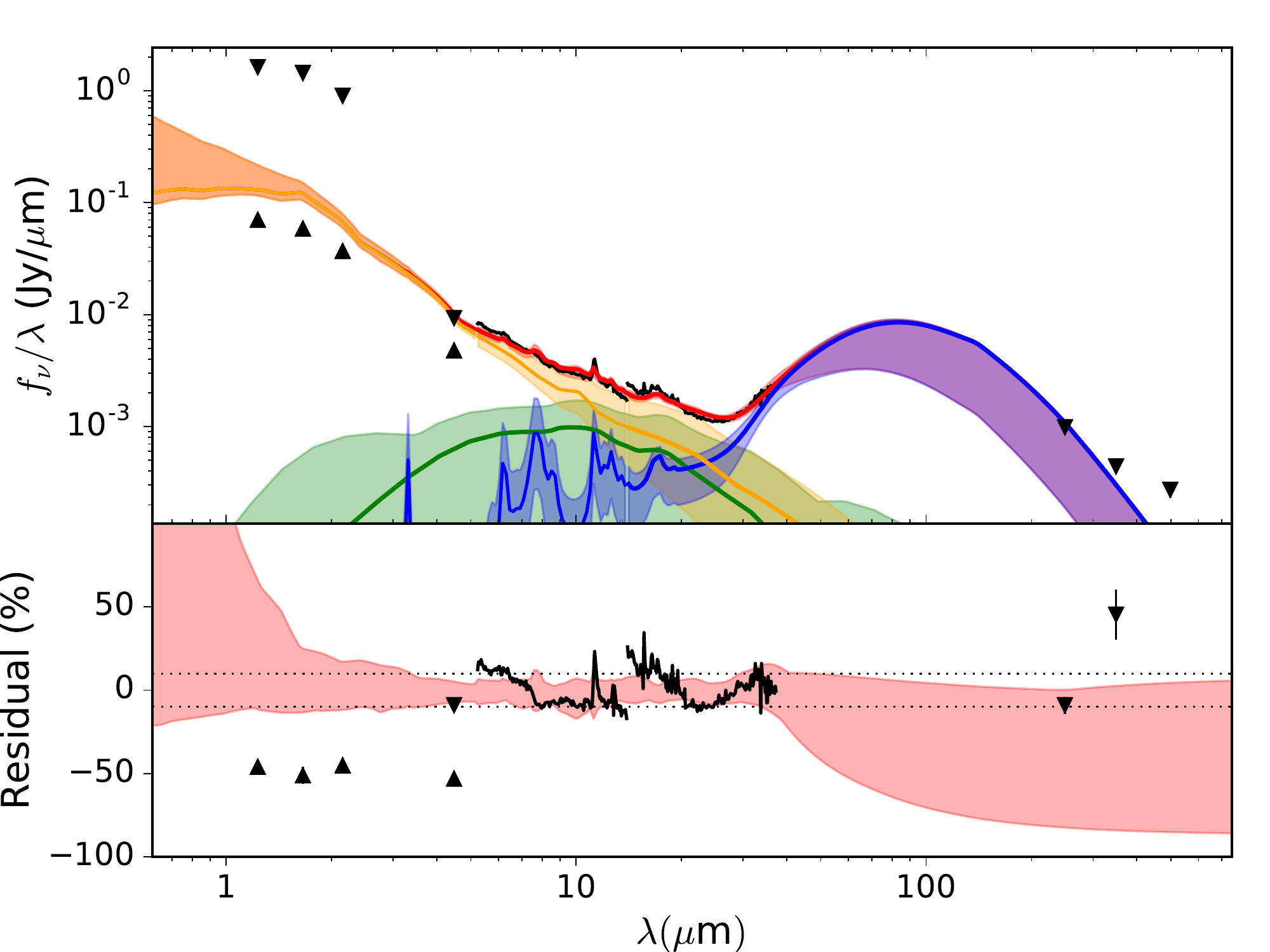} 
\caption{Continuum fit for M 84 forcing a torus component (green). To ensure maximal torus component we excluded the NLR and hot dust components seen in Figure \ref{fig:SED-M84-nbdoa}.}
\label{fig:SED-M84-cdoa}
\end{figure}
\end{center}

\begin{table}
\centering
\begin{tabular}{ccccc}
Component & Wavelength & Best fit & Minimum & Maximum\\
 & ($\mu$m) & (\%) & (\%) & (\%)\\
\hhline{=====}
\multirow{4}{1.5cm}{Torus} & 5.0 & 9.5 & 0.0 & 17.2\\
 & 15.0 & 33.6 & 0.0 & 67.4\\
 & 30.0 & 14.1 & 0.0 & 46.5\\
 & 60.0 & 0.2 & 0.0 & 3.0\\
\hline
\multirow{4}{1.5cm}{Diffuse ISM} & 5.0 & 0.3 & 0.1 & 0.8\\
 & 15.0 & 16.0 & 7.0 & 21.4\\
 & 30.0 & 65.7 & 45.1 & 80.8\\
 & 60.0 & 98.9 & 46.1 & 100\\
\hline
\multirow{4}{1.5cm}{Stars} & 5.0 & 90.2 & 82.5 & 100\\
 & 15.0 & 50.4 & 21.0 & 95.1\\
 & 30.0 & 20.3 & 3.7 & 46.6\\
 & 60.0 & 0.9 & 0.1 & 2.2\\
\hline
\end{tabular}
\caption{Fractional contributions of the various model components to the overall continuum in M 84 with a forced torus component.\label{tab:M84_frac_cdoa}}
\end{table}

\subsection{M~87}
\label{sec:M87cont}
\FloatBarrier

\begin{center}
\begin{figure}[ht]
\centering
\includegraphics[width=0.8\linewidth]{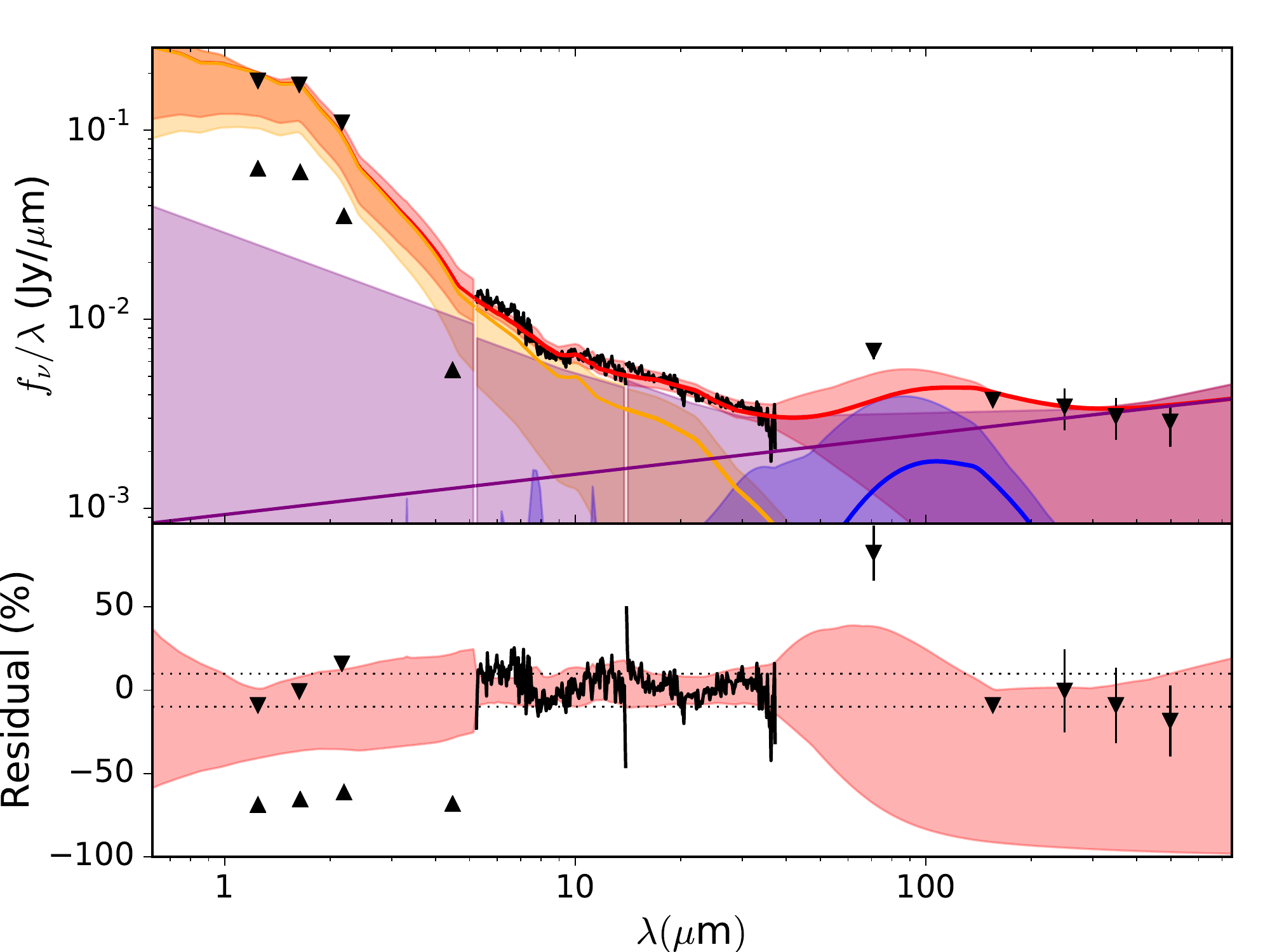} 
\caption{Best continuum fit for M 87. Model consists of a power law (purple), a simple stellar population (yellow), and diffuse ISM (blue). The total fit is shown in red although in the NIR it appears as orange due to overlap with the stellar population. We have applied aperture correction between the SL and LL regions. Upright triangles represent lower limits from photometry and inverted triangles represent upper limits.}
\label{fig:SED-M87-dopa}
\end{figure}
\end{center}

\begin{center}
\begin{figure}[ht]
\centering
\includegraphics[width=0.8\linewidth]{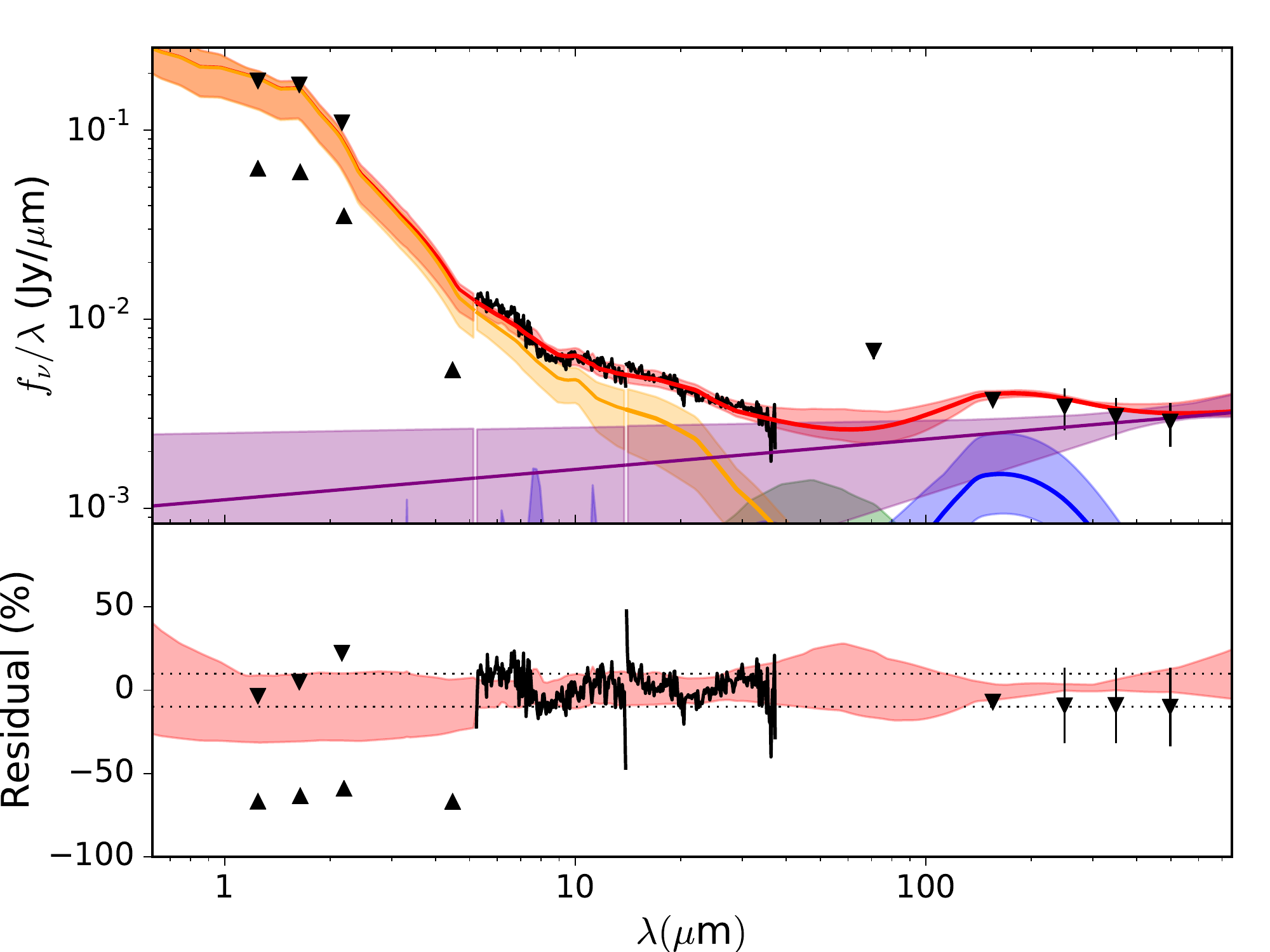} 
\caption{Continuum fit for M 87 forcing a torus component (green). To ensure maximal torus component we excluded the NLR component seen in Figure \ref{fig:SED-M87-dopa}.}
\label{fig:SED-M87-cdopa}
\end{figure}
\end{center}

Although every source in our sample has a radio jet and a couple have large scale optical jets, M 87 is the only source in which the red power-law synchrotron spectrum of the inner jet provides a significant contribution to the IR flux. This power law has a best-fit index of $\alpha = 1.2$ (recall that we use the $L_\nu \propto \nu^{-\alpha}$ convention) which is consistent with the optical spectral indices of $\alpha_o \sim 1.0-1.2$ \cite{M87jet} found for inter-knot regions of the inner jet. The model with a ``forced" torus component, shown in Figure \ref{fig:SED-M87-cdopa}, actually has a marginally lower BIC (392.0) than the otherwise ``best" fit (398.7), shown in Figure \ref{fig:SED-M87-dopa}, but this must be a fitting artifact as it contributes no flux. This lack of a torus and strong power law spectrum is consistent with findings by \cite{Whysong04}, previously mentioned in Section \ref{sec:evidence}, that M 87 is weak in IR and that its nuclear IR emission is consistent with a synchrotron spectrum from the base of its jet. Similarly, the absence of a torus is consistent with the upper limits on MIR thermal emission established by \cite{Perlman07}.

\begin{table}
\centering
\begin{tabular}{ccccc}
Component & Wavelength & Best fit & Minimum & Maximum\\
 & ($\mu$m) & (\%) & (\%) & (\%)\\
\hhline{=====}
\multirow{4}{1.5cm}{Diffuse ISM} & 5.0 & 0.0 & 0.0 & 0.4\\
 & 15.0 & 1.3 & 0.0 & 8.2\\
 & 30.0 & 4.5 & 0.0 & 45.1\\
 & 60.0 & 26.1 & 0.0 & 89.5\\
\hline
\multirow{4}{1.5cm}{Stars} & 5.0 & 90.3 & 41.5 & 98.4\\
 & 15.0 & 65.1 & 12.8 & 84.2\\
 & 30.0 & 36.4 & 7.2 & 46.5\\
 & 60.0 & 8.1 & 1.6 & 10.0\\
\hline
\end{tabular}
\caption{Fractional contributions of the various model components to the overall continuum in the best fit model of M~87. \label{tab:3C274_frac_dopa}}
\end{table}

\begin{table}
\centering
\begin{tabular}{ccccc}
Component & Wavelength & Best fit & Minimum & Maximum\\
 & ($\mu$m) & (\%) & (\%) & (\%)\\
\hhline{=====}
\multirow{4}{1.5cm}{Torus} & 5.0 & 0.0 & 0.0 & 1.0\\
 & 15.0 & 0.0 & 0.0 & 5.7\\
 & 30.0 & 0.2 & 0.0 & 31.8\\
 & 60.0 & 0.4 & 0.0 & 46.2\\
\hline
\multirow{4}{1.5cm}{Diffuse ISM} & 5.0 & 0.0 & 0.0 & 0.4\\
 & 15.0 & 0.9 & 0.6 & 7.4\\
 & 30.0 & 3.2 & 2.0 & 23.4\\
 & 60.0 & 7.3 & 4.2 & 25.6\\
\hline
\multirow{4}{1.5cm}{Stars} & 5.0 & 89.0 & 63.0 & 100\\
 & 15.0 & 64.4 & 37.8 & 83.3\\
 & 30.0 & 36.7 & 19.9 & 47.0\\
 & 60.0 & 10.3 & 4.8 & 12.9\\
\hline
\end{tabular}
\caption{Fractional contributions of the various model components to the overall continuum in M 87 with a forced torus component.\label{tab:3C274_frac_cdopa}}
\end{table}

\subsection{NGC~7052}
\label{sec:NGC7052cont}
\FloatBarrier

The model of NGC~7052 with the lowest BIC and the model with a forced torus have very similar goodness-of-fit with the best fit having a BIC of 425.7 and the forced-torus fit having a BIC of 429.8. However, inspection of the forced torus fit shows that it is unusually cold for a circumnuclear torus. Indeed, \emph{HST} images  of NGC~7052 by \cite{VerdoesKleijn99} show a promising candidate for such a cold thermal source in the form of a kpc-scale dust disk  which crosses the core region just off to the side of our line-of-sight into the nucleus. Overall, we conclude that a warm thermal component, such as a torus, is unnecessary to explain the IRS flux.

\begin{center}
\begin{figure}[ht]
\centering
\includegraphics[width=0.8\linewidth]{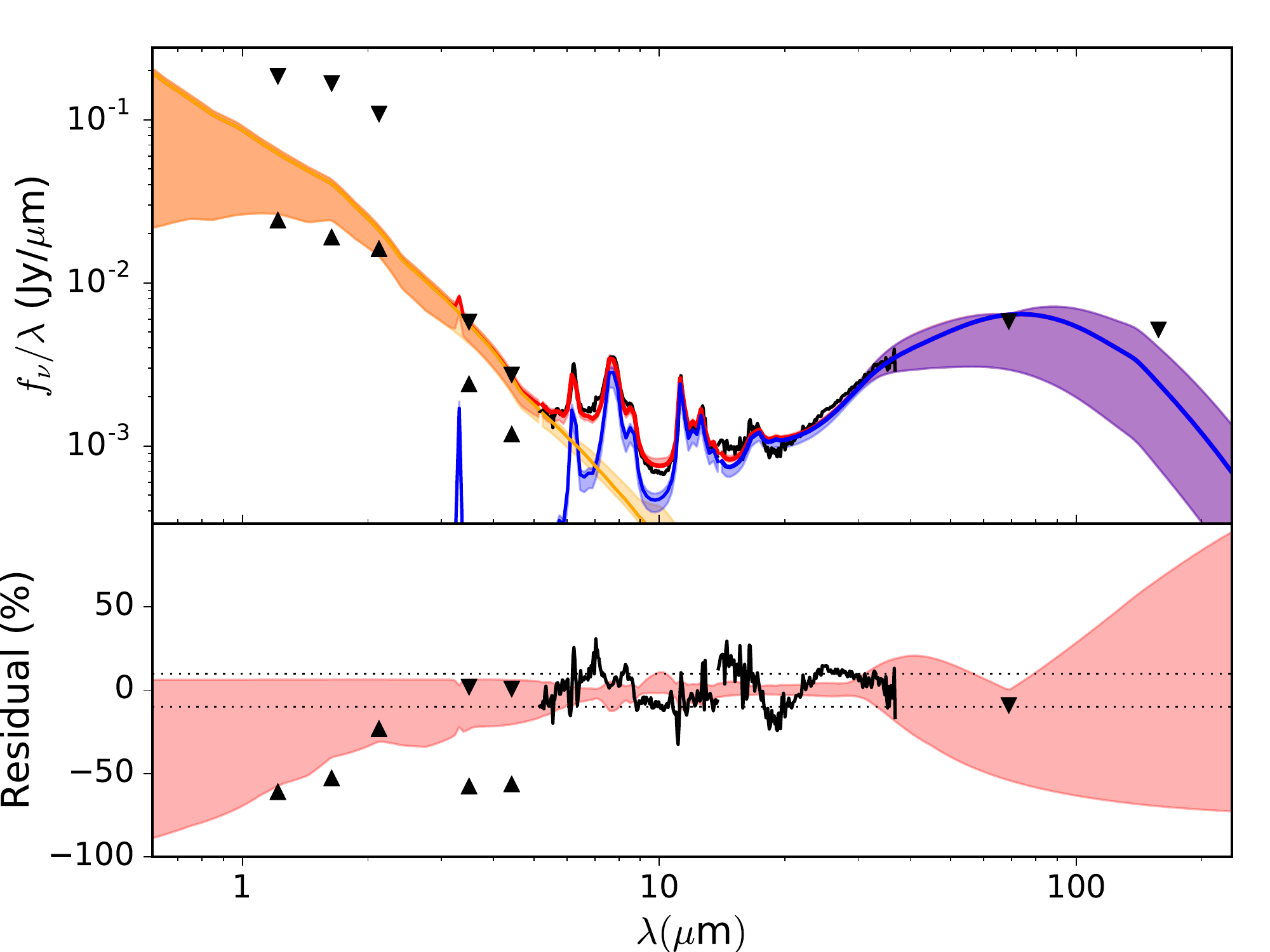} 
\caption{Best continuum fit for NGC 7052. Model consists of a simple stellar population (yellow) and diffuse ISM (blue). The total fit is shown in red although in the NIR it appears as orange due to overlap with the stellar population. Upright triangles represent lower limits from photometry and inverted triangles represent upper limits.}
\label{fig:SED-NGC7052-do}
\end{figure}
\end{center}

\begin{table}
\centering
\begin{tabular}{ccccc}
Component & Wavelength & Best fit & Minimum & Maximum\\
 & ($\mu$m) & (\%) & (\%) & (\%)\\
\hhline{=====}
\multirow{4}{1.5cm}{Diffuse ISM} & 5.0 & 5.3 & 4.3 & 5.7\\
 & 15.0 & 90.8 & 79.3 & 92.6\\
 & 30.0 & 99.6 & 92.4 & 100\\
 & 60.0 & 100.0 & 50.8 & 100\\
\hline
\multirow{4}{1.5cm}{Stars} & 5.0 & 94.7 & 77.6 & 100\\
 & 15.0 & 9.2 & 8.5 & 21.9\\
 & 30.0 & 0.4 & 0.4 & 3.9\\
 & 60.0 & 0.0 & 0.0 & 0.5\\
\hline
\end{tabular}
\caption{Fractional contributions of the various model components to the overall continuum in the best fit model for NGC 7052.\label{tab:NGC7052_frac_do}}
\end{table}

\begin{center}
\begin{figure}[ht]
\centering
\includegraphics[width=0.8\linewidth]{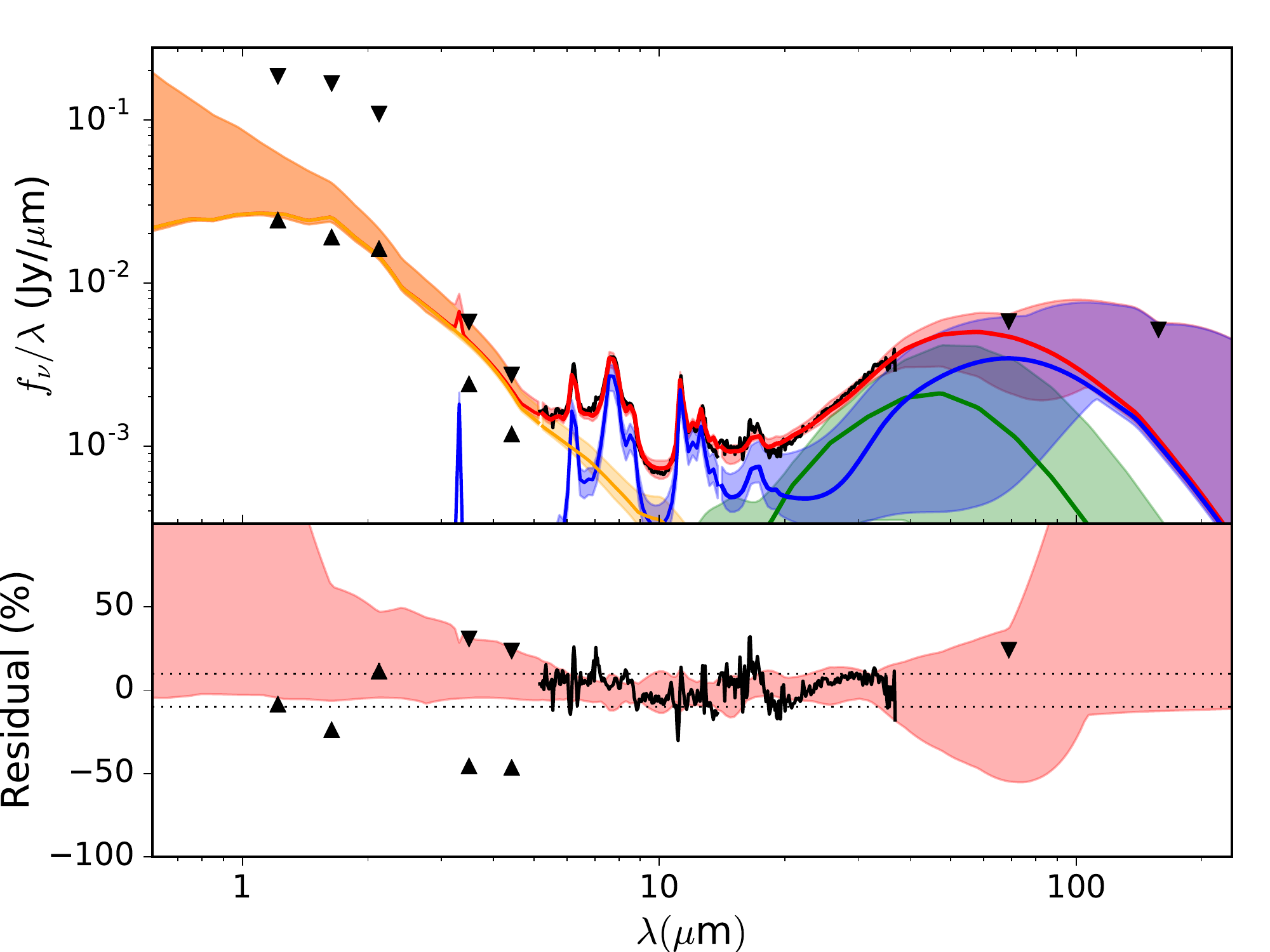} 
\caption{Continuum fit for NGC 7052 forcing a torus component (green). Model is otherwise the same as in Figure \ref{fig:SED-NGC7052-cdo}.}
\label{fig:SED-NGC7052-cdo}
\end{figure}
\end{center}

\begin{table}
\centering
\begin{tabular}{ccccc}
Component & Wavelength & Best fit & Minimum & Maximum\\
 & ($\mu$m) & (\%) & (\%) & (\%)\\
\hhline{=====}
\multirow{4}{1.5cm}{Torus} & 5.0 & 3.9 & 0.0 & 9.5\\
 & 15.0 & 33.0 & 5.1 & 52.3\\
 & 30.0 & 60.9 & 14.8 & 92.4\\
 & 60.0 & 32.3 & 3.0 & 79.2\\
\hline
\multirow{4}{1.5cm}{Diffuse ISM} & 5.0 & 5.3 & 4.1 & 6.3\\
 & 15.0 & 52.4 & 42.5 & 73.6\\
 & 30.0 & 35.8 & 14.6 & 83.8\\
 & 60.0 & 67.2 & 8.9 & 100\\
\hline
\multirow{4}{1.5cm}{Stars} & 5.0 & 90.9 & 87.2 & 100\\
 & 15.0 & 14.6 & 7.2 & 26.1\\
 & 30.0 & 3.3 & 0.4 & 4.2\\
 & 60.0 & 0.5 & 0.0 & 0.6\\
\hline
\end{tabular}
\caption{Fractional contributions of the various model components to the overall continuum in NGC 7052 with a forced torus component.\label{tab:NGC7052_frac_cdo}}
\end{table}

\subsection{Continuum fitting summary}
\label{sec:summaryCont}
\FloatBarrier

Some of the stellar age parameters for the best fits, shown in Table \ref{tab:stellarParams}, are unusually low for the various SFR estimates. This could be explained by those sources having recently quieted after a starburst, indeed \cite{Das05} claim something similar for NGC~3801, but it seems improbable that so many of such a small sample would be within 2~Gyr of the end of a starburst and NGC~3801 is not one of the sources with these low stellar ages. The fits with a forced torus, shown in Table \ref{tab:stellarParamsTorus}, are better in this regard with only 3C~66B and M~87 having this unusual combination.  We do not put much stock in this result because stellar ages are generally poorly constrained by infrared data since young stars are brightest at shorter wavelengths and, as mentioned earlier, the GRASIL stellar population model assumes a single Gyr-long starburst followed by passive evolution which may not be a good approximation in these galaxies. This would likely be clearer with more NIR spectrometry and optical spectroscopy (especially in blue) would help characterize the population of young stars.

\begin{table}
\centering
\begin{tabular}{ccc}
Source & Stellar Age & SFR \\
 & (Gyr) & ($\mathrm{M}_{\odot}\, \mathrm{yr}^{-1}$)\\
\hhline{===}

NGC 315 & ${7.7}_{-4.9}^{+3.0}$ & ${0.24}_{-0.081}^{+0.15}$\\

3C 31 & ${1.7}_{-0.7}^{+0.4}$ & ${0.88}_{-0.28}^{+0.14}$\\

NGC 541 & ${2.3}_{-0.4}^{+0.4}$ & ${0.12}_{-0.017}^{+0.034}$\\

3C 66B & ${1.0}_{-0.2}^{+0.3}$ & ${0.017}_{-0.0035}^{+0.0038}$\\

NGC 3801 & ${9.4}_{-5.0}^{+3.0}$ & ${0.35}_{-0.074}^{+0.087}$\\

NGC 3862 & ${0.9}_{-0.3}^{+0.2}$ & ${0.20}_{-0.038}^{+0.084}$\\

3C 270 & ${1.0}_{-0.2}^{+0.2}$ & ${0.044}_{-0.014}^{+0.0058}$\\

M 84 & ${0.4}_{-0.0}^{+0.1}$ & ${0.034}_{-0.0099}^{+0.0057}$\\

M 87 & ${1.0}_{-0.2}^{+0.3}$ & ${0.0026}_{-0.0021}^{+0.0031}$\\

NGC 7052 & ${0.1}_{-0.0}^{+0.0}$ & ${0.67}_{-0.061}^{+0.052}$\\
\hline
\end{tabular}
\caption{Best fit parameters for the stellar population component. \label{tab:stellarParams}}
\end{table}

\begin{table}
\centering
\begin{tabular}{ccc}
Source & Stellar Age & SFR \\
 & (Gyr) & ($\mathrm{M}_{\odot}\, \mathrm{yr}^{-1}$)\\
\hhline{===}

3C 31 & ${7.9}\pm {3.5}$ & ${0.7}\pm {0.2}$\\

NGC 541 & ${7.4}_{-3.6}^{+4.1}$ & ${0.12}_{-0.01}^{+0.03}$\\

3C 66B & ${1.0}_{-0.2}^{+0.2}$ & ${0.0005}_{-0.0005}^{+0.0022}$\\

NGC 3801 & ${6.7}_{-6.4}^{+5.6}$ & ${0.3}_{-0.04}^{+0.09}$\\

NGC 3862 & ${2.5}_{-1.5}^{+6.8}$ & ${0.21}_{-0.04}^{+0.08}$\\

3C 270 & ${8.2}_{-3.8}^{+3.6}$ & ${0.03}\pm 0.01$\\

M 84 & ${3.2}_{-0.9}^{+1.5}$ & ${0.022}_{-0.003}^{+0.004}$\\

M 87 & ${1.1}_{-0.3}^{+0.2}$ & ${0.011}\pm 0.001$\\

NGC 7052 & ${9.8}_{-2.9}^{+2.2}$ & ${0.47}_{-0.06}^{+0.10}$\\
\hline
\end{tabular}
\caption{Best fit parameters for the stellar population component in fits forcing a torus. \label{tab:stellarParamsTorus}}
\end{table}

The diffuse ISM components of the two model sets, the best fit models shown in Table \ref{tab:ISMparams} and the forced-torus models shown in Table \ref{tab:ISMparamsTorus}, show lower limits on their  interstellar radiation fields, PAH fractions, and ISM luminosities which are generally consistent between the two sets with the exception of M~87 which shows an improbably weak radiation field and significantly higher ISM luminosity in the forced-torus model.

\begin{table}
\centering
\begin{tabular}{cccc}
Source & $u_\mathrm{min}$ & $q_\mathrm{PAH}$ & ISM Luminosity\\
 & & (\%) & ($10^{41}\, \mathrm{erg} \, \mathrm{s}^{-1}$)\\
\hhline{====}

NGC 315 & ${8.}_{-5.}^{+10.}$ & ${2.46}_{-0.9}^{+1.24}$ & ${52.0}_{-18.0}^{+34.0}$\\

3C 31 & ${3.}_{-1.}^{+2.}$ & ${2.6}_{-0.3}^{+1.0}$ & ${200}_{-60}^{+30}$\\

NGC 541 & ${11}_{-5}^{+6}$ & ${3.8}_{-0.8}^{+0.6}$ & ${26.}_{-4.}^{+8.}$\\

3C 66B & ${6.}_{-3.}^{+5.}$ & ${1.1}_{-0.4}^{+0.8}$ & ${3.8\pm 0.8}$\\

NGC 3801 & ${1.9}_{-0.8}^{+2.4}$ & ${3.4}_{-0.6}^{+0.8}$ & ${80 \pm 20}$\\

NGC 3862 & ${9.}_{-6.}^{+10.}$ & ${3.5}_{-0.9}^{+0.7}$ & ${44.}_{-8.}^{+19.}$\\

3C 270 & ${15.}_{-8.}^{+6.}$ & ${1.3}_{-0.4}^{+0.7}$ & ${10.}_{-3.}^{+1.}$\\

M 84 & ${10.}_{-3.}^{+8.}$ & ${1.6}_{-0.3}^{+0.6}$ & ${8.}_{-2.}^{+1.}$\\

M 87 & ${15.}_{-12.}^{+7.}$ & ${2.}\pm 1.$ & ${0.6}_{-0.5}^{+0.7}$\\

NGC 7052 & ${16.}_{-4.}^{+5.}$ & ${2.4\pm 0.2}$ & ${150\pm 10}$\\

\hline
\end{tabular}
\caption{ISM component best fit parameters. Recall that $q_\text{PAH}$ is the mass fraction of polycyclic aromatic hydrocarbons relative to the total dust mass and $U_\text{min}$ is the lower limit of the interstellar radiation field as a scaling of that in the solar neighborhood. \label{tab:ISMparams}}
\end{table}

\begin{table}
\centering
\begin{tabular}{cccc}
Source & $u_\mathrm{min}$ & $q_\mathrm{PAH}$ & ISM Luminosity\\
 & & (\%) & ($10^{41}\, \mathrm{erg} \, \mathrm{s}^{-1}$)\\
\hhline{====}

3C 31 & ${3.}_{-1.}^{+4.}$ & ${3.1}_{-0.6}^{+1.1}$ & ${160\pm +50}$\\

NGC 541 & ${15.}_{-8.}^{+4.}$ & ${4.0}_{-0.6}^{+0.4}$ & ${26.}_{-3.}^{+6.}$\\

3C 66B & ${7.}_{-4.}^{+7.}$ & ${2.}_{-1.}^{+2.}$ & ${0.1}_{-0.1}^{+0.5}$\\

NGC 3801 & ${4.\pm 2.}$ & ${3.9}_{-0.8}^{+0.5}$ & ${67.}_{-9.}^{+19.}$\\

NGC 3862 & ${10.}_{-6.}^{+9.}$ & ${3.5}_{-0.9}^{+0.7}$ & ${46.}_{-9.}^{+18.}$\\

3C 270 & ${14.\pm -8.}$ & ${2.1}_{-0.9}^{+1.2}$ & ${6.}_{-2.}^{+3.}$\\

M 84 & ${18.}_{-5.}^{+4.}$ & ${1.3}_{-0.3}^{+0.4}$ & ${4.9}_{-0.6}^{+1.0}$\\

M 87 & ${0.1}_{-0.0}^{+0.0}$ & ${1.6}_{-0.8}^{+1.1}$ & ${2.5}\pm {0.2}$\\

NGC 7052 & ${17.}_{-8.}^{+4.}$ & ${3.8\pm 0.6}$ & ${100}_{-10}^{+20}$\\

\hline
\end{tabular}
\caption{ISM component best fit parameters with a forced torus component. \label{tab:ISMparamsTorus}}
\end{table}

The fit parameters for the torus, when present (forced-torus models and NGC 315) are shown in Tables \ref{tab:torusparams} and \ref{tab:torusparams2}. Of particular interest is the low inclination of NGC~3801, the large half-opening angle and high escape probability of NGC~3801, NGC~3862, and M~84. These are unusual for sources in which we can't see broad lines from the central engine. The masses of obscuring material in the forced-torus fits to NGC~3801 and M~84 are remarkably small compared to the others with both being $\sim 10^{-3}\,\mathrm{M}_\odot$. And finally the power law indices of the cloud distributions in the forced-torus models are inconsistent with the silicate line ratios in several sources, especially three previously mentioned in this paragraph: NGC~3801, NGC~3862, and M~84. When considered together these issues suggest that the forced-torus fits to NGC~3801, NGC~3862, and M~84 are inconsistent with independent observations.

\begin{table}
\centering
\begin{tabular}{cccccccc}
Source & $\sigma$ & $Y$ & $N_0$ & $\tau_V$ & $q$ & Inclination \\
 & (deg) & & & & & (deg) \\
\hhline{=======}

NGC 315* & ${60.}_{-10.}^{+8.}$ & ${21.}_{-7.}^{+25.}$ & ${12.}_{-3.}^{+2.}$ & ${130}_{-30}^{+50}$ & ${1.3}_{-0.9}^{+0.6}$ & ${60}\pm {20}$\\

3C 31 & ${50}_{-20}^{+10}$ & ${30}_{-20}^{+40}$ & ${11.}_{-4.}^{+3.}$ & ${70}_{-40}^{+120}$ & ${1.9}_{-1.0}^{+0.6}$ & ${50}\pm {30}$\\

NGC 541 & ${50}_{-20}^{+10}$ & ${40}_{-20}^{+40}$ & ${10.}_{-4.}^{+3.}$ & ${150}_{-90}^{+100}$ & ${1.7}_{-1.0}^{+0.8}$ & ${50}_{-30}^{+20}$\\

3C 66B & ${60\pm 10}$ & ${24.}_{-6.}^{+22.}$ & ${11.}_{-4.}^{+3.}$ & ${220}_{-70}^{+50}$ & ${1.00}_{-0.68}^{+0.70}$ & ${70}_{-20}^{+10}$\\

NGC 3801 & ${40\pm 20}$ & ${50}_{-30}^{+40}$ & ${4.}_{-2.}^{+5.}$ & ${30}_{-10}^{+140}$ & ${2.4}_{-1.1}^{+0.5}$ & ${30}_{-20}^{+30}$\\

NGC 3862 & ${40\pm 20}$ & ${50}_{-30}^{+40}$ & ${7.}_{-4.}^{+5.}$ & ${100}_{-70}^{+120}$ & ${2.2}_{-1.2}^{+0.5}$ & ${40\pm 30}$\\

3C 270 & ${58.}_{-13.}^{+8.}$ & ${16.}_{-5.}^{+36.}$ & ${12.}_{-3.}^{+2.}$ & ${140}_{-40}^{+80}$ & ${1.6}_{-1.0}^{+0.7}$ & ${60}_{-30}^{+20}$\\

M 84 & ${40}\pm {20}$ & ${30}_{-20}^{+40}$ & ${7.}_{-4.}^{+5.}$ & ${40}_{-30}^{+120}$ & ${2.5}_{-0.7}^{+0.4}$ & ${40}\pm {30}$\\

M 87 & ${60}_{-20}^{+10}$ & ${80}\pm {20}$ & ${11.}_{-4.}^{+2.}$ & ${200}\pm {70}$ & ${0.5}_{-0.3}^{+0.4}$ & ${70}_{-30}^{+20}$\\

NGC 7052 & ${64.}_{-7.}^{+4.}$ & ${90.}_{-10.}^{+7.}$ & ${12.\pm 2.}$ & ${60\pm 20}$ & ${0.3}_{-0.2}^{+0.3}$ & ${60}_{-20}^{+20}$\\

\hline
\end{tabular}
\caption{Best fit parameters for forced torus components. $\sigma$ is the angular scale height of the torus, $Y$ is the radial extent of the torus in units of the dust sublimation radius, $N_0$ is the number of clouds along an equatorial line-of-sight, $\tau_V$ is the optical depth of a single cloud and $q$ is the power law index of the radial distribution of clouds. The inclination is defined such that $0^\circ$ is pole-on and $90^\circ$ is edge-on. *Recall that NGC 315 favors a torus component; it is not forced. \label{tab:torusparams}}
\end{table}

\begin{table}
\centering
\begin{tabular}{cccccc}
 & Half- & Escape & & & AGN \\
 Source & Opening Angle  &  Probability & Mass  & Radius  & Luminosity \\
 & (deg) & (\%) & ($\mathrm{M}_{\odot}$) & (pc) & ($10^{41} \, \mathrm{erg} \,  \mathrm{s}^{-1}$)\\
 \hhline{======}
NGC 315* & ${10.}_{-3.}^{+6.}$ & ${0.02}_{-0.0}^{+0.60}$ & ${2.2}_{-0.9}^{+1.4}$ & ${0.9}_{-0.3}^{+1.2}$ & ${110}_{-20}^{+30}$\\ 

3C 31 & ${15.}_{-6.}^{+16.}$ & ${0.8}_{-0.7}^{+24.5}$ & ${0.04}_{-0.03}^{+0.09}$ & ${0.4}_{-0.3}^{+0.5}$ & ${9.}_{-3.}^{+5.}$\\ 

NGC 541 & ${18.}_{-9.}^{+21.}$ & ${1.}_{-1.}^{+40.}$ & ${0.04}_{-0.03}^{+0.11}$ & ${0.3}_{-0.2}^{+0.4}$ & ${4.}_{-2.}^{+3.}$\\ 

3C 66B & ${13.}_{-5.}^{+10.}$ & ${0.02}_{-0.0}^{+0.55}$ & ${0.4}_{-0.2}^{+0.3}$ & ${0.31}_{-0.09}^{+0.27}$ & ${9.}_{-2.}^{+3.}$\\ 

NGC 3801 & ${40}\pm 20$ & ${70}_{-50}^{+30}$ & ${0.001}_{-0.001}^{+0.005}$ & ${0.3}_{-0.2}^{+0.4}$ & ${4.\pm 3.}$\\ 

NGC 3862 & ${40\pm 20}$ & ${30}_{-30}^{+70}$ & ${0.03}_{-0.02}^{+0.11}$ & ${0.6}_{-0.5}^{+0.9}$ & ${20\pm 20}$\\ 

3C 270 & ${11.}_{-3.}^{+8.}$ & ${0.03}_{-0.00}^{+1.37}$ & ${0.09}_{-0.04}^{+0.06}$ & ${0.19}_{-0.08}^{+0.47}$ & ${9.}_{-2.}^{+3.}$\\ 

M 84 & ${30\pm 20}$ & ${20}_{-20}^{+80}$ & ${0.0008}_{-0.0005}^{+0.0012}$ & ${0.13}_{-0.09}^{+0.21}$ & ${1.2}_{-0.6}^{+0.8}$\\ 

M 87 & ${13.}_{-4.}^{+16.}$ & ${0.04}_{-0.00}^{+1.44}$ & ${0.1}_{-0.1}^{+0.3}$ & ${0.1\pm 0.1}$ & ${0.2}_{-0.2}^{+0.6}$\\ 

NGC 7052 & ${8.}_{-1.}^{+2.}$ & ${0.02}_{-0.00}^{+0.15}$ & ${14.}_{-5.}^{+9.}$ & ${2.3\pm 0.3}$ & ${41.}_{-7.}^{+8.}$\\ 

\hline
\end{tabular}
\caption{More best fit parameters for forced torus components. *Recall that NGC 315 favors a torus component; it is not forced. \label{tab:torusparams2}}
\end{table}

Overall, we find no evidence for a warm obscuring torus in six of our ten sources and we find convincing evidence for a torus in one of the ten (NGC 315). We have three more sources in which we find strong evidence for some warm MIR structure but cannot conclude that said structure is consistent with the \cite{nenkova02} clumpy torus model. We summarize our conclusions as to the presence or absence of each of our model components for each source in Table \ref{tab:Overall} and we compare the BIC of the best-fit models to those with a forced torus component in Table~\ref{tab:BICs}. The main reason for inconclusive results is the fitting degeneracy between torus, NLR, and hot dust.

\begin{table}[htp]
\centering
\begin{tabular}{cccccc}
Source & Evidence & Evidence & Evidence & Evidence  & Warm \\
 & for torus & for power law & for NLR & for hot dust  & Component? \\
 \hhline{======}
 NGC 315 & Y & N & N & N & Y\\
 3C 31 & N & N* & N & N & N\\
 NGC 541 & N & N & N & N & N\\
 3C 66B & I & N* & I & I & Y\\
 NGC 3801 & N & N & I & N & I \\
 NGC 3862 & N & N* & N & N & N \\
 3C 270 & I/Y & N & I/N & N & Y\\
 M 84 & I & N & I & I & Y\\
 M 87 & N & Y* & N & N & N\\
 NGC 7052 & N & N & N & N & N \\
 \hline
\end{tabular}
\caption{Summary of best fit results. Results in which we are reasonably confident are marked with ``Y" (for yes) if present or ``N" (for no) if not. Results which are more ambiguous are  denoted with ``I" (for inconclusive), although we follow the 3C 270 results with a marker describing the model which we favor. The NLR and torus are both considered a warm component. *These sources have known optical or IR jets so were checked for a power-law component. \label{tab:Overall}}
\end{table}

\begin{table}
\centering
\begin{tabular}{ccc}
Source & Best fit & Forced torus \\
 & BIC & BIC \\
 \hhline{===}
 NGC 315 & 420.3 & - \\
 3C 31 & 403.5 & 425.7 \\
 NGC 541 & 380.5 & 408.7 \\
 3C 66B & 472.3 & 506.8 \\
 NGC 3801 & 423.5 & 453.7 \\
 NGC 3862 & 391.9 & 420.3 \\
 3C 270 & 410.7 & 424.9 \\
 M 84 & 386.9 & 444.4 \\
 M 87 & 398.7 & 392.0 \\
 NGC 7052 & 425.7 & 429.8  \\
 \hline
\end{tabular}
\caption{BIC for the the best fit and forced torus fit in each source. Recall that we do not favor the presence of a torus in M 87 despite the lower BIC because the torus component does not contribute to the flux when forced. \label{tab:BICs}}
\end{table}

\FloatBarrier
\section{Discussion}
\label{sec:discuss}
Our results support the existence of a warm dusty structure in NGC~315, 3C~66B, 3C~270, and M~84. Of those, only NGC~315 is clearly consistent a clumpy obscuring torus. One more source, NGC~3801, benefits from the inclusion of a warm component. However, the best fit to the warm component in NGC~3801 is unusually dim relative to the ISM and stellar population compared to the other sources with a warm component.

A few of the sources in which we find no thermal component show significant silicate absorption which may be due to larger-scale dust structures which would be too cool to appear in our continuum spectra. All of the thermal sources produce some silicate emission, as do two non-thermal sources. The silicate emission from non-thermal sources is harder to reconcile, though. Perhaps those sources may be consistent with silicate emission from ADAF models or other non-torus possibilities which we have not explored because modelling those is beyond the scope of this paper.

We find that the thermal MIR excess \cite{Leipski09} found in 3C~270 is better described by a NLR component. However, due to the bias in which the relative simplicity of the NLR component compared to the torus component biases the BIC against a torus component, this disagreement is quite minor. They also found a MIR excess in other sources in which we find a warm thermal component, 3C~66B and M~84, however they attribute these excesses to other effects, non-thermal emission in 3C~66B and starburst activity in M~84. We rule out a starburst origin for the MIR excess in M~84 because all our estimates of the SFR in M~84 agree that $\mathrm{SFR} \lesssim 0.1\,\mathrm{M}_{\odot}$ (see Tables~\ref{tab:SFR} and \ref{tab:stellarParams}). Recall that \cite{Wu06} give a star formation rate estimate of $0.20\,\mathrm{M}_{\odot}$ for M~84 which is higher than our estimates but is still inconsistent with a starburst. \cite{Leipski09} found a small MIR excess in 3C~31 as well, which they attribute to star formation. We don't find any of our thermal model components in 3C~31 but we believe that our star formation rates listed in Tables~\ref{tab:SFR} and \ref{tab:stellarParams} are sufficient to be loosely consistent with their results. Thus, the results of \cite{Leipski09} seem consistent with our detection of a warm thermal component in 3C~270 and M~84 but not 3C~66B and our overall conclusion is similar to theirs despite some different conclusions regarding individual objects. We believe our analysis is more effective at differentiating between MIR thermal than that performed by \cite{Leipski09} because MCMC is known for its efficiency at exhaustive searches through large and complex parameter spaces (see \citealt{clumpyDREAM}) but our more specific results rest on the validity of the \cite{nenkova08a,nenkova08} torus models which are not unique.

Since the luminosity of a given high-ionization line will be dependent on both the degree of ionization and the overall abundance of the element in question (see \cite{Spinoglio92}), we show the line ratio of our fitted $[\mathrm{O\,IV}]$ luminosities to archival $[\mathrm{O\,III}]$ luminosities for each of our sample galaxies in Table \ref{tab:OxyRatio}. This ratio is independent of the oxygen abundance in the line-emitting gas and as such provides a better indication of the ionization state of the gas at the cost of greater sensitivity to reddening than line rations with more similar wavelengths e.g., H$\beta/[\mathrm{O\,III}]$. However, most of the ratios for our sample are upper limits because no $[\mathrm{O\,IV}]$ was detected in those sources. The two FR-I sources in which we detected $[\mathrm{O\,IV}]$ emission are inconsistent with broad-line radio galaxies and quasars in the \cite{Haas05} sample and consistent with obscured sources (e.g., Seyfert 2) in the \cite{Baum10} Seyfert sample.

\begin{table}
\centering
\begin{tabular}{ccccc}
Source& $L_{[\mathrm{OIV}]}$ 26 $\mu$m  & $L_\mathrm{[O\, III]}$ & $\frac{L_\mathrm{[O\, IV]}}{L_\mathrm{[O\, III]}}$ & [O III] reference \\
 & ($10^{39}\,\mathrm{erg}\,\mathrm{s}^{-1}$) & $(10^{39}\,\mathrm{erg}\,\mathrm{s}^{-1})$ &  &  \\
\hhline{=====}
NGC 315 & $<4.98$ & $2.75$ & $<1.81$ & \cite{315OIII} \\
3C 31 & $3.5_{-0.8}^{+0.7}$ & $2.88$ & ${1.2}_{-0.3}^{+0.2}$ & \cite{3C31OIII}\\
NGC 541 & $<6.28$ & $0.886$ & $<7.09$ & \cite{541OIII} \\
3C 66B & $<3.10$ & $11.2$ & $<0.28$ & \cite{3C31OIII} \\
NGC 3801 & $<3.20$ & $0.146$ & $<22.0$ & \cite{3801OIII} \\
NGC 3862 & $<4.12$ & $1.58$ & $<2.60$ & \cite{3C31OIII} \\
3C 270 & $<0.924$ & $0.794$ & $<1.16$ & \cite{3C270OIII} \\
M 84* & $0.32_{-0.06}^{+0.04}$ & $0.88$ & ${0.36}_{-0.07}^{+0.04}$ & \cite{M84OIII} \\
M 87 & $<0.751$ & $1.17$ & $<0.64$ & \cite{315OIII} \\
NGC 7052 & $<7.60$ & $2.75$ & $<2.76$ & \cite{7052OIII} \\
\hline
\end{tabular}
\caption{Oxygen line ratios in our sample of FR-I sources. *\cite{M84OIII} uses a different cosmology from ours so we have corrected their $L_\mathrm{[OIII]}$ to our distance. \label{tab:OxyRatio}}
\end{table}

\FloatBarrier

To put together all our lines of evidence we summarize in Table \ref{tab:summary} all our results and those we found in the literature. We note that most sources which show at least one test with an inconclusive or positive result also show inconclusive or positive results on other tests. In particular, all sources which show evidence of a warm component are backed up by at least one additional line of evidence. 

\begin{table}
\centering
\begin{tabular}{ccccc}
Source & Warm component  & Warm component & Broad polarized & High \\
 & (ours) & (others) & H$\alpha$ & [OIV]/[OIII] \\
\hhline{=====}
NGC 315 & Y & - & Y  & N \\
3C 31 & N & N  & Y & Y \\
NGC 541 & N & - & N & N \\
3C 66B & Y & I  & Y & N \\
NGC 3801 & I & - & N & N\\
NGC 3862 & N & N  & N & N \\
3C 270 & Y & Y  & I & N \\
M 84 & Y & Y  & N & Y \\
M 87 & N & N  & N & N \\
NGC 7052 & N & - & N & N \\
\hline
\end{tabular}
\caption{Summary of evidence for a warm obscuring structure in each source. ``Others" includes classifications by \cite{Leipski09} except for M~87; classification for M~87 by \cite{Whysong04} and \cite{Perlman07}. Broad polarized H$\alpha$ references: NGC 315 \citep{Barth99}; 3C 31 \citep{541OIII}; 3C 66B \citep{541OIII}; 3C 270 (Antonucci, private communication). \label{tab:summary}}
\end{table} 

Recall the suggestion by \cite{Elitzur06} that hydromagnetic disk winds off sources with bolometric luminosities $L_\mathrm{bol} \lesssim 10^{42}\,\mathrm{erg}\,\mathrm{s}^{-1}$ could not sustain a clumpy obscuring torus. In the torus model parameters continuum fits with a forced torus component, shown in Table \ref{tab:torusparams2}, NGC~315, NGC~3862, and NGC~7052 have best-fit AGN luminosities $> 10^{42}\,\mathrm{erg}\,\mathrm{s}^{-1}$ and 3C~31, 3C~66B, and 3C~270 come in just under that at $\sim 9\times 10^{41}\,\mathrm{erg}\,\mathrm{s}^{-1}$. If \cite{Elitzur06} are correct, this suggests that the other four sources which all have bolometric luminosities significantly below $10^{42}\,\mathrm{erg}\,\mathrm{s}^{-1}$, including M~84 which we previously considered inconclusive, are not likely to host a clumpy AGN-wind-driven torus. Note, however, that this does not necessarily prohibit other obscuring structures.

\cite{3C31OIII} proposed various line ratios with which to distinguish between low-excitation and high-excitation sources. The measure with the clearest distinction betwen the two populations was the Excitation Index (EI), defined as
\begin{equation}
\label{eq:excitIndex}
\mathrm{EI} = \log \frac{\mathrm{[OIII]}}{\mathrm{H}\beta}-\frac{1}{3}\bigg(\log \frac{\mathrm{[NII]}}{\mathrm{H}\alpha}+\log \frac{\mathrm{[SII]}}{\mathrm{H}\alpha}+\log \frac{\mathrm{[OI]}}{\mathrm{H}\alpha}\bigg).
\end{equation}
 For low-excitation sources $\mathrm{EI} \lesssim 0.95$ and for high excitation sources $\mathrm{EI} \gtrsim 0.95$. \cite{Hu16} calculated excitation indices for the 3C catalog using line fluxes measured by \cite{Buttiglione09} and we include those 3C sources which are also in our sample in Table \ref{tab:exciteClass} and find that all but NGC~315 and NGC~541 are low-excitation sources. We classify those remaining two as extremely low excitation radio galaxies (ELERGs) based on their unusually low  $\mathrm{[OIII]}/\mathrm{H}\beta$ of 0.57 and 0.67, respectively, which is consistent with the $\mathrm{[OIII]}/\mathrm{H}\beta \sim 0.5$ indicative of extremely low-excitation sources. \cite{Capetti11} describe extremely low-excitation galaxies as ``radio relics" i.e., sources which recently experienced a drop in nuclear activity --- perhaps this is the case in NGC~315 and NGC~541. We see no issues with this interpretation for NGC~541 but NGC~315 shows a warm thermal MIR component. This warm thermal component in NGC~315 is well fit by a clumpy torus, and is thus indicative of a high-excitation galaxy, yet its optical emission line ratios are those of an extremely low-excitation galaxy. We hypothesize that the warm dust cooling is occurring on a longer time-scale than [OIII] recombination, perhaps because the warm dust structure is much larger than the emission-line nucleus.

\begin{table}[htp]
\centering
\begin{tabular}{cccccccc}
Source & $L_\mathrm{H\alpha}$ & $L_\mathrm{H\beta}$ & $L_\mathrm{[NII]}$ & $L_\mathrm{[SII]}$ & $L_\mathrm{[OI]}$ & EI & Class\\
\hhline{========}
NGC 315* & 4.19 & 4.80 & 7.70 & 2.12 & 2.46 & -0.154 & ELERG \\
3C 31 & 6.76 & 1.01 & 6.69 & 4.66 & 0.946 & 0.797 & LERG \\
NGC 541 & 4.58 & 1.32 & 2.97 & 0.57 & 1.10 & -0.484 & ELERG \\
3C 66B & 12.88 & 1.54 & 31.16 & 7.21 &3.35 & 0.746 & LERG \\
NGC 3801 & 0.98 & 0.06 & 0.96 & 0.57 & 0.17 & 0.721 &  LERG \\
NGC 3862 & 4.79 & 1.29 & 6.94 & 3.16 & 1.05 & 0.313 & LERG \\
3C 270$^\dagger$ & 0.14 & $<0.15$ & 0.10 & $<0.12$ & 0.07 & 0.048 & LERG \\
M 84 & 0.83 & 0.08 & 1.06 & 0.72 & 0.19 & 0.478 & LERG \\
M 87 & 3.16 & 0.54 & 7.34 & 4.58 & 1.14 & 0.233 & LERG \\
NGC 7052 & 7.62 & 1.65 & 8.86 & 3.62 & 0.93 & 0.612 &  LERG \\
\hline
\end{tabular}
\caption{Classification based on excitation indices. All luminosities in units of $10^{39}\,\mathrm{erg}\,\mathrm{s}^{-1}$. $L_\mathrm{[O\,III]}$ given in Table \ref{tab:OxyRatio}. Line luminosities and excitation indices for galaxies in 3C (3C~31, 3C~66B, NGC~3862, 3C~270, M~84,  and M~87) by \cite{Hu16}. Line luminosities for NGC~541, NGC~3801, and NGC~7052 by \cite{541OIII}.   H$\alpha$ and [O~I] line luminosities for NGC~315 by \cite{Fernandes04}, H$\beta$ by \cite{M84OIII}, and [N~II] and [S~II]  by \cite{315OIII}.  Classification cutoff given by \cite{3C31OIII} as $\mathrm{EI}\lesssim 0.95$ for low excitation galaxies. \cite{Hu16} calculated EI using line fluxes by \cite{Buttiglione09}. *\cite{Fernandes04} and \cite{M84OIII} use a different cosmology than us, so we have corrected their luminosities for our distance. Additionally, since the luminosities are from different instruments, systematic errors in our calculations are likely significant and difficult to quantify.  $^\dagger$\cite{3C31OIII} removed this source from their data due to a misplaced SDSS fiber. We have included it for completeness but it is unreliable. \label{tab:exciteClass}}
\end{table}

We compare our continuum fitting results to $1.4\,\mathrm{GHz}$ radio luminosity density $L_{\nu,\,1.4\,\mathrm{GHz}}$, estimated black hole mass $M_\mathrm{BH}$, Eddington luminosity $L_\mathrm{Edd}$, and core-extended brightness ratio $R_c$ in Table \ref{tab:radio}. $R_c$ is a indicator for AGN orientation based on relativistic beaming of the radio jet \citep{OrrBrowne82,Zirbel95}. The presence or absence of a MIR thermal component appears to be independent of $L_{\nu,\,1.4\,\mathrm{GHz}}$, $R_c$ as well as redshift but there may be some dependence on black hole mass. No source in the sample with an estimated black hole mass below $7\times 10^{8}\,\mathrm{M}_\odot$ has a MIR thermal component and only two sources (3C~31 and M~87) without a MIR thermal component have black hole masses above that. However, this should be viewed with caution due to the small sample size.

\begin{table}[htp]
\centering
\begin{tabular}{ccccc}
Source & $L_{\nu,\,1.4\,\mathrm{GHz}}$ & $M_\mathrm{BH}$ & $L_\mathrm{Edd}$ & $\log_{10} R_c$ \\
 & $(10^{31}\,\mathrm{erg}\,\mathrm{s}^{-1}\,\mathrm{Hz}^{-1})$ & $(10^8\,\mathrm{M}_\odot)$ & $(10^{47}\,\mathrm{erg}\,\mathrm{s}^{-1})$ & \\
 \hhline{=====}
 NGC 315 & 1.26 & 14.6 & 1.84 & -0.39 \\
 3C 66B & 8.71 & 18.6 & 2.34 & -1.29 \\
 M 84 & 0.224 & 7.30 & 0.920 & -1.18 \\
 3C 270 & 2.51 & 7.75 & 0.976 & -1.44 \\
 \hline
 3C 31 & 3.24 & 9.24 & 1.16 & -1.34 \\
 NGC 541 & 0.871 & 2.03 & 0.226 & - \\
 NGC 3801 & 0.309 & 1.53 & 0.193 & - \\
 NGC 3862 & 5.62 & 4.66 & 0.587 & -1.00 \\
 M 87 & 7.94 & 22.5 & 2.84 & -1.24 \\
 NGC 7052 & 0.110 & 3.62 & 0.456 & -0.22 \\
 \hline
\end{tabular}
\caption{Similar to Table \ref{tab:radioSamp}, we show the $1.4\,\mathrm{GHz}$ radio luminosity density, black hole mass estimate, Eddington luminosity, and logarithm of core-extended brightness ratio (an indicator for AGN orientation) for FR-I AGN in our sample. Here we have divided the sample based on the presence (top) or abscence (bottom) of a MIR thermal component in the best fit for each source.  $L_{\nu,\,1.4\,\mathrm{GHz}}$ by \cite{Condon88},  $M_\mathrm{BH}$ by \cite{Noel07} based on a relation by \cite{Merritt01}, and $\log_{10} R_c$ by \cite{Kharb04}. We notice no dependence of the presence of a thermal MIR component on $L_{\nu,\,1.4\,\mathrm{GHz}}$, $R_c$, or redshift (see Table \ref{tab:sample}). Sources with a thermal MIR component seem to have larger $M_\mathrm{BH}$ as a group however 3C~31 and M~87 do not fit with that distinction and we can think of no clear reason why this should be the case. \label{tab:radio}}
\end{table}


We also compare the best-fit AGN bolometric  luminosity output by the forced-torus clumpyDREAM runs (and NGC~315, see Table \ref{tab:torusparams2}) to bolometric luminosity estimates based on archival $[\mathrm{O\,III}]$ luminosities (see Table \ref{tab:OxyRatio}) using a calibration of 
\begin{equation}
\label{eq:Heckmanbolcalib}
\frac{L_\mathrm{bol}}{L_\mathrm{[O\,III]}}=3500\pm 0.38\,\mathrm{dex}
\end{equation}
by \cite{Heckman04}. This calibration was developed for type 1 AGN so $L_\mathrm{bol}$ is likely to be underestimated in type 2 AGN due to $\mathrm{[O\,III]}$ extinction by dust. We show the comparison in Table \ref{tab:OxyBolRatio} and note we only find $L_\mathrm{clumpy}/L_\mathrm{bol}\gtrsim 1$ in NGC~315. Even considering the large uncertainties on the bolometric luminosity estimate from [O~III] emission only NGC~315, NGC~3801, and NGC~3862 are consistent with $L_\mathrm{clumpy} = L_\mathrm{bol}$ assuming the \cite{Heckman04} calibration holds. This is consistent with our finding that NGC~315 has the best fit of a clumpy torus to its continuum infrared spectral energy distribution. All sources except NGC~315 have lower emission in the IR than would be expected given their [O~III] luminosity. This is consistent with a lack of dusty circumnuclear gas in these LERG FR-Is. \cite{Dicken10} note that low-redshift broad-line radio galaxies show enhanced [O~III] emission and warmer mid- and far-infrared colors than narrow-line radio galaxies at similar redshift but that compares broad-line objects to hidden broad-line objects and therefore says little about narrow-line objects without a hidden broad-line region. 

\begin{table}
\centering
\begin{tabular}{ccccc}
Source& $L_{[\mathrm{O\,III}]}$  & $L_\mathrm{bol}$ & $L_\mathrm{clumpy}$ & \multirow{ 2}{1cm}{$\frac{L_\mathrm{clumpy}}{L_\mathrm{bol}}$} \\
 & ($10^{39}\,\mathrm{erg}\,\mathrm{s}^{-1}$) & $(10^{41}\,\mathrm{erg}\,\mathrm{s}^{-1})$ & $(10^{41}\,\mathrm{erg}\,\mathrm{s}^{-1})$  &  \\
\hhline{=====}
NGC~315*	& $	2.75	 $ & $	96.3	\pm	84.2	 $ & $ {	110.0	}_{	21.0	}^{	34.0	} $ & ${	1.14	}_{	1.14	}^{	1.35	} $ \\
3C~31	& $	2.88	 $ & $	101	\pm	88.2	 $ & $ {	9.1	}_{	2.8	}^{	4.9	} $ & ${	0.090	}_{	0.11	}^{	0.13	} $ \\
NGC~541	& $	0.886	 $ & $	31.0	\pm	27.1	 $ & $ {	4.5	}_{	1.5	}^{	2.8	} $ & ${	0.15	}_{	0.18	}^{	0.22	} $ \\
3C~66B	& $	11.2	 $ & $	392	\pm	343	 $ & $ {	9.1	}_{	1.6	}^{	3.3	} $ & ${	0.023	}_{	0.024	}^{	0.029	} $ \\
NGC~3801	& $	0.146	 $ & $	5.11	\pm	4.47	 $ & $ {	4.0	}_{	2.7	}^{	3.2	} $ & ${	0.78	}_{	1.21	}^{	1.31	} $ \\
NGC~3862	& $	1.58	 $ & $	55.3	\pm	48.4	 $ & $ {	20.0	}_{	18.0	}^{	20.0	} $ & ${	0.36	}_{	0.64	}^{	0.68	} $ \\
3C~270	& $	0.794	 $ & $	27.8	\pm	24.3	 $ & $ {	9.0	}_{	1.8	}^{	2.9	} $ & ${	0.32	}_{	0.35	}^{	0.39	} $ \\
M~84	& $	0.88	 $ & $	30.8	\pm	27.0	 $ & $ {	1.2	}_{	0.6	}^{	0.8	} $ & ${	0.039	}_{	0.052	}^{	0.061	} $ \\
M~87	& $	1.17	 $ & $	41.0	\pm	35.8	 $ & $ {	0.21	}_{	0.20	}^{	0.58	} $ & ${	0.0051	}_{	0.0094	}^{	0.019	} $ \\
NGC~7052	& $	2.75	 $ & $	96.3	\pm	84.2	 $ & $ {	41.0	}_{	6.6	}^{	8.2	} $ & ${	0.4	}_{	0.4	}^{	0.5} $ \\
\hline
\end{tabular}
\caption{Comparison of bolometric luminosity estimate from $\mathrm{[O\,III}]$ emission $L_\mathrm{bol}$ (see Table \ref{tab:OxyRatio}) to the best fit luminosity from forced-torus clumpyDREAM runs (see Table \ref{tab:torusparams2}). *Recall that the best fit to NGC~315 includes a torus component. \label{tab:OxyBolRatio}}
\end{table}

In Table~\ref{tab:eddRatio} we show estimates of the Eddington ration using black hole mass estimates by \cite{Noel07} and our two bolometric luminosity estimates. All of these sources are extremely sub-Eddington with ratios of $<10^{-3.5}$. We do not see a clear distinction between  sources with warm thermal components in their best clumpyDREAM fits and those without. This suggests that Eddington ratio is not the driving factor separating the two modes. 

\begin{table}
\centering
\begin{tabular}{ccc}
Source & $\log_{10}\Big(\frac{L_\mathrm{clumpy}}{L_\mathrm{Edd}}\Big)$ & $\log_{10}\Big(\frac{L_\mathrm{bol}}{L_\mathrm{Edd}}\Big)$\\
\hhline{===}
NGC~315	& $-4.22$ & $-4.28$ \\
3C~31	& $-5.10$ & $-4.06$ \\
NGC~541	& $-4.70$ & $-3.86$ \\
3C~66B	& $-5.41$ & $-3.77$ \\
NGC~3801	& $-4.68$ & $-4.58$ \\
NGC~3862	& $-4.47$ & $-4.02$ \\
3C~270	& $-5.04$ & $-4.54$ \\
M~84 & $-5.88$ & $-4.48$ \\
M~87	 & $-7.13$ & $-4.84$ \\
NGC~7052	 & $-4.05$ & $-3.68$ \\
\hline
\end{tabular}
\caption{Estimated Eddington ratios for each source based on the \cite{Noel07} black hole mass estimates listed in Table~\ref{tab:radio} and the bolometric luminosity estimates listed in Table~\ref{tab:OxyBolRatio}. We do not see a clear distinction between sources with a best-fit warm thermal component and those without.\label{tab:eddRatio}}
\end{table}

\FloatBarrier

\section{Conclusion}
\label{sec:conc}

We used fitting codes based on MCMC algorithms to examine the IR spectra, primarily from \emph{Spitzer}/IRS, of ten nearby FR-I radio galaxies (NGC~315, 3C~31, NGC~541, 3C~66B, NGC~3801, NGC~3862, 3C~270, M~84, M~87, and NGC~7052) in order to determine whether there is significant contribution in the MIR from warm dust. This dust would serve to obscure the central engine and thus explain why AGN with high radio luminosities can resemble quiescent galaxies in optical wavelengths. The leading alternative model for a FR-I is powered by a ADAF which puts the vast majority of its output power into a radio jet.


We see evidence for the IR signature of a warm dusty structure in only four of these ten LERGs. The fact that we do not see evidence for a clumpy torus in six of the LERGs suggests that not only is the central engine different at these low luminosities (e.g., an ADAF), but the circumnuclear absorbing material is also different than in high excitation sources. This difference in nuclear environment is consistent with previous results obtained using different methods (e.g., \citealt{Capetti05,Leipski09,VanderWolk10}). All sources except NGC~315 have lower emission in the IR than would be expected given the observed [O~III] luminosity using the calibration of \cite{Heckman04}. This is also consistent with a lack of dusty circumnuclear gas. The accretion rates onto the central black holes in LERGs are thought to be much less that those in the high excitation sources (e.g., \cite{Baum95,Hardcastle09,3C31OIII,BestHeckman12}). The difference in circumnuclear environment is thought to be related to the much lower accretion rates in the LERGs. Our Eddington ratio estimates do not support this as they are all quite low including the one in which we find evidence of an obscuring torus and all the ones where we found some thermal component.

However, four of the LERGs (NGC~315, 3C~66B, 3C~270, and M~84) have infrared spectral energy distributions consistent with the presence of a significant amount of warm dust (one of which is consistent with a \cite{nenkova08a,nenkova08} clumpy torus), and three of the four (NGC~315, 3C~66B, 3C~270) show evidence for polarized H$\alpha$ emission consistent with a hidden Type 1 nucleus. This is puzzling. Perhaps these are objects in transition, possibly with either growing or declining accretion rate or possibly in a semi-stable intermediate state in a more complex transition. In any case, this suggests that the central engines in LERGs are not yet understood and further work is needed. 

\vfill

\appendix

\acknowledgements
\section{Acknowledgements}
\noindent This research has made use of the NASA/IPAC Extragalactic Database (NED),
which is operated by the Jet Propulsion Laboratory, California Institute of Technology,
under contract with the National Aeronautics and Space Administration.

SB and CO acknowledge support from the Natural Sciences and Engineering Research Council of Canada (NSERC) grant 318679-352700-2000.

This work is based in part on observations made with the Spitzer Space Telescope, which is operated by the Jet Propulsion Laboratory, California Institute of Technology under a contract with NASA.

SPIRE has been developed by a consortium of institutes led by Cardiff University (UK) and including Univ. Lethbridge (Canada); NAOC (China); CEA, LAM (France); IFSI, Univ. Padua (Italy); IAC (Spain); Stockholm Observatory (Sweden); Imperial College London, RAL, UCL-MSSL, UKATC, Univ. Sussex (UK); and Caltech, JPL, NHSC, Univ. Colorado (USA). This development has been supported by national funding agencies: CSA (Canada); NAOC (China); CEA, CNES, CNRS (France); ASI (Italy); MCINN (Spain); SNSB (Sweden); STFC, UKSA (UK); and NASA (USA).

{\it Herschel} is an ESA space observatory with science instruments provided by European-led Principal Investigator consortia and with important participation from NASA.


\pagebreak
\bibliographystyle{apj}
\bibliography{refs}
\end{document}